\documentclass[
11pt, 
english, 
singlespacing, 
headsepline,                  
]{ClassFile}                  

\usepackage[utf8]{inputenc}   
\usepackage[T1]{fontenc}      

\usepackage{lmodern}          
\usepackage[T1]{fontenc}

\usepackage{physics}
\usepackage{amssymb}
\usepackage{siunitx}
\usepackage{mathtools}

\usepackage{mwe}
\usepackage{subfig}       %
\usepackage{makecell}     

\usepackage[backend=bibtex,style=nature,natbib=true]{biblatex} 

\usepackage[autostyle=true]{csquotes} 

\usepackage{subfig}    

\usepackage{titlesec, blindtext, color}
\definecolor{gray75}{gray}{0.75}

\newcommand{\hsp}{\hspace{20pt}}
\titleformat{\chapter}[hang]{\Huge\bfseries}{\thechapter\hsp\textcolor{gray75}{|}\hsp}{0pt}{\Huge\bfseries}

\thesistitle{Dialing back time on Timepix3} 
\supervisor{Dr. James \textsc{Smith}} 
\examiner{} 
\degree{Doctor of Philosophy} 
\author{Robbert Geertsema} 
\addresses{} 

\university{\href{http://www.university.com}{University Name}} 
\department{Nikhef - National Institute for Subatomic Physics} 
\group{Detector Research and Development} 
\faculty{\href{http://faculty.university.com}{Faculty Name}} 

\AtBeginDocument{
\hypersetup{pdftitle=\ttitle} 
\hypersetup{pdfauthor=\authorname} 
\hypersetup{pdfkeywords=\keywordnames} 
}

\usepackage{lineno}

\usepackage{verbatim}   

\begin{document}

\def\subsectionautorefname{Section}     
\def\sectionautorefname{Section}
\def\chapterautorefname{Chapter}
\def\equationautorefname{Eq.}

\frontmatter 

\pagestyle{plain} 


\begin{titlepage}
\begin{center}

\begin{center}
\large
\includegraphics[width=.87\linewidth]{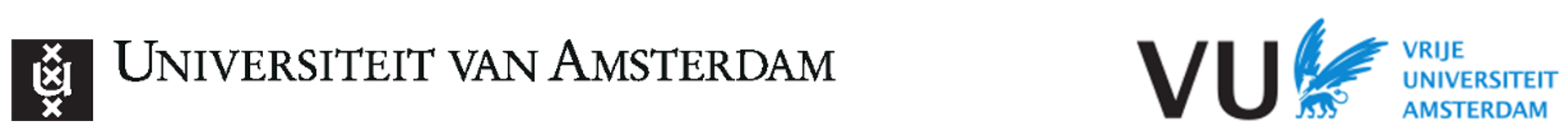}  
\end{center}

\vspace*{.06\textheight}
{\scshape\LARGE MSc Physics and Astronomy\par}\vspace{.2cm} 
{\scshape\Large Advanced Matter and Energy Physics \par}\vspace{1.2cm}
\textsc{\Large Master Thesis}\\[0.8cm] 

\HRule \\[0.4cm] 
{\huge \bfseries \ttitle \par}\vspace{0.2cm} 
{\Large  A study on the timing performance of Timepix3 \par}\vspace{0.4cm}
\HRule \\[1.5cm] 
 
\textit{by}\\[0.2cm]

Robbert Geertsema \\
10757805\\[0.4cm]
60 EC\\
September 2018 - July 2019\\[1.5cm]
 
\begin{minipage}[t]{0.4\textwidth}
\begin{flushleft} \large
\emph{Daily Supervisor:}\\
dr. D. Hynds 
\end{flushleft}
\end{minipage}
\begin{minipage}[t]{0.4\textwidth}
\begin{flushright} \large
\emph{Examiner:} \\
dr. H.L. Snoek\\ 
\emph{Second assessor:} \\
dr. M. Vreeswijk

\end{flushright}
\end{minipage}\\[3cm]
 
\vfill


Detector Research and Development \\[.2cm]
 
\includegraphics[scale=.3]{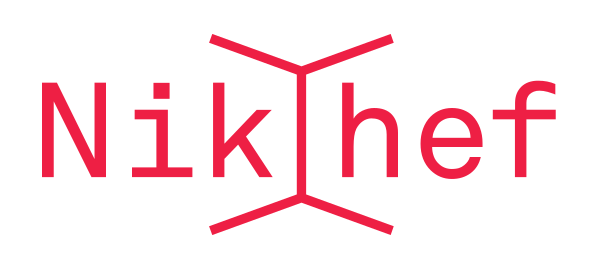}

\vfill

\vfill
\end{center}
\end{titlepage}

\hypersetup{allcolors=.}


\begin{abstract}
\addchaptertocentry{\abstractname} 

For the advancement of the understanding of timing systematics in pixelized readout chips and for the benefit of future fast timing detectors to aid 4D tracking technology in the HL-LHC, we have performed detailed studies of the timing properties of the Timepix3, a hybrid pixel detector developed by the Medipix collaboration. These studies use three different measurement techniques to investigate the timing systematics of this detector: test pulses, testbeam, and a laser setup that we build for this work. The average delay over the pixel matrix is determined with the testbeam and the laser and shows the same structure for different Timepix3 chips. The difference in delay results in a maximum difference over the pixel matrix of around \SI{4}{\nano\second}, which is large compared to the time bins of \SI{1.56}{\nano\second}. By correcting for this difference on a per-pixel level, the time resolution is on average improved with \SI{159}{\pico\second} depending on the sensor. The best time resolution that is achieved after a timewalk correction and a correction for the difference in the average delay per pixel is $686.4\pm0.2$ \si{\pico\second} for a single Timepix3 chip with a \SI{200}{\micro\meter} planar silicon sensor, compared to a time resolution of $892.9\pm0.3$ \si{\pico\second} without any correction. However, this improved time resolution is not yet the naively expected time resolution of \SI{451}{\pico\second}. The origin of this delay is determined with this laser setup, and is due to a combination of signal propagation and a difference of the start-up time of the fast oscillator within the pixels. This work indicates that it is vital for next generation pixel detectors to correct for these systematic timing effects in order to reach a better time resolution to aid 4D tracking technology in the HL-LHC.

\end{abstract}

\tableofcontents 




\begin{abbreviations}{ll} 

\textbf{ASIC} & \textbf{A}pplication \textbf{S}pecific \textbf{I}ntegrated \textbf{C}ircuit\\
\textbf{CMOS} & \textbf{C}omplementary \textbf{M}etal \textbf{O}xide \textbf{S}emiconductor\\
\textbf{FBG} & \textbf{F}iber \textbf{B}ragg \textbf{G}rating\\
\textbf{fToA} & \textbf{f}ine \textbf{T}ime-\textbf{o}f-\textbf{A}rrival\\
\textbf{Laser} & \textbf{L}ight \textbf{a}mplification by \textbf{s}timulated \textbf{e}mission of \textbf{r}adiation\\
\textbf{LHC} & \textbf{L}arge \textbf{H}adron \textbf{C}ollider\\
\textbf{MIP} & \textbf{M}inimum \textbf{I}onizing \textbf{P}article\\
\textbf{NA} & \textbf{N}umerical \textbf{A}perture\\
\textbf{PLL} & \textbf{P}hase-\textbf{L}ocked \textbf{L}oop\\
\textbf{PnR} & \textbf{P}lace a\textbf{n}d \textbf{R}oute\\
\textbf{SPIDR} & \textbf{S}peedy \textbf{PI}xel \textbf{D}etector \textbf{R}eadout\\
\textbf{TDC} & \textbf{T}ime-to-\textbf{D}igital \textbf{C}onverter\\
\textbf{ToA} & \textbf{T}ime-\textbf{o}f-\textbf{A}rrival\\
\textbf{ToT} & \textbf{T}ime-\textbf{o}ver-\textbf{T}hreshold\\
\textbf{TPX3} & \textbf{T}ime\textbf{p}i\textbf{x}\textbf{3}\\
\textbf{TSV} & \textbf{T}hrough-\textbf{S}ilicon \textbf{V}ia\\
\textbf{SPS} & \textbf{S}uper \textbf{P}roton \textbf{S}ynchrotron\\
\textbf{VCO} & \textbf{V}oltage-\textbf{C}ontrolled \textbf{O}scillator\\

\end{abbreviations}


\mainmatter 

\pagestyle{thesis} 

\chapter{Introduction} 
\label{sec:Introduction}

Particle detectors are becoming increasingly important in both science and medicine. The development of particle detectors has been driven by particle physics since the origin of these detectors, which dates back to the discovery of X-rays using a photographic plate in 1896 by W.C. R\"ontgen \citep{rontgen1896new}. The application of particle detectors, in this case photographic plates, started directly after the discovery of R\"ontgen. The use of these plates was initially limited to different types of pathology such as bone fractures.

In 1940 R.S. Ohl \citep{ohl1946light} discovered that a silicon crystal he was working on changed current considerably when exposed to bright light. This discovery proved to be the foundation of the p-n junction, and subsequently the photovoltaic cell \citep{perlin1999space}. However, the application of this discovery in particle detectors was not until the 1970s. At this point it became a necessity for particle physics experiments to have a high spatial resolution which could not be fulfilled by the particle detectors then in existence. Bubble chambers and emulsions were able to meet the desired spatial resolution, though the data taking procedures, especially analysis rates, were too laborious \citep{hall1994semiconductor}. At this point silicon particle detectors proved superior to the previous detection mechanisms, largely because of their spatial resolution and their reliability. 

From that point on silicon particle detectors where further developed and pixelated silicon particle detectors emerged. These pixel detectors proved especially useful in a range of applications. They are used in medicine for the traditional x-ray images and CT scans, as well as in applications such as ion mass spectrometry \citep{jungmann2011high}. Besides their use in medicine, they are also used in a wide range of scientific applications such as x-ray diffraction imaging \citep{lodola2017pixelated} and electron diffraction imaging \citep{van2016ab}. Improvement to these detectors are at present driven mainly by particle physics, while these other applications benefit from the new technologies developed. 

\begin{figure}
\centering
\captionsetup{width=.75\linewidth}
\includegraphics[width=.45\linewidth]{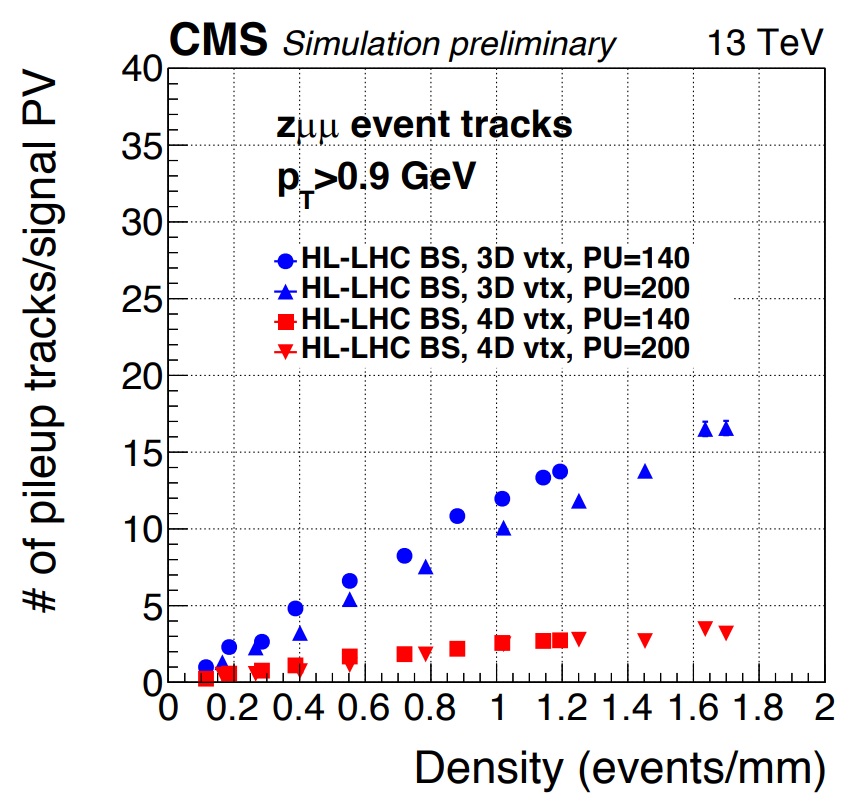}
\captionof{figure}{A simulation from the CMS collaboration of the number of pileup events per signal primairy vertex (PV). Two different scenarios are indicated (PU of 140 or 200) with two methods of tracking: 3D (in space) and 4D (in space and time, with a time resolution of \SI{30}{\pico\second}). This Figure is taken from \citep{Gray2017}.}
\label{fig:Introduction/CMSSimulation}
\end{figure}

At the Large Hadron Collider (LHC) at Organisation Européenne pour la Recher\-che Nucléaire (CERN) proton-proton collisions are used to probe the physics of the standard model of particle physics. One of the plans of the LHC in the comming years is to increase the luminosity such that the average number of events per bunch crossing increases from around 50 to around 150-200 \citep{Sadrozinski_2017} in 2026 when the high luminosity LHC \citep{apollinari2015high} will start operating. Current simulations from CMS show that, with a vertex separation resolution of 250-300 \si{\micro\meter} and reconstruction methods that solely rely on spatial information (3D tracking), up to 16 pileup tracks are associated to the signal primary vertex \citep{Gray2017}. This may impact physics analysis significantly. The results from this simulation are shown in \autoref{fig:Introduction/CMSSimulation}. The number of misassociation pileup tracks to the primary vertex can be decreased using timing information in the tracking, and thus using reconstruction methods that rely on spatial information as well as time information (4D tracking). The same simulations from CMS indicate that with a time resolution of \SI{30}{\pico\second} per track, the number of pileup tracks per signal primary vertex decreases to \textasciitilde3 at the highest event density, resulting in an acceptable number of pileup tracks per signal primary vertex. Multiple layers of silicon pixel detectors are used in constructing the tracks from these vertices. Therefore it is possible to reach the desired \SI{30}{\pico\second} time resolution by a time measurement at multiple planes that have a decreased time resolution. If a simple system of four pixel detectors layers is assumed, the required time resolution decreases to \SI{60}{\pico\second} (naively requiring a time bin of \SI{208}{\pico\second}) per plane in order to achieve a final time resolution of \SI{30}{\pico\second}. It should be noted that it is envisioned to achieve this timing precision while keeping accurate spatial resolution, hence a small pixel area at the same time.

To achieve this time resolution of around \SI{60}{\pico\second}, there is an ongoing drive to increase the timing performance of current silicon particle detectors to cope with this new requirement. However, the systematic effects of these detectors are progressively determining what can be achieved with them. One of the problems is that systematic variations in the electronics are now almost at the same level as the timing precision of the detector as a whole. As a result it is starting to become important to quantify and correct for these systematic variations in order to push the next generation of silicon pixel detectors to their limits.

\begin{figure}
\centering
\captionsetup{width=.75\linewidth}
\includegraphics[width=.5\linewidth]{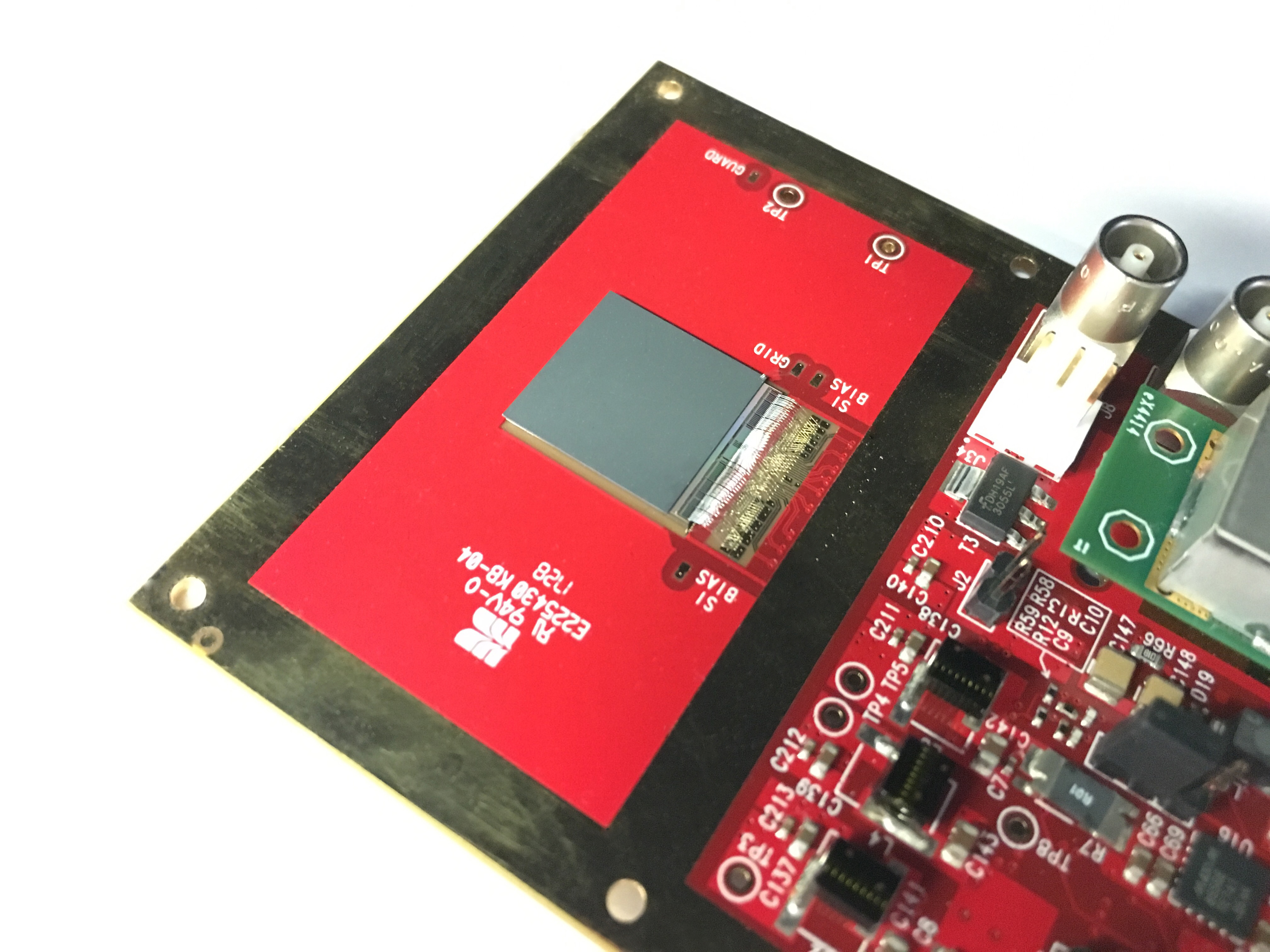}
\captionof{figure}{A picture of a Timepix3 pixel detector. The square in the middle of the picture is the Timepix3 chip measuring \SI{1.4}{\centi\meter}$\times$\SI{1.4}{\centi\meter}.}
\label{fig:Introduction/Timepix3Picture}
\end{figure}

One of the pixelated ASICs that has been developed by this drive from particle physics is the Timepix3 \citep{poikela2014timepix3} (shown in \autoref{fig:Introduction/Timepix3Picture}). The Timepix3 provides a pitch of \SI{55}{\micro \meter} as well as a time measurement with a precision of \SI{1.56}{\nano\second}, resulting in a maximum time resolution of \SI{451}{\pico\second}. This precise time resolution has not yet been achieved with Timepix3. This is partly due to systematic variations in timing precision over the pixel matrix of the ASIC. An improvement of the time resolution can be achieved by a detailed characterisation of the time delay structures in the pixel matrix of Timepix3.

\section*{The characterisation of time structures in the Timepix3}

There are different methods to characterise a silicon pixel detector. These methods range from simulations of the electronics to intricate testbeam facilities. The latter is one of the three methods that are generally used to characterise existing silicon pixel detectors, while simulations are generally performed in the design stage of a pixel detector and are not used to correct for systematic offsets in detectors. Three of these methods are used in this work to characterise the Timepix3: test pulses, testbeam facility, and a laser setup. 

The Timepix3 ASIC has the ability to inject a charge within its own electronic front-end circuitry, via a process named test pulses. This way the behaviour of the ASIC can be investigated for a range of charges. However, the method to inject this charge partly relies on the same electronics that are used to measure the signal and thus the results from the test pulse studies could be biased by this limitation. Therefore, test pulses are generally not used to investigate timing related aspects of the Timepix3 ASIC.

The testbeam facility at CERN has been used to characterise several detectors \citep{buchanan2017lhcb} for the VELO upgrade at LHCb \citep{Collaboration:1624070}. This testbeam facility utilizes charged particles to investigate several aspect of a detector. 
A particle accelerator is used to generate a large number of particles with a high momentum. Besides the characterisation studies for the VELO upgrade, this method has already been employed several times in the past to investigate other detectors \citep{akiba2019arxiv, pitters2018time}, and is carried out at the Super Proton Synchrotron (SPS) at CERN.

\begin{figure}
\centering
\captionsetup{width=.95\linewidth}
\includegraphics[width=.95\linewidth]{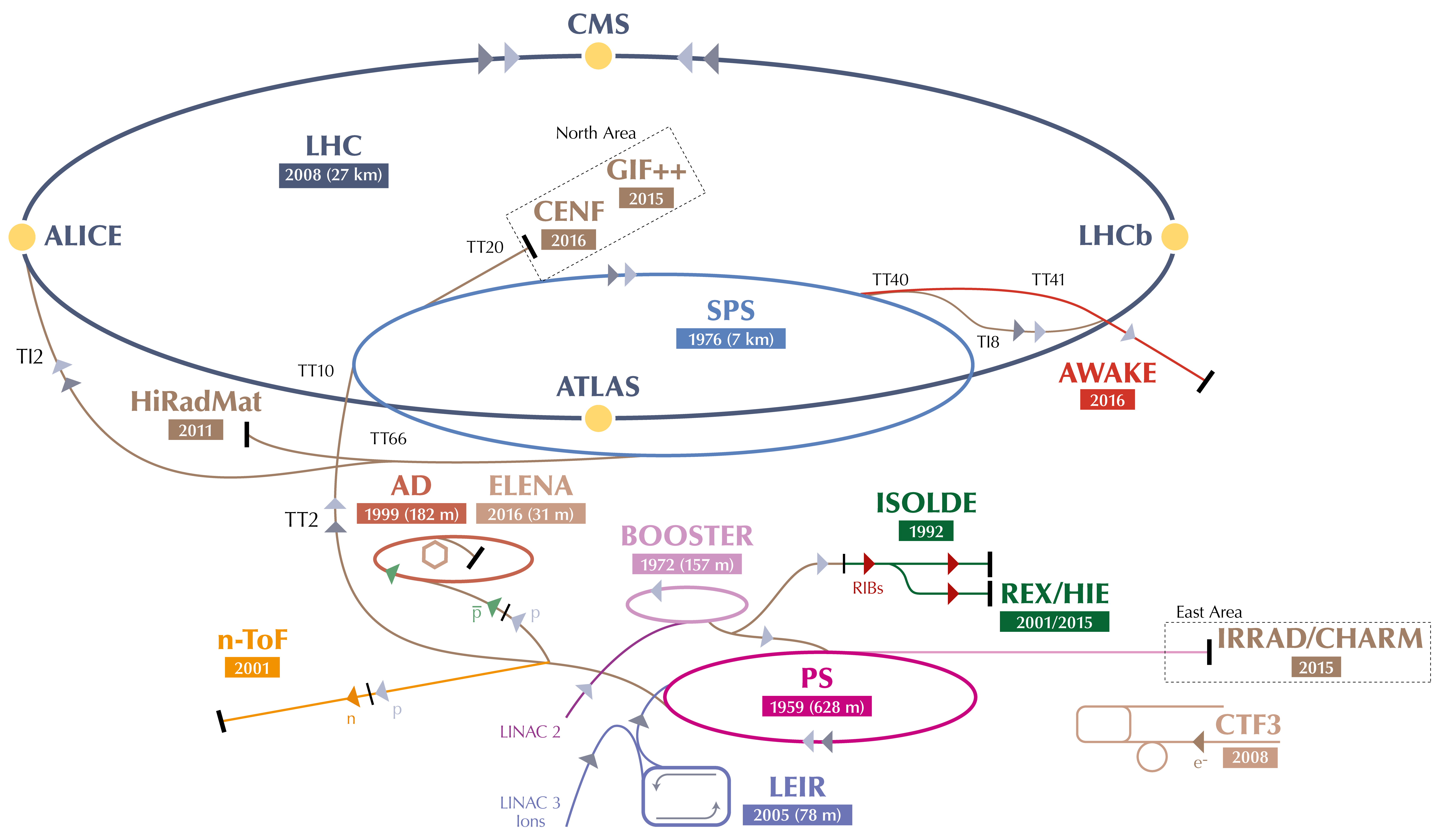}
\captionof{figure}{The accelerator complex at CERN. Figure taken from \citep{Mobs:2197559}.}
\label{fig:Introduction/CERN}
\end{figure}

At CERN, located at the border between France and Switzerland near Geneva, multiple particle accelerators are used to accelerate protons up an energy of \SI{6.5}{\tera\eV}, giving a centre-of-mass energy of \SI{13}{\tera\eV} for a single proton-proton collision. The various accelerators present at CERN are shown in \autoref{fig:Introduction/CERN}. The LHC itself will accelerate the protons from an energy of \SI{450}{\giga \eV} to \SI{6.5}{\tera \eV}. To achieve the starting energy of \SI{450}{\giga \eV}, a different accelerator is required that injects the protons at this energy in the LHC. This acceleration is primarily achieved by the SPS, which is capable of an acceleration from \SI{26}{\giga \eV} to \SI{450}{\giga \eV}, and has previously been used as the main accelerator for particle physics experiments before the construction of the LHC was finished in 2008. Besides the role as an injector for the LHC, the SPS is still used as an accelerator that provides a proton beam for fixed-target experiments located around it. These experiments include among others COMPASS, NA61/SHINE and NA62. There are four targets located at the North-Area of CERN: T2, T4, T6, and T10. Of these four targets, only three (T2, T4, and T6) are directly connected to the SPS, while T10 is located behind T6. T2 and T4 are used to provide a beam to the North-Area testbeam facility, which is the location where the measurements presented in \autoref{sec:Results} were taken. T2 and T4 produce in total four secondary beam lines that are directed to the testbeam facilities: H2, H4, H6, and H8. The momentum range of the secondary beam at H2, H4, and H8 is \SI{10}{\giga \eV} to \SI{400}{\giga \eV}, with the possibility of a primary proton beam at an energy of \SI{400}{\giga \eV}. H6 operates in the range of \SI{5}{\giga \eV} to \SI{205}{\giga \eV}, without the possibility of a primary proton beam.

A setup (Timepix3 telescope \citep{akiba2019arxiv}) located at beam line H8 is developed to provide both position and timing information of the individual particles traversing the beam line. Using this telescope the pixel matrix of the Timepix3 can be investigated. The telescope consists of specialised equipment and uses an entire accelerator system to generate the particles that are used to characterise the Timepix3.

Another method to investigate and characterise the Timepix3 is to use photons generated by a pulsed laser diode to generate hits on the Timepix3. Such a method requires little specialized equipment and does not rely on the use of an accelerator complex. Therefore such a method can easily be realised as a \textit{tabletop} experiment in any lab. Another advantage of such a setup compared to the use of a testbeam is that the time of arrival and intensity of the photons can easily be adjusted to the specifications that are needed for the specific application. Such a freedom can even be achieved within a setup by adjusting the drive sequence of the laser diode that is use to generate the photons. Previous experiments have already successfully used such a setup to investigate the behaviour of silicon detectors. Examples of this are the End-cap semiconductor tracker of ATLAS, which has been tested using a laser \citep{dolevzal2007laser}, and the characterisation of silicon microstrip detectors \citep{abt1999characterization} using an infrared laser.

These three different methods to investigate the systematics of the Timepix3 ASIC will be discussed and the measurements of these methods will be compared to each other alongside an overview of the origin of these systematics.

\section*{Thesis Outline}
This Thesis studies the systematic effects of the Timepix3 using three different methods, using test pulses, using the data taken at a testbeam setup, and using a laser. In \autoref{Theory} the timing aspects of silicon detectors are explained. In \autoref{Timepix3Chapter}, the Timepix3 ASIC is discussed in detail and various aspects of the ASIC are discussed. After this Chapter, the Timepix3 telescope is discussed which is used during the testbeam to provide position and timing information in order to analyse the Timepix3 results. \autoref{sec:LaserSetup} discusses the laser setup that we build to investigate the timing of Timepix3.  \autoref{sec:Results} shows the results obtained from both the testbeam and the laser setup. In this Chapter the various observations and measurements from these two setups are discussed, as well as a comparison of the measurements from the two setups. Finally, in \autoref{sec:Conclusion} the results are summarized and the different aspects that can be improved further are discussed.

\chapter[Timing aspects of silicon detectors]{Timing aspects of \\silicon detectors}

\label{Theory}

A silicon hybrid pixel detector relies on a few basic principles in order to detect hits. A charged particle or a photon traversing the silicon detector can liberate charge carriers within a silicon crystal connected to an ASIC. The principle of creating charge carriers within a semiconductor are discussed in \autoref{sec:Theory/CreationCharge}. After the creation of the charge carriers in the silicon, these charge carriers need to travel towards the implant within the silicon such that the charge can be collected by the ASIC. This movement of charge carriers within the silicon is discussed in \autoref{sec:Theory/pnJunction}, the drift time of the charge carriers is discussed in \autoref{Theory/EField}, and in \autoref{sec:Theory/DiffusionChargeCloud} the lateral drift of the charge cloud in the silicon is discussed. After the charge carriers reach the ASIC, the signal will be amplified and a hit is detected as soon as this amplified signal will reach threshold. However, it takes time to reach threshold, and this time depends on the charge of the initial signal. This effect is called timewalk and is discussed in a later Chapter (\autoref{sec:Timepix3/Timewalk}). As soon as this signal crosses threshold the time measurement of the hits starts. 

The time resolution of the detector depends on both the time resolution of the sensor, as well as the time resolution of the ASIC. Therefore it is important to understand the timing aspects of the silicon sensor in order to deconvolute these timing aspects from the timing aspects of the ASIC. The time resolution from the ASIC is discussed later on (\autoref{sec:Timepix3/ToAToT}). 

Both timewalk and the time resolution of the ASIC depend on the charge of the signal. Therefore, in the next Section the charge profile generated by both a charged particle and photons is discussed.

\section{Energy deposition in matter}
\label{sec:Theory/CreationCharge}

Charged particles and photons are able to liberate electron-hole pairs within certain materials. The number of these electron-hole pairs that they are able to liberate are crucial in order to understand both the lateral drift as well as the time it takes for the signal to reach threshold. Therefore in this Section the energy spectrum that is created by both charged particles and photons is discussed.

All solids can be classified in three different groups by their electronic band structure: metals, semiconductors, and insulators \citep{hook2013solid}. Of these three different materials, the bandgap of semiconductors is the only one which is in the energy range of thermal excitations ($k_b T$). Therefore, in semiconductors, electrons are easily excited from the valance to the conduction band by thermal excitations or photons, and thus makes semiconductors suitable for the detection of charged particles and photons. However, pure semiconductors are rarely used as sensor for solid state detectors, but are first doped to create extra energy states near either the valance or conduction band \citep{hook2013solid}.


\begin{figure}
\centering
\begin{minipage}{.47\textwidth}
\centering
\includegraphics[width=1\linewidth]{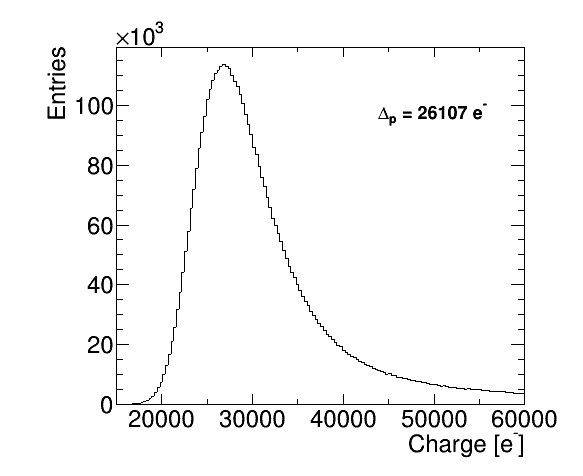}
\end{minipage}\qquad
\begin{minipage}{.47\textwidth}
\centering
\includegraphics[width=1\linewidth]{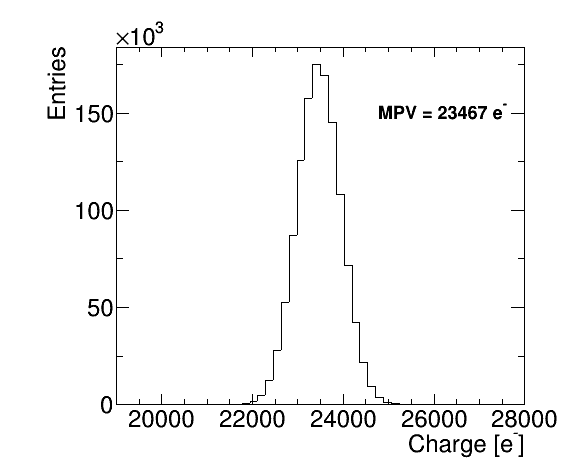}
\end{minipage}


\begin{minipage}[t]{.47\textwidth}
\centering
\captionsetup{width=.9\linewidth}
\captionof{figure}{A typical charge deposition profile of MIP's for a \SI{300}{\micro\meter} silicon sensor. A fit to \textit{Langaus} distribution is used to determine $\Delta_p$. The value for of $\Delta_p$ is determined to be 26107 e$^{-}$.}
\label{fig:Theory/Languas300umPlane1}
\end{minipage}\qquad
\begin{minipage}[t]{.47\textwidth}
\centering
\captionsetup{width=.9\linewidth}
\caption{A typical charge deposition profile of the \SI{680}{\nano\meter} laser for a measurement with a \SI{200}{\micro\meter} silicon sensor. A fit to a normal distribution is used to determine the MPV. The MPV value for this spectrum is determined to be 23467 e$^{-}$.}
\label{fig:Theory/MPVLaserEnergy}
\end{minipage}
\end{figure}

The charge within a silicon sensor is usually induced by a particle, however this charge can also be induced by optical photons. The difference between these two processes and the resulting charge spectrum is important to understand effects that are different for different number of charges, such as timewalk (see \autoref{sec:Timepix3/Timewalk}). First the charge spectrum of a charged particles is discussed after which the charge spectrum of a pulse of optical photons is discussed.

When a minimum ionizing particle (MIP) traverses a doped silicon crystal it deposits some of its energy within the crystal by exciting an electrons from the valance band to the conduction band; creating an electron-hole pairs. The number of generated electron-hole pairs during an interaction with a MIP does not follow a normal distribution. To understand the energy spectrum of MIP's in silicon, one first needs to understand the energy deposited by a MIP in a silicon sensor. The rate of energy loss (in \si{\mega\eV \gram^{-1} \centi\meter \squared}) of a MIP in the range $0.1\lesssim \beta\gamma \lesssim 1000$ is well described by \citep{bichsel2004passage}
\begin{equation}
	-\eval{\dv{E}{x}}_{T < T_{cut}}= K z^2 \frac{Z}{A}\frac{1}{\beta^2} \left[ \frac{1}{2} \ln(\frac{2 m_e c^2 \beta^2 \gamma^2 T_{cut}}{I^2}) - \frac{\beta^2}{2} \left( 1+\frac{T_{cut}}{T_{max}} \right) -\frac{\delta}{2}\right].
	\label{eq:Theory/bichel}
\end{equation}
Here $I$ is the mean excitation energy, Z is the atomic number of the material, $\delta$ is the density effect correction, and $T_{max}$ is the maximum kinetic energy that can be given to a free electron in a single collision. $T_{max}$ is given by \citep{bichsel2004passage} 
\begin{equation}
    T_{max} = \frac{2 m_e c^2 \beta^2 \gamma^2}{1+\frac{2\gamma m_e}{M}+\left( \frac{m_e}{M} \right)^2}.
\end{equation}
\autoref{eq:Theory/bichel} also assumes a limit on the energy transfer of $T\leq T_{cut}\leq T_{max}$.

The energy loss probability function $f(\Delta ;\beta\gamma,x)$ associated to this energy loss function is described by a highly-skewed Landau-Vavilov distribution \citep{vavilov1957ionization}. This function can also be written as a convolution of a Landau distribution with a gaussian distribution. Such a convolution is known as a \textit{Langaus}. The most probable total energy loss in the detector is given by \citep{bichsel1988straggling}
\begin{equation}
	\Delta_{p}=\xi\left[ \ln(\frac{2 m c^2 \beta^2 \gamma^2}{I}) + \ln(\frac{\xi}{I}) + j - \beta^2 -\delta\left( \beta \gamma \right) \right].
\end{equation}
For a detector with a thickness $x$, $\xi$ is given by $\frac{K}{2} \left\langle \frac{Z}{A} \right \rangle \left(\frac{x}{\beta^2}\right)$ (in \si{\mega\eV}), and $j=0.200$ \citep{vavilov1957ionization}. The distribution $f(\Delta ;\beta\gamma,x)$ of a \SI{180}{\giga\eV} positively charged pion beam traversing a silicon detector with a width of \SI{300}{\micro\meter} is shown in \autoref{fig:Theory/Languas300umPlane1}. The corresponding value for $\Delta_p$ is shown in this figure. A detailed explanation of the processes that occur and the corresponding formulas for the passage of particles through matter can be found in \citep{bichsel2004passage}.

The shape and position of this charge distribution gives an indicating of the range of charges that will be processed by the ASIC, and can thus give an indication of the origin of effects within the ASIC such as timewalk (see \autoref{sec:Timepix3/Timewalk}).

\begin{figure}
\centering
\includegraphics[width=.75\linewidth]{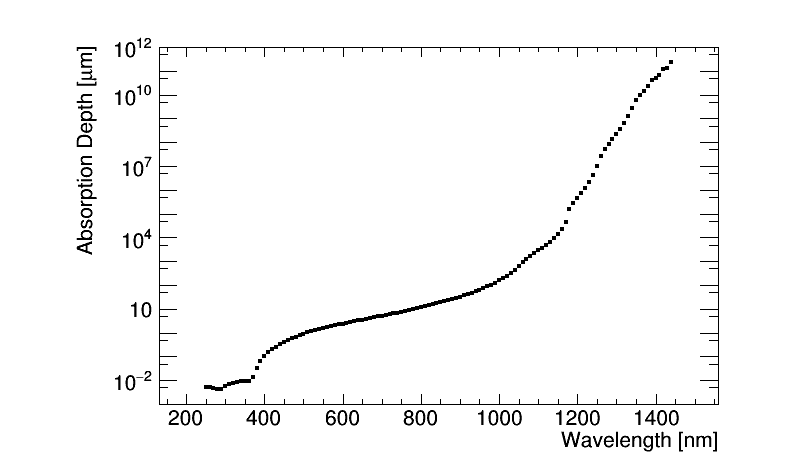}
\captionof{figure}{The absorption depth of optical photons with various wavelengths in the (near-)optical range. The absorption coefficients to calculate the absorption depth are taken from \citep{green1995optical}.}
\label{fig:Theory/SiAbsDepth}
\end{figure}

Besides MIPs, solid state detectors with a doped silicon crystal can also measure optical photons. However, this process is slightly different from that of MIPs. The main difference is that a single optical photon can only excite a single electron-hole pair and thus a single optical photon cannot be detected using traditional silicon sensors. Due to the lower energy of optical photons compared to MIPs, an optical photon will generally not traverse the complete depth of the sensor. The absorption depth $\lambda_{abs}$ of optical photons in silicon is calculated by \citet{green1995optical}, and is shown in \autoref{fig:Theory/SiAbsDepth}. The absorption depth is defined as the depth at which the intensity has dropped to a factor $1/e$ of the original intensity. The bandgap energy of silicon is \SI{1.12}{\eV}, which is the same energy as a photon with a wavelength of \SI{1107}{\nano\meter}. Photons with a wavelength of more than \SI{1107}{\nano\meter} are therefore less likely to be absorbed by the silicon. This explains the rapidly increasing absorption depth at higher wavelengths.

At each depth in the silicon, the remaining light has a chance to excite an electron-hole pair. Therefore, a certain fraction of the intensity of the light will be converted to electron-hole pairs. Note that this fraction only include the photons that are transmitted on the silicon-air interface, and does not include the photons that are reflected at this interface. Due to these effects, the number of electrons that are generated follow a normal distribution centered around the most probable value (MPV). An example of an charge spectrum created by \SI{680}{\nano\meter} photons is shown in \autoref{fig:Theory/MPVLaserEnergy}. The MPV is shown in the top right corner of the Figure.

\section{P-N junction}
\label{sec:Theory/pnJunction}
By combining a p-type doped region with an n-type doped region, a p-n junction is formed. In a p-n junction the electrons will diffuse towards the p-region, while the holes will diffuse towards the n-region due to electrostatic repulsion. The formation of these two space charge regions creates an electric field within the p-n junction which counteracts this diffusion of the charge carriers. The presence of this electric field creates a region which is free of any mobile charge carriers, which is called the depletion region \citep{hook2013solid}. Only in this depletion region the charge carriers excited by a particle can be detected. Besides the depth of the depletion region, it is also important to understand the electric field that is created in the p-n junction in order to understand the collection time of charge carriers. Therefore, this region is an important part of the solid state detector. The depletion region is discussed in detail in this Section.

\begin{figure}
\centering
\begin{minipage}{.47\textwidth}
\centering
\includegraphics[width=1\linewidth]{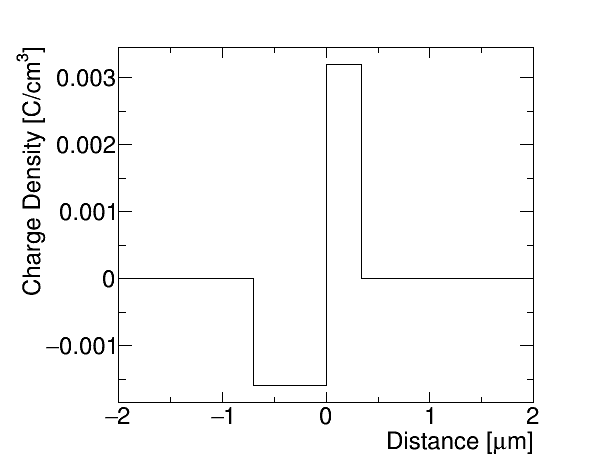}
\end{minipage}\qquad
\begin{minipage}{.47\textwidth}
\centering
\includegraphics[width=1\linewidth]{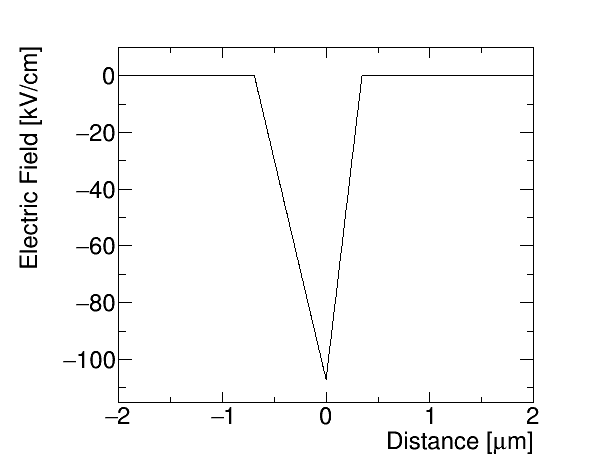}
\end{minipage}


\begin{minipage}[t]{.47\textwidth}
\centering
\captionsetup{width=.9\linewidth}
\captionof{figure}{The charge density profile along the depth of a semiconductor sensor with $N_a=10^{16} \si{cm^{-3}}$, $N_d=2\cdot10^{16} \si{cm^{-3}}$, and a bias voltage of \SI{25}{\volt}. These are approximately the same values as for the silicon sensors used in this work.}
\label{fig:Theory/ChargeDensity}
\end{minipage}\qquad
\begin{minipage}[t]{.47\textwidth}
\centering
\captionsetup{width=.9\linewidth}
\caption{The electric field profile along the depth of a semiconductor sensor with $N_a=10^{16} \si{cm^{-3}}$, $N_d=2\cdot10^{16} \si{cm^{-3}}$, and a bias voltage of \SI{25}{\volt}. These are approximately the same values as for the silicon sensors used in this work.}
\label{fig:Theory/ElectricField}
\end{minipage}
\end{figure}

For the calculations in this Section an abrupt doping profile is assumed. In reality this doping profile is not infinitely sharp, however in many cases it can be approximated by an abrupt doping profile \citep{hook2013solid}. This doping profile assumes a constant acceptor density $N_a$ in the region $-x_p \leq x \leq 0$, and a constant donor density $N_d$ in the region $0 \leq x \leq x_n$. An example of a charge profile that is used in this Section is shown in \autoref{fig:Theory/ChargeDensity}. Both $N_a$ and $N_d$ have been chosen such that they are approximately the values for a doped silicon sensor.

Assuming that the doping concentration for a doped silicon crystal is much higher than the electron and hole concentration ($p-n+N_d-N_a\approx N_d-N_a$). The electrostatic potential $V$ can be expressed as \citep{tsopelas2016silicon}:

\begin{equation}
    V(x)=
    \begin{cases}
        \begin{aligned}
            &V_p+q\frac{N_a}{2\epsilon_0\epsilon_{Si}}\left( x+x_p \right)^2 & -x_p\leq &x \leq 0\\
            &V_n-q\frac{N_d}{2\epsilon_0\epsilon_{Si}}\left( x-x_n \right)^2 & 0\leq &x \leq x_n
        \end{aligned}
    \end{cases}.
    \label{eq:Theory/potential2}
\end{equation}

The electric field inside the p-n junction is given by the first derivative of the electrostatic potential with respect to $x$, and is given by \citep{green2000physics}
\begin{equation}
    E(x)=
    \begin{cases}
        \begin{aligned}
            &-\frac{qN_a}{\epsilon_0\epsilon_{Si}}\left( x+x_p \right) & -x_p\leq &x \leq 0\\
            &\frac{qN_a}{\epsilon_0\epsilon_{Si}}\left( x-x_n \right) & 0\leq &x \leq x_n
        \end{aligned}
    \end{cases}.
    \label{eq:Theory/Efield2}
\end{equation}
The electric field for a p-n junction with a voltage of $25V$ is shown in \autoref{fig:Theory/ElectricField}. For both $N_a$ and $N_d$ in this Figure, a value has been chosen that is approximately the same as in a doped silicon sensor.

Due to the p- and n-type regions in the junction there is a build-in voltage $V_{bi}$ between these two regions which defined as the difference between $V_p$ and $V_n$ when there is no external bias voltage applied to the junction. This build-in voltage is given by \citep{spieler2005semiconductor}

\begin{equation}
    V_{bi}=\frac{kT}{q}\ln(\frac{N_aN_d}{n_i^2}),
    \label{eq:Theory/BuildIn}
\end{equation}
where the intrinsic carrier concentration $n_i$ is given by \citep{misiakos1993accurate}

\begin{equation}
    n_i\left( T \right) = 5.28\cdot10^{19} \left( \frac{T}{300} \right)^{2.54} \exp(\frac{-6726}{T}).
\end{equation}

The depth $w$ of the depletion region is given by the sum of $x_p$ and $x_n$. This depth can be expressed as a function of the build-in voltage \citep{spieler2005semiconductor}

\begin{equation}
    w=x_n+x_p=\sqrt{\frac{2\epsilon_0\epsilon_{Si}V_{bi}}{q}\left( \frac{1}{N_a} + \frac{1}{N_d} \right)}
\end{equation}

When an external bias $V_{bias}$ is applied to the p-n junction, the potential inside the junction is increased by $V_{bias}$. Therefore $V_{bi}\xrightarrow{}V_{bi}+V_{bias}$. In practise however, this built-in voltage is small compared to the applied voltage, and is thus often neglected. Therefore the depth of the depletion region can be expressed as a function of $V_{bias}$

\begin{equation}
    w=\sqrt{\frac{2\epsilon_0\epsilon_{Si}\left(V_{bi}+V_{bias}\right)}{q}\left( \frac{1}{N_a} + \frac{1}{N_d} \right)}
    \approx \sqrt{\frac{2\epsilon_0\epsilon_{Si}V_{bias}}{q}\left( \frac{1}{N_a} + \frac{1}{N_d} \right)}.
    \label{eq:Theory/DepletionWidth}
\end{equation}

If a charged particle traverses this depletion region, the liberated electrons will drift to n-side of the junction, and the liberated holes will drift to the p-side \citep{lutz1999semiconductor}. These charge carriers are responsible for inducing the signal that is measured. Note that this drift only happens in the depleted region, and not in the un-depleted region due to the absence of an electric field. Therefore, depending on the type of sensor (n-in-p or p-in-n), the depletion region might not extend for enough to reach the implants (which collect the charge carriers), and thus no signal can be collected.

\section{Effect of the electric field}
\label{Theory/EField}

As discussed in the previous Section, charged particles can only be detected in the depleted volume of the silicon. To increase the depleted region of the silicon p-n junction, an external bias voltage $V_{bias}$ can be applied, and thus increase the total amount of charge carriers collected from a charged particle. The bias voltage at which full depletion is reached is given by \citep{tsopelas2016silicon}

\begin{equation}
    V_{fd}=\frac{qN_a w^2}{2\epsilon}-V_{bi}\ \ \ ;\ \ \ \epsilon\equiv\epsilon_0\epsilon_{Si}.
\end{equation}

Two different situations can be distinguished: partial and full depletion. At partial depletion, the electrons will drift towards the collection pad until they are outside the depletion region where $E=0$. At partial depletion the electric field within the p-n junction is given by \citep{spieler2005semiconductor}

\begin{equation}
    E(x)=E_0\left( 1-\frac{x}{w} \right)\ \ \ ;\ \ \ E_0=\frac{2V}{w}.
\end{equation}

For full depletion there is a non-zero electric field throughout the depth of the detector. After the full-depletion is reached, the additional voltage that is applied, will linearly increase the overall electric field in the detector. The electric field for full-depletion consists of two parts $E_d$, and $E_1$. $E_d$ is the electric field that is build up due to the voltage that is required to reach full-depletion, and $E_1$ is build up due to the additional voltage that is applied. The electric field at position $x$ within the silicon for $V_{bias} > V_{fd}$ is given by \citep{green2000physics}
\begin{equation}
    E(x)=E_d\left( 1-\frac{x}{d} \right) + E_1\ \ \ ;\ \ \ E_d=\frac{2 V_d}{d}\ \ \ ;\ \ \ E_1=\frac{V+V_d}{d},
    \label{eq:Theory/EFieldCompleteDepl}
\end{equation}
where $d$ is the thickness of the silicon.

The absolute strength of the electric field is important for the velocity of the charge carriers, because the drift velocity induced by the electric field linearly depends on the local electric field. The mobility of electrons is around 1400 \si{cm^2.V^{-1}.s^{-1}}, and for holes around 450 \si{cm^2.V^{-1}.s^{-1}} (see \citep{ioffeRu} for more information on the mobility of electrons and holes in silicon). This implies that holes travel approximately three times slower in silicon compared to electrons. This difference therefore implies that faster solid state detectors can be achieved by collecting electrons instead of holes.

\begin{figure}
\centering
\includegraphics[width=.75\linewidth]{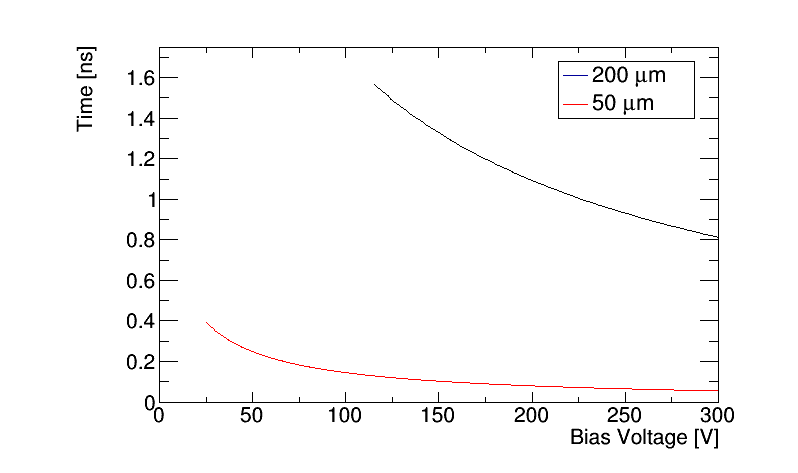}
\captionof{figure}{The time it takes charge carriers to traverse a silicon sensor of thickness \SI{200}{\micro\meter} in black (depleting at \SI{115}{\volt}), or \SI{50}{\micro\meter} in red (depleting at \SI{25}{\volt}) for a range of bias voltages.}
\label{fig:Theory/TimePerThickness}
\end{figure}

Another important aspect of a silicon pixel detector is the time it takes the charge carriers to traverse the depth of the silicon, or in other words how much time there is between the generation and the collection of the charge carriers. This time is given by \citep{bergmann20173d}
\begin{equation}
    t=\frac{d^2 \ln(\frac{U_b+U_d}{U_b-U_d})}{2 \mu U_d}.
    \label{eq:Theory/DriftTime}
\end{equation}
For two electron-collecting detectors with a different thickness the time it takes the charge carriers to traverse the complete detector as a function of bias voltage are plotted in \autoref{fig:Theory/TimePerThickness}. The curves are only plotted at bias voltages that exceed the bias voltage that is required to reach full-depletion. The major difference for the two detectors is the difference thickness. For the thinner detector, the electrons need to travel a smaller distance before they can be measured, implying that the faster collection of electrons decreases the time before a hit is measured and thus improving the timing performance that can be achieved. The fastest collection time of these detectors can be achieved with the thinner detector at a bias voltage that is as high as possible. However, when the same bias voltage applied to both detectors this results in a different electric field within the sensor due to the different thickness. The magnitude of the electric field will be six times as high in the \SI{50}{\micro\meter} sensor than it will be in the \SI{200}{\micro\meter}, solely due to the smaller depth of the sensor. This higher electric field in turn will result in a higher drift velocity of the electrons and thus decrease the time it takes the electrons to drift towards the electronics. As mentioned before, a time resolution of \SI{60}{\pico\second} is desired for a four plane silicon tracking system in order to achieve a final time resolution of \SI{30}{\pico\second}. A faster collection of charge leads to a larger signal in less time, making it possible with current noise and threshold levels to measure the time of arrival more accurately. Therefore, a faster collection of charge, and generally thus a thinner detector can lead to an increase of the time resolution. Thin detectors are therefore possible candidates to reach the desired time resolution of \SI{60}{\pico\second}. 

\section{Diffusion of charge cloud}
\label{sec:Theory/DiffusionChargeCloud}

For most silicon detectors in high energy physics, good spatial resolution is essential. This spatial resolution is primarily achieved by charge sharing, the process in which the liberated charge cloud is detected by more than one pixel due to the lateral diffusion of the charge cloud. If this is the case, the charged weighted centre of the charge cloud can be calculated which gives a better spatial resolution than a single pixel can. However, making the sensor faster can potentially harm the spatial resolution of them. To give an indication of this, in this Section the lateral radius of the charge cloud is discussed.

Besides the downwards drift of the charge carriers due to the electric field, there are also two effects that spread the cloud of charge carriers in the lateral direction. The first effect is the diffusion of the charge carriers due to thermal effects. The standard deviation of the charge cloud due to thermal effects is given by \citep{bosma2012cutting}
\begin{equation}
    \sigma=\sqrt{2Dt_{drift}}=\sqrt{2\frac{kT\mu}{q}t_{drift}}=\sqrt{\frac{kT}{q}\frac{d^2}{U_d}\ln(\frac{U_d+U_b}{U_d-Ub})}.
    \label{eq:Theory/ThermalDiffusion}
\end{equation}

\begin{figure}
\centering
\includegraphics[width=.75\linewidth]{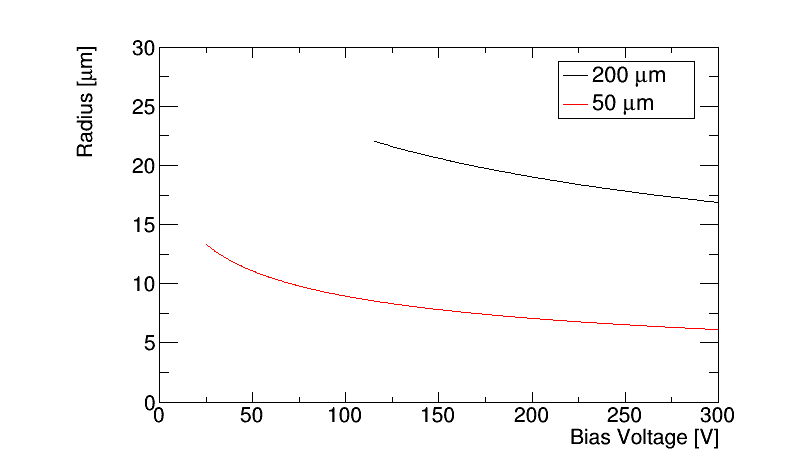}
\captionof{figure}{The radius of the charge cloud at the front of the detector assuming that the radius of the charge cloud is negligible at the back of the detector. The black curve indicates the radius of the charge cloud for a \SI{200}{\micro\meter} silicon sensor (depleting at \SI{115}{\volt}), and the red curve indicates the radius of the charge cloud for a \SI{50}{\micro\meter} silicon sensor (depleting at \SI{25}{\volt}).}
\label{fig:Theory/SpreadInDetector}
\end{figure}

The second effect is the lateral diffusion due to the electrostatic repulsion of the charge cloud itself. If a lot of charge is deposited in the sensor, the total charge of this cloud is high enough to significantly push the cloud itself outward. The radius of the charge cloud due to this effect is given by \citep{gatti1987dynamics}
\begin{equation}
    r_0(t)=\left( 3 \frac{q N_e d^2 \ln(\frac{U_b+U_d}{U_b-U_d})}{ 8\pi \epsilon U_d} \right)^{1/3}.
    \label{eq:Theory/ElectrostaticDiffusion}
\end{equation}
Where $N_e$ is the number of electrons that are present in the charge cloud. Both of these effects are independent of the mobility of the charge carriers, due to the mobility dependence of $t_{drift}$. The spread of the charge cloud due to thermal diffusion and electrostatic repulsion for two different detectors is shown in \autoref{fig:Theory/SpreadInDetector}. 

A thicker detector means that the spread of the charge cloud is bigger due to a longer drift time and thus allowing more time for the charge carriers to drift in the lateral direction. By decreasing the total drift time of the electrons the radius of the charge cloud will also become smaller as is visible in \autoref{fig:Theory/SpreadInDetector}. This also implies that charge sharing in a thinner detector is also less likely to occur. Therefore the spatial resolution may suffer under the increase in time resolution, and thus for the \SI{50}{\micro\meter} thick sensor the charge cloud is too narrow to allow charge sharing with the currently feasible pixel dimensions.

\chapter{Timepix3 readout chip} 

\label{Timepix3Chapter}

The Timepix3 readout chip is an ASIC developped by the Medipix collaboration. It has a $55\times 55$ \si{\micro\meter} pixelated structure and has $256\times 256$ pixels. It has the unique capability of reading out the charge and hit arrival time of in bins of \SI{1.56}{\nano\second} simultaneously. This makes this readout chip an ideal chip to study timing properties of silicon detectors.

\section{Timepix and Medipix family}

The first use of large area hybrid pixel detectors dates back to the WA97 experiment in 1995. The WA97 experiment demonstrated that the use of photon-counting detectors can be of great use in high-energy physics due to the noise free single-photon counting properties of these detectors \citep{knudsen1997presentation}, as well as their ability to measure highly energetic charged particles. These detectors were able to count the number of hits above a certain threshold. Over time, an increasing number of high energy physics experiments adopted the use of hybrid pixel detectors for vertex detection \citep{sauvage1995pixel,heijne1994first}. At this time, the development of dedicated pixelated photon-counting hybrid chips started, which led to the creation of the Medipix collaboration \citep{Delpierre_2014}. In 1997 this collaboration produced its first pixelated chip: the Medipix \citep{campbell1997readout}, aimed at medical applications. The Medipix consists of 64$\times$64 pixel, each measuring \SI{170}{\micro\meter}$\times$\SI{170}{\micro\meter} and employing a photon-counting mode. After the release of the Medipix, it quickly became apparent that there was a great potential for hybrid pixel detectors within the medical field.

New developments in commercial sub-micron CMOS-processing technologies led to the Medipix2 \citep{llopart2007design} in 1999. The Medipix2 consists of 256$\times$256 pixels, each measuring just \SI{55}{\micro\meter}$\times$\SI{55}{\micro\meter}, and employing two simultaneous thresholds. Most of this downscaling was made possible by the use of \SI{0.25}{\micro\meter} CMOS process. In 2006, a chip based on the Medipix2 was created: the Timepix \citep{llopart2007timepix}. This chip also employs the same number of pixels and the same dimension as the Medipix2. However, next to measuring the number of photon counts per pixel, the Timepix can also measure the time of arrival (\SI{10}{\nano\second} precision) and the charge of the individual hits, though it cannot measure them simultaneously.

Continuing advancements in the ASIC design, along with further improvement in CMOS-processing technologies led to the Medipix3 \citep{ballabriga2011medipix3} in 2010. Medipix3 is fabricated in a \SI{130}{\nano\meter} CMOS process, employing the same dimensions and pixel size as the Medipix2. In 2014 a successor to the Timepix chip was developed by the Medipix3 collaboration: the Timepix3 \citep{poikela2014timepix3}. This Timepix3 chip provides a more precise time of arrival (of \SI{1.56}{\nano\second}) and charge measurement, additionally carrying out both, while still having the possibility to be used as a photon counting chip. In 2016 the Medipix4 collaboration was initiated, aiming to develop two new read-out chips that are fully prepared for through-silicon via (TSV) processing and can be tiled on all four sides; the Medipix4 and the Timepix4 ASICs. This will enable the seamless tiling of large areas, which was until then not possible due to the inactive periphery of the previous chips. The current plans for the Medipix4 will increase the size of the pixel slightly to \SI{70}{\micro\meter}$\times$\SI{70}{\micro\meter}. The Timepix4 will still employ the pixel dimensions of the Timepix3, however the precision of the time of arrival will increase to \SI{200}{\pico\second}.

All of the chips arisen from the Medipix, Medipix2, and Medipix3 collaborations are commercially available, though some older chips are currently not available any more.

\section{The main characteristics of the ASIC}
Timepix3 is an ASIC, used in pixel detectors, and is part of the Medipix chip family. The Timepix3 ASIC is designed in \SI{130}{\nano\meter} CMOS technology and consists of 256$\times$256 pixels each measuring \SI{55}{\micro\meter}$\times$\SI{55}{\micro\meter}. An inactive region called the periphery which reads out the ASIC is located on one side, making it buttable on three sides. The Timepix3 can operate in three different modes: event-counting, time-of-arrival only, and time-of-arrival and charge simultaneously. The first mode is an event counting mode. In this mode the pixel will count the number of times the voltage at the input crosses a predefined threshold, and integrates the time the voltage is above the threshold for all the hits, returning the integrated charge (iToT) and the number of hits per pixel (PC). The second mode is time-of-arrival only mode, and the third mode performs a measurement of the time-of-arrival and charge simultaneously.

\begin{figure}
\centering
\captionsetup{width=.95\linewidth}
\includegraphics[width=.55\linewidth]{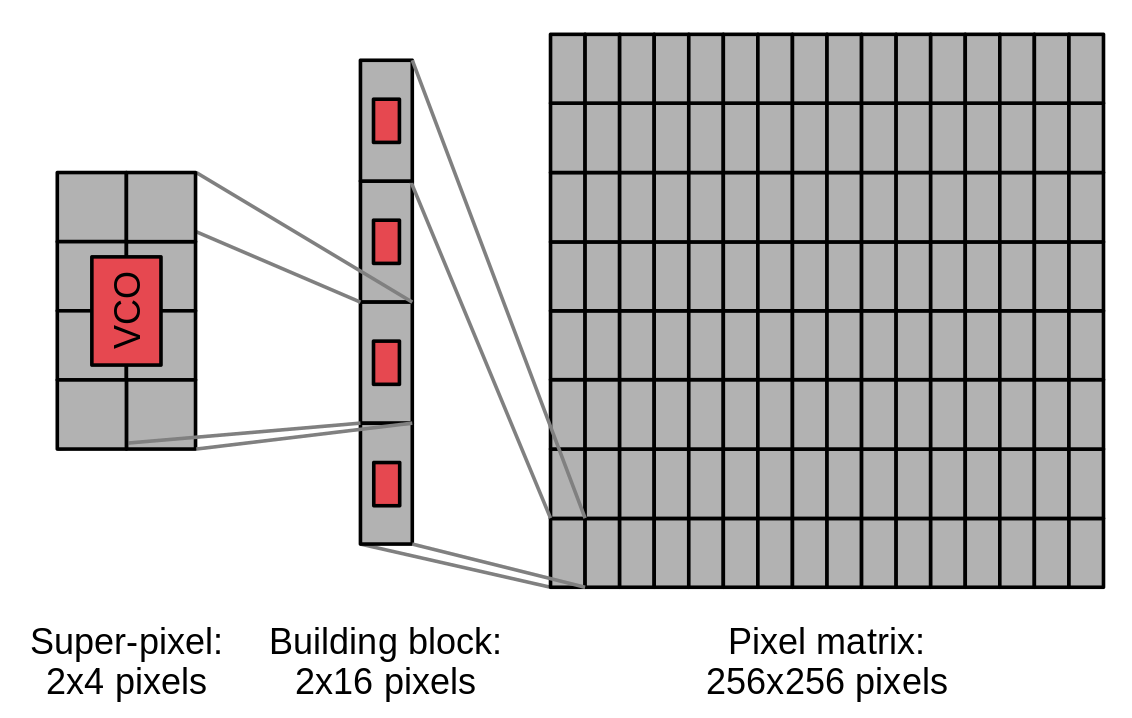}
\captionof{figure}{The structure of the pixel matrix. The smallest part of the pixel matrix is the building block which consists of four separate super-pixels. This building block is copied to form the complete pixel matrix of the Timepix3.}
\label{fig:Timepix3/MatrixStructure}
\end{figure}

\begin{figure}
\centering
\captionsetup{width=.95\linewidth}
\includegraphics[width=.95\linewidth]{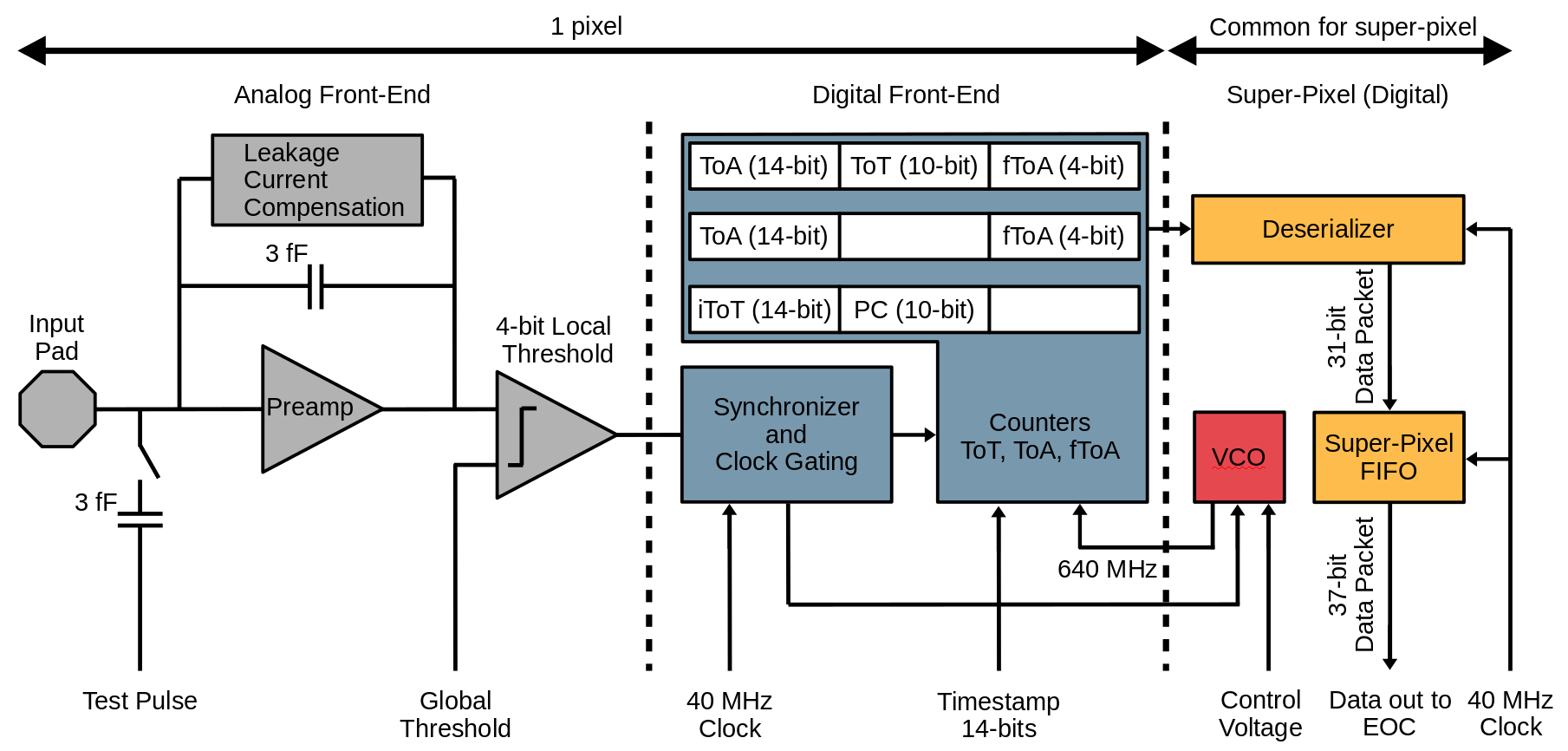}
\captionof{figure}{A schematic diagram of the various electrical and digital components in a pixel of the Timepix3. This diagram is adapted from \citep{poikela2014timepix3}.}
\label{fig:Timepix3/PixelLogic}
\end{figure}

Each pixel in the pixel matrix contains both analogue and digital circuitry. The analogue part of the pixel is based on a charge preamplifier with Krummenacher feedback \citep{krummenacher1991pixel}. This includes a preamplifier with leakage current compensation, a local threshold adjustment, and a discriminator. The digital part of each pixel includes a time-over-threshold counter, a coarse time counter, and a fine time counter. Each group of eight pixels (a super-pixel, discussed in the next paragraph) shares a local start-stop voltage-controlled ring oscillator (VCO) that generates a fast clock (\SI{640}{\mega\hertz}) which is used to determine the fine time for each pixel cell within the super-pixel structure. The different components of a single pixel are shown in \autoref{fig:Timepix3/PixelLogic}. In this Figure the divide between digital and analogue electrons is  indicated by a dashed line.

The pixel matrix of Timepix3 consists of small buttable building blocks that make up the complete pixel matrix (see \autoref{fig:Timepix3/MatrixStructure}). Column-wise the chip is divided into pairs of columns called double-columns, and row-wise the chip is divided into groups of four pixels. Such a block of eight pixels is called a super-pixel (see \autoref{fig:Timepix3/Timepix3Layout}) and shares some common logic amongst pixels. Four of these super-pixels placed in the column direction share a common clock buffer, making this two-by-sixteen pixel block the basic building block within the pixel matrix. In the design process of Timepix3, automated Place-and-Route (PnR) tools were used to wire the block of four super-pixels. Due to this, the routing of various signals within these four super-pixels are not completely but nearly identical.

In the next Section the time-of-arrival and the charge measurement of the Time\-pix3 will be explained.

\subsection{Time of arrival and time over threshold measurement}
\label{sec:Timepix3/ToAToT}
The measurement of the hit arrival time is split in two parts, a coarse time and a fine time. The coarse time derives from a \SI{40}{\mega\hertz} clock (also referred to as the system clock) and gives a timestamp accurate to \SI{25}{ns}, while the fine time relies on a \SI{640}{\mega \hertz} clock and gives a timestamp accurate up to \SI{1.56}{\nano \second}. Both timestamps are needed to generate the full ToA; the fine time only measures until the next period of the system clock. The coarse time is extended with the fine time in order to provide a ToA with time bins of \SI{1.56}{\nano\second}. A further explanation of the two main clocks used in Timepix3 is given in \autoref{Theory/Timepix3/Clocks}.

\begin{figure}
\centering
\captionsetup{width=.95\linewidth}

\includegraphics[width=.95\linewidth]{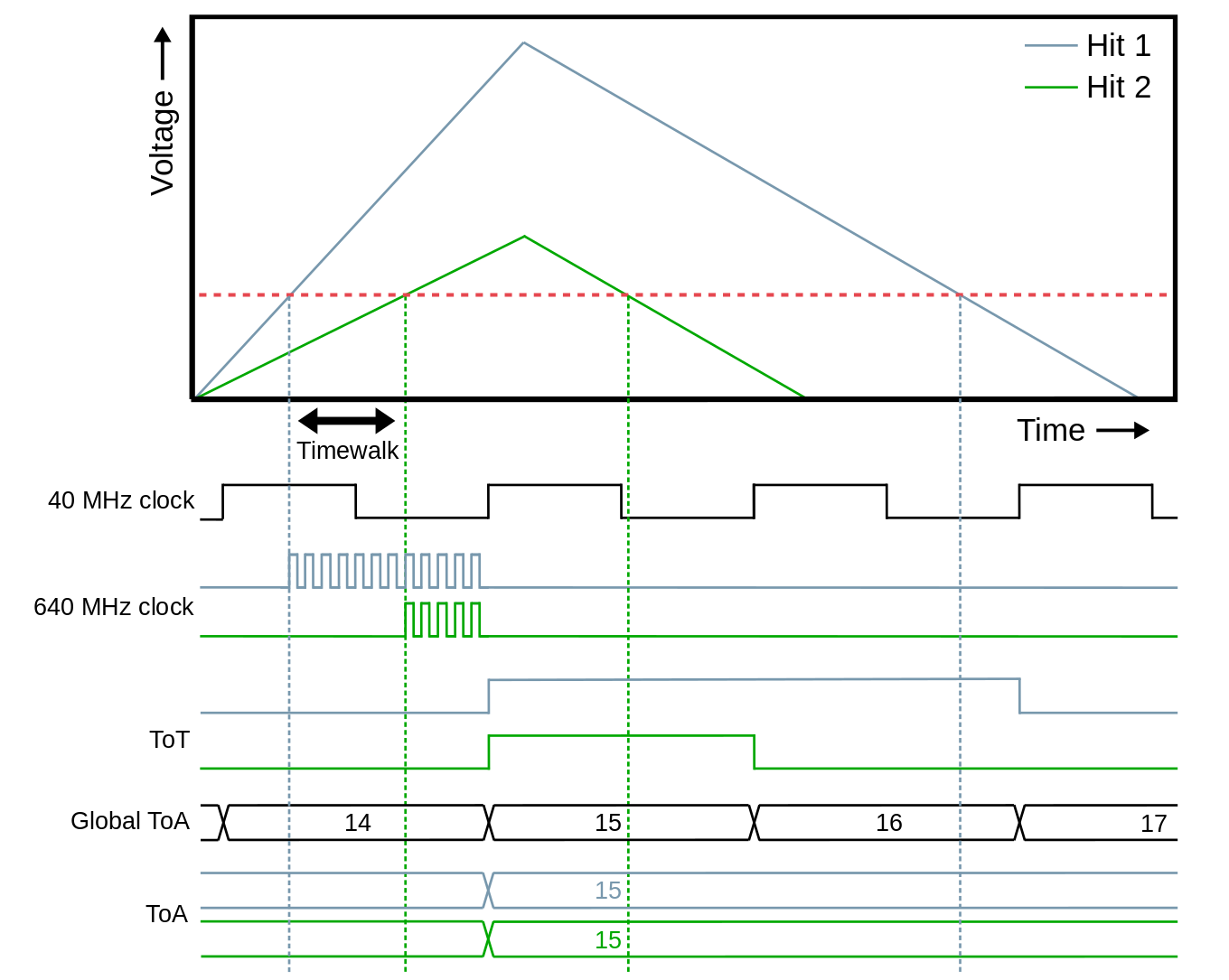}
\captionof{figure}{A schematic overview of the different clocks and effects in a measurement with Timepix3. The top window displays voltage of the amplifier output over time and shows two different hits (the high-voltage hit called hit 1 and the low-voltage hit is called hit 2). The corresponding fToA, ToT and fToA measurement are depicted below the graph. The clocks are colour coded according to the two different hits and the dashed red line indicates within the graph is the threshold. This Figure is taken and adapted from \citep{poikela2014timepix3}.}
\label{fig:Timepix3/ToTDiagram}
\end{figure}

As discussed in previously, when a charged particle traverses the silicon, charge carriers are generated that drift towards the implant in the silicon. The implant is directly connected to the input pad on the ASIC by means of a bump bond. The input pad is connected to a preamplifier that will amplify the signal. At this point the amplifier output arrives at the discriminator, for which the threshold can be set, and converts the charge into a voltage pulse. When the amplifier output crosses threshold, the voltage-controlled oscillator (VCO, which provides the \SI{640}{\mega\hertz}) starts. At this moment the coarse time is determined (also called the time-of-arrival, ToA, and shown in \autoref{fig:Timepix3/ToTDiagram}) and the VCO is started. After the start-up of the VCO, a counter in the pixel starts counting the number of periods since the threshold crossing until there is a rising edge of the \SI{40}{\mega \hertz} clock. The number of periods passed is called the fine time-of-arrival (fToA). Note that the fToA is inversely proportional to the time, as it measures the number of periods until between the start-up of the VCO and the rising edge of the \SI{40}{\mega \hertz} clock. As soon as the rising edge of the \SI{40}{\mega \hertz} clock is reached, the VCO is stopped. What happens next depends on whether or not the amplifier output is still above threshold. If the amplifier output is below threshold, the ToT counter is not incremented, and it is not considered as a hit. If the amplifier output is still above threshold, the ToT counter is incremented by one, and will continue counting the number of rising edges of the \SI{40}{\mega \hertz} until the charge is below threshold again.

After the amplifier output drops below threshold, both the ToA and the ToT for that hit are determined. At this point a data packet is generated for that pixel; this data packet is 28 bits long and the layout depends on the operating mode of the Timepix3. The data layout can be seen in \autoref{fig:Timepix3/PixelLogic}. From the pixel, the data packet is first sent to the deserializer located within the super-pixel. After this, the data packet is transferred to the super-pixel FIFO and from here is transferred to the end-of-column FIFO via a shared column bus common to all pixels in the double-column. At the end-of-column FIFO, a double-column ID is added to the data packet which then enters the end-of-chip logic. This double-column ID is needed to determine the position of the pixel within the pixel matrix. In total, the maximum transfer rate is limited to 2.56 Gbps \citep{gromov2011development}, which translates to a maximum hit rate of 80 Mhits/s/ASIC. The maximum hit rate for a super-pixel is \SI{400}{\kilo\hertz}, with a maximum rate of \SI{50}{\kilo\hertz} per pixel within a super-pixel \citep{poikela2015readout}. 

\begin{figure}
\centering
\captionsetup{width=.75\linewidth}
\includegraphics[width=.95\linewidth]{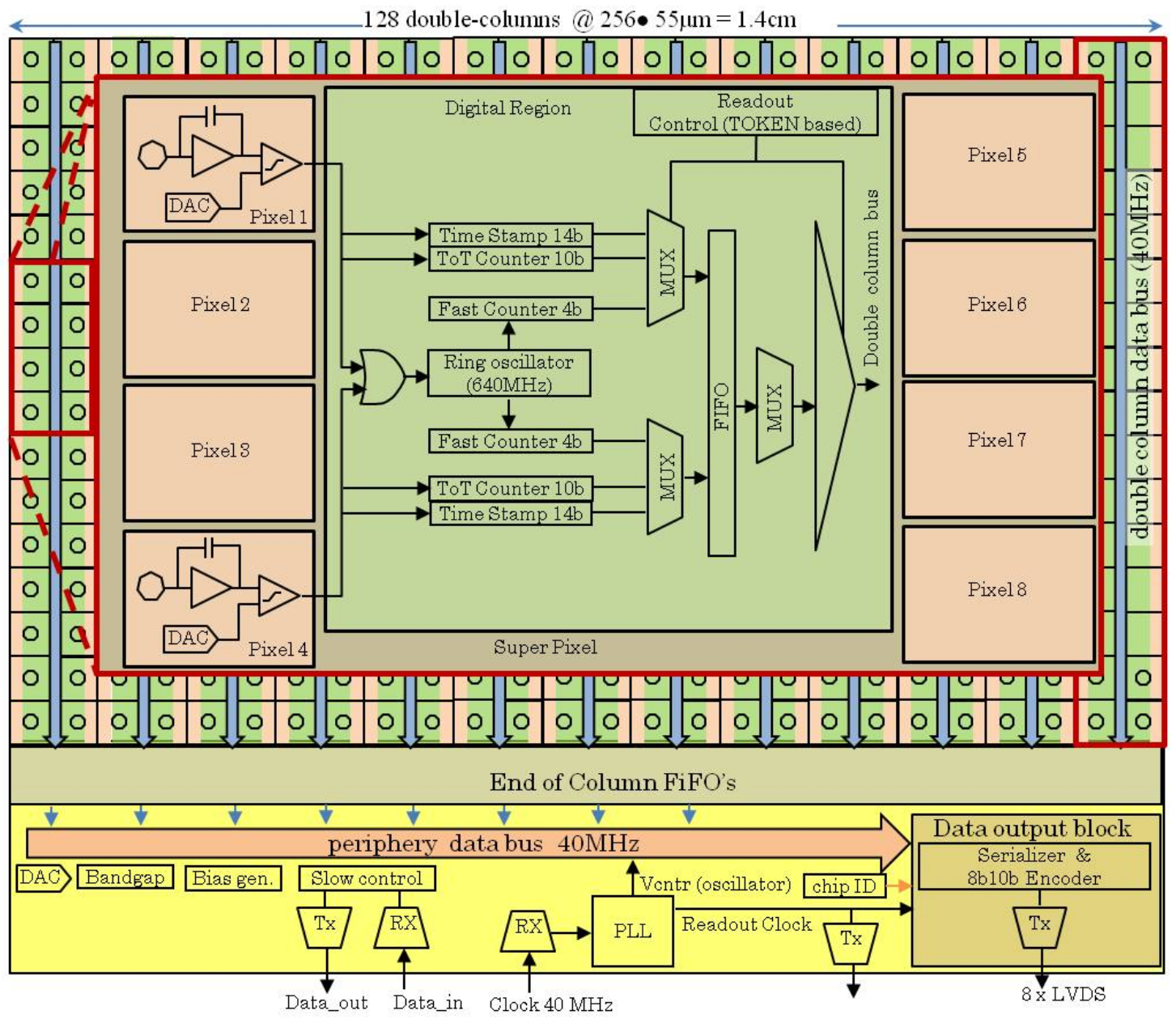}
\captionof{figure}{The top level block diagram of the layout Timepix3. This Figure was taken from \citet{gromov2011development}.}
\label{fig:Timepix3/Timepix3Layout}
\end{figure}

An overview of this process and the position of the different components within the Timepix3 ASIC is shown in \autoref{fig:Timepix3/Timepix3Layout}.

\section{Timewalk correction functions}
\label{sec:Timepix3/Timewalk}

Given a constant peaking time of the preamplifier, output pulses from the preamplifier of different magnitudes take a different time to cross threshold. This effect is known as timewalk \citep{llopart2007timepix} and is shown in \autoref{fig:Timepix3/ToTDiagram}. In this Figure the (higher) blue curve indicates a pulse with a high magnitude and the (lower) green curve indicates a pulse with a low magnitude. Especially at low energies, timwalk can significantly degrade the timing resolution of the chip. Therefore it is import to correct for timewalk. A correction for this effect can be calculated with the use of the internal test pulse function of the chip. To do so, test pulses with different magnitudes are used. These test pulses with different charges are injected at the same time with respect to the \SI{40}{\mega \hertz} clock. This enables the possibility to determine the delay due to timewalk by determining the difference in the fToA for different charges.

\begin{figure}
\begin{minipage}{.5\linewidth}
\centering
\subfloat[]{\label{fig:Timepix3/var1Timewalk}\includegraphics[width=1\linewidth]{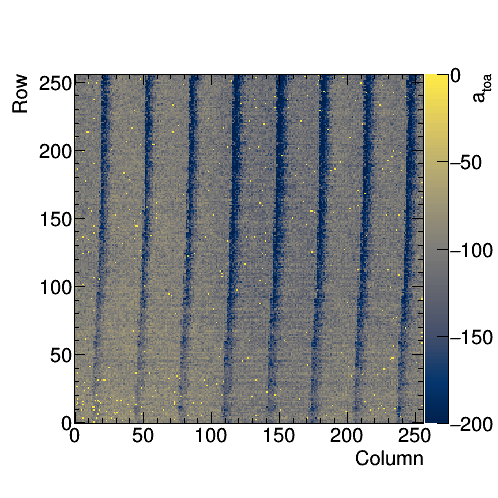}}
\end{minipage}%
\begin{minipage}{.5\linewidth}
\centering
\subfloat[]{\label{fig:Timepix3/var2Timewalk}\includegraphics[width=1\linewidth]{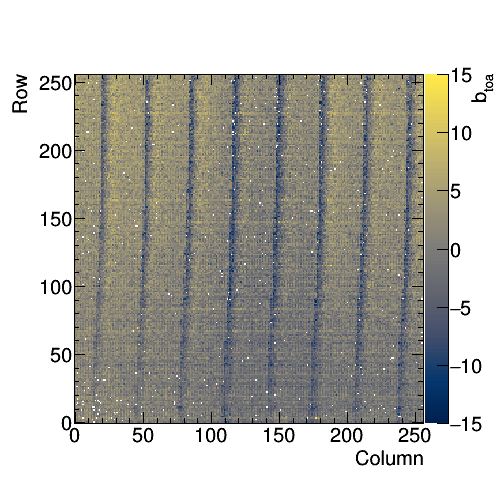}}
\end{minipage}\par\medskip
\begin{minipage}{.5\linewidth}
\centering
\subfloat[]{\label{fig:Timepix3/var3Timewalk}\includegraphics[width=1\linewidth]{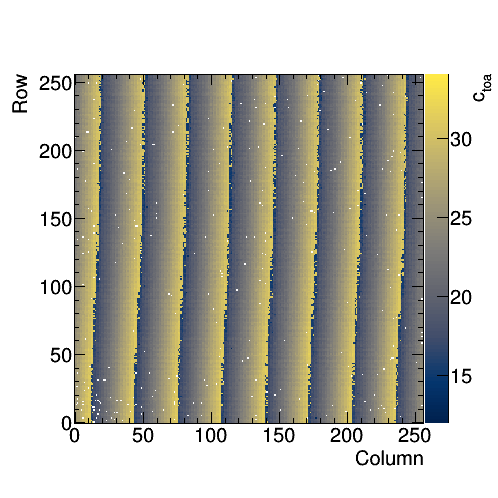}}
\end{minipage}%
\begin{minipage}{.5\linewidth}
\centering
\subfloat[]{\label{fig:Timepix3/TimewalkExample}\includegraphics[width=1\linewidth]{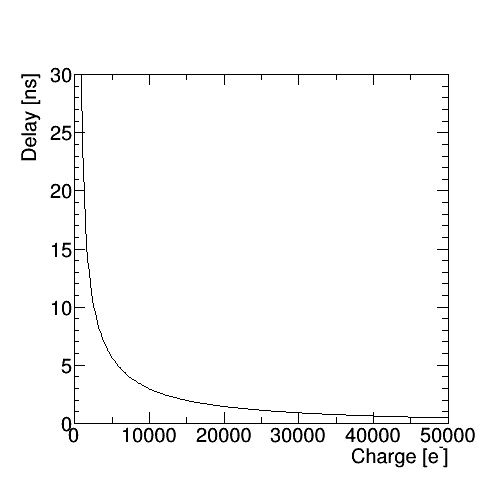}}
\end{minipage}\par\medskip

\captionsetup{width=.95\linewidth}
\caption{The parameter values of a timewalk calibration for the pixel matrix of a Timepix3 chip. Figure A) shows the values for $a_{toa}$, Figure B) shows the values for $b_{toa}$, and Figure C) shows the values for $c_{toa}$. The vertical lines in the values for the parameters is due to the propagation of the test pulse throughout the chip. Figure D) shows an the charge dependent time response of a single pixel of the delay for different injected charges. This curve shows the timewalk effect}
\label{fig:main}
\end{figure}

The shift in time due to timewalk can be modelled as \citep{pitters2018time}
\begin{equation}
	f_{toa}(Q)=\frac{a_{toa}}{Q-b_{toa}}+c_{toa},
\end{equation}
where Q is the charge of the pulse, and $a_{toa}$, $b_{toa}$ and $c_{toa}$ are free parameters. This curve can be fitted for each pixel, after which a per-pixel timewalk correction can be applied to every data set. The values for $a_{toa}$, $b_{toa}$, and $c_{toa}$ of such a per-pixel timewalk correction are shown respectively in \autoref{fig:Timepix3/var1Timewalk}, \autoref{fig:Timepix3/var2Timewalk}, and \autoref{fig:Timepix3/var3Timewalk}. The vertical lines observed in all three variables are due to the propagation of the test pulse signal throughout the chip.

\subsection*{Suggested improvement for an advanced timewalk correction}

The ToT measurement only begins and ends on the rising edge of the \SI{40}{\mega\hertz} clock, therefore the time resolution for the start- and end-points of the ToT measurement is \SI{25}{\nano\second}. Due to this, it is on average more likely for a hit to have a lower ToT when the hit arrives early in the \SI{25}{\nano\second} period. The more accurate time of arrival information of the fToA can be used to correct part of this effect. By extending the ToT value by 1/16th for each fToA value, the standard deviation can be lowered from \SI{10.2}{\nano\second} (two times a standard deviation of one \SI{25}{\nano\second} bin: $\left[25/\sqrt{12}\right]\oplus \left[25/\sqrt{12}\right]$) to \SI{7.2}{\nano\second} (a combination of a bin size of \SI{1.56}{\nano\second} and a bin size of \SI{25}{ns}: $\left[25/\sqrt{12}\right]\oplus \left[1.56/\sqrt{12}\right]$). This increased accuracy will directly provide a more precise charge measurement.

The timewalk correction discussed in this Chapter relies on the charge of the signal to correct for timewalk. A more precise determination of the charge will thus yield an increased precision in the correction for timewalk. However, this also means that the correction that should be applied is dependent on the fToA as well as the ToT. Sixteen timewalk curves should therefore be made with test pulses per pixel, one for each fToA bin. This process is very time consuming, and has not yet been performed in this work. To increase the timing performance of the Timepix3 in the future, this could prove to be a valuable next step.

\section{The clock structure}
\label{Theory/Timepix3/Clocks}

A combination of two different clocks in the Timepix3 are needed to produce a timestamp with a bin size of \SI{1.56}{\nano \second}. These are the \SI{40}{\mega\hertz} clock (also referred to as the system clock) and the \SI{640}{\mega\hertz} clock. The \SI{40}{\mega\hertz} clock gives a time resolution of \SI{25}{\nano\second} which is called the coarse time, and the \SI{640}{\mega\hertz} clock gives a time resolution of \SI{1.56}{\nano \second}, which is called the fine time. The reason that there is not only a \SI{640}{\mega\hertz} is because of power consumption. If all super-pixels would have the VCO running all the time, the power consumption of the complete chip would be large, which causes further problems in the chip. Therefore the \SI{40}{\mega\hertz} clock is used, which is only distributed through the chip, and not locally generated like the \SI{640}{\mega\hertz}. The \SI{40}{\mega\hertz} can either be generated internally or supplied externally, making it possible to easily synchronise multiple Timepix3 modules or synchronise the Timepix3 to external equipment.

\begin{figure}
\centering
\captionsetup{width=.95\linewidth}
\includegraphics[width=.75\linewidth]{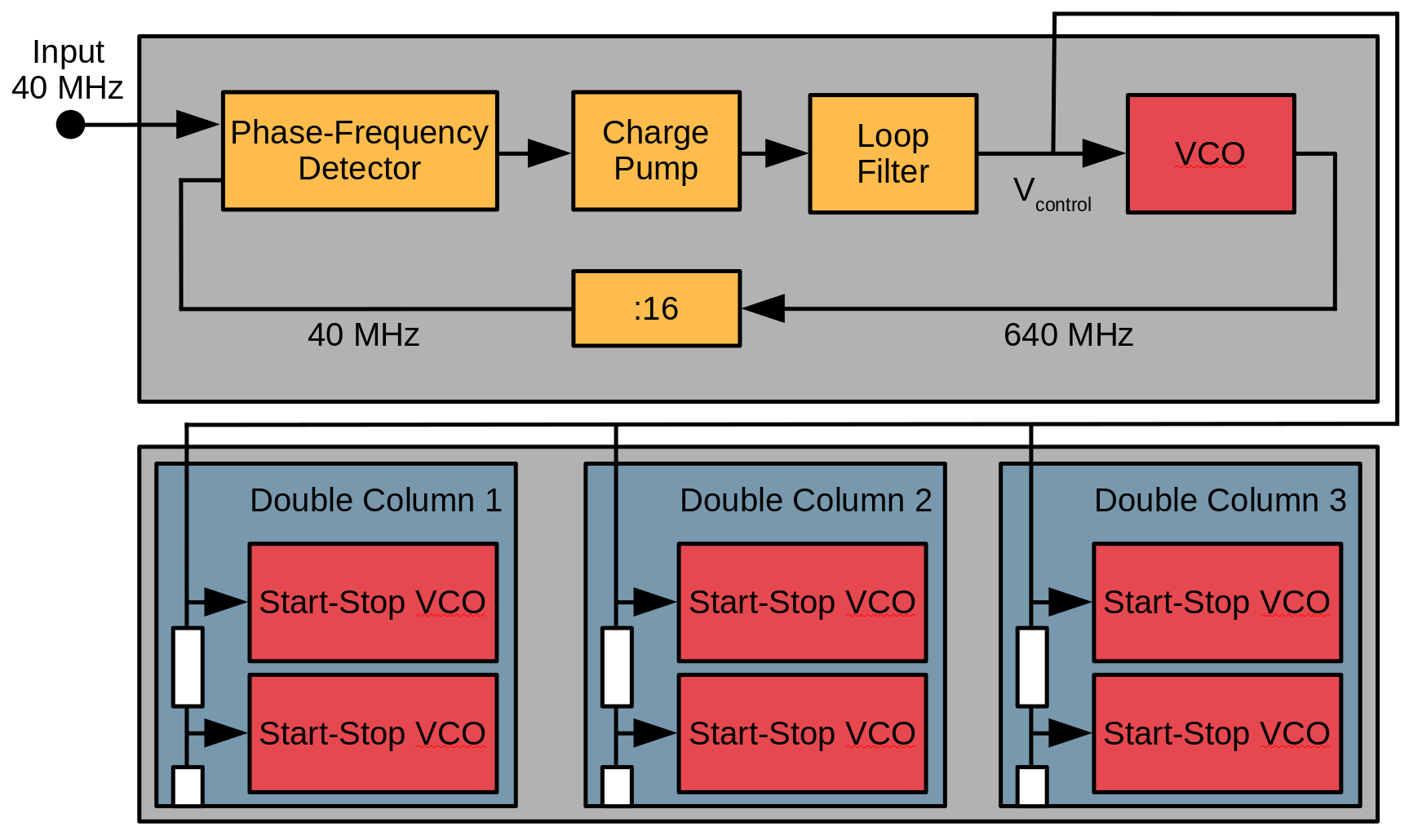}
\captionof{figure}{A schematic overview of the generation and distribution of the control voltage that determines the frequency of the VCO in each super-pixel.}
\label{fig:Timepix3/PLL}
\end{figure}

To achieve an accuracy of \SI{1.56}{\nano \second}, each pixel has a high resolution Time-to-Digital Converter (TDC), which employs a start-stop ring oscillator. To ensure a frequency of \SI{640}{\mega\hertz} throughout the chip, a Phase-Locked Loop (PLL) in the periphery of the chip is used. In this PLL, a control voltage is generated from the \SI{40}{\mega\hertz} clock. This control voltage is supplied to a VCO, which generates an arbitrary frequency in a range around \SI{640}{\mega \hertz} depending on the magnitude of the control voltage. The signal from the VCO is divided by sixteen (\SI{640}{\mega\hertz} / 16 = \SI{40}{\mega\hertz}), and is compared to the input \SI{40}{\mega \hertz} clock. If either the phase or the frequency of the generated signal does not match the \SI{40}{\mega \hertz} clock, the control voltage is changed such that the generated signal matches the original clock. The control voltage can now be distributed throughout the chip, such that all VCO in the super-pixels generate the same \SI{640}{\mega \hertz} clock, assuming that the production variations in each VCO are negligible. A diagram of the generation and distribution to each super-pixel of this control voltage is shown in \autoref{fig:Timepix3/PLL}.

The VCO in a super-pixel can be started by each pixel, and is distributed to each pixel within the super-pixel. The VCO can be started by the output of the discriminator of a pixel at any time within the \SI{25}{\nano \second} period, and is stopped at the rising edge of the \SI{40}{\mega\hertz} clock. 

Besides the distribution of the control voltage throughout the chip, the system \SI{40}{\mega \hertz} clock is distributed as well. This clock is distributed to each pixel to provide the coarse time measurement. It is also distributed among a few common logic components that are shared by a super-pixel. Each double-column employs a clock-tree \citep{poikela2015readout} to ensure minimal bottom-to-top skew of this clock. The clock-tree is divided into sections of sixteen pixels (four super-pixels); after the first two super-pixels the clock is buffered with a relatively large buffer which uses transistors to limit the delay. Smaller buffers are used to buffer the clock to each super-pixel.

\section{Equalization of the Timepix3}

\begin{figure}
\centering
\captionsetup{width=.95\linewidth}
\includegraphics[width=1\linewidth]{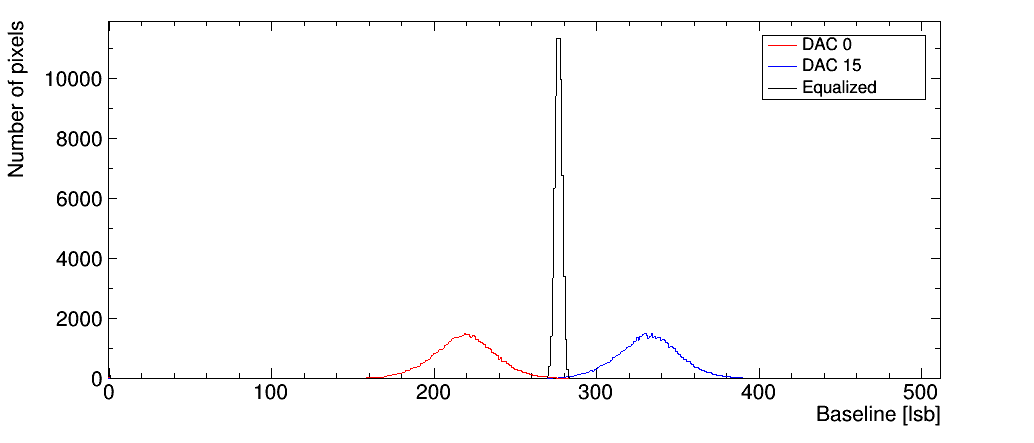}
\captionof{figure}{The baseline value of the preamplifiers for the pixels of the pixel matrix. The left curve indicates the baseline for a local threshold of 0, the right curve indicates the baseline for a local threshold of 15, and the middle curve indicates the baseline value for an equalized pixel matrix.}
\label{fig:Timepix3/DACSEqualization}
\end{figure}

As is shown in the pixel schematic (see \autoref{fig:Timepix3/PixelLogic}), the discriminator accepts both a global threshold and a local threshold adjustment. The global threshold can be set in a range from 0 to 512 DAC values and is the same for each pixel. However, slight variations between pixels leads to a difference in the applied threshold, hence a 4-bit local threshold adjustment can be set for each pixel individually. This local threshold adjustment ensures that the threshold is the same for each pixel independent of the per-pixel variations. To determine the local threshold for the each pixel, the baseline value of the preamplifier of each pixel is determined for local threshold adjustment 0 and 15. The distribution of this global threshold value for all pixels is plotted in \autoref{fig:Timepix3/DACSEqualization} for local threshold adjustment DAC 0 and local threshold adjustment DAC 15. After the baseline value is known for each pixel at both local threshold adjustment 0 and 15, the baseline for each value of the local threshold can be extrapolated. This way the local threshold can be determined for each pixel in such a way that the baseline value for all the pixels is around the same global threshold value. With these local threshold values the chip is equalized in such a way that all the pixels have the baseline value at around the same global threshold value. This equalization reduces the threshold dispersion from around 19 LSB (190 e$^-$) to 2.9 LSB (29 e$^-$) \citep{de2014design}. The distribution of the baseline value of the preamplifiers after an equalization is shown in black in \autoref{fig:Timepix3/DACSEqualization}.

\section{The SPIDR readout board}
The Timepix3 is connected via a FMC connector to a SPIDR (Speedy PIxel Detector Readout) board \citep{visser2015spidr} which is used to read out the Timepix3. The SPIDR board can be connected to a server or PC via either a 1 Gb Ethernet connection or a 10 Gb Ethernet connection. In addition to reading out the Timepix3, an external trigger signal of up to \SI{3.3}{\volt} can be supplied to a TDC located directly on the SPIDR. This signal will be timestamped by the TDC, and is processed offline along with the data from the Timepix3. The binning process employed by this TDC is different from that in the Timepix3; from the same system clock used by the Timepix3 a single \SI{320}{\mega\hertz} clock is produced, similar to the generation of the \SI{640}{\mega\hertz} clock in the ASIC periphery of the Timepix3. Six copies of this \SI{320}{\mega\hertz} are made, of which five are shifted in phase such that the phase difference between the individual clocks is $\pi/3$. When a trigger signal is registered, the SPIDR will determine for all six copies if the signal is either high or low. This way, twelve different combinations of the individual states of the clocks can be distinguished throughout a single period of the \SI{320}{\mega\hertz} clock. This produces time bins with a width of \SI{260}{\pico \second} for the trigger signal, and is thus more accurate than the time bin of the fToA of the Timepix3. 

\section{Charge calibration using test pulses}
\label{sec:Timepix3/ChargeCalibration}
Timepix3 has the ability to inject a known number of charge carriers in the analogue front-end circuitry before the preamplifier. These test pulses, reassemble the signal that would be created by a charged particle traversing the detector, and can be varied in both magnitude and relative phase with respect to the \SI{40}{\mega \hertz} clock, making them useful to determine the charge as well as the timewalk calibration of the chip. However, the injection of the pulses relies on the same clock which is subsequently used to measure them, therefore a shift or frequency change in this clock, even at the per-pixel level, can never be measured. 

\begin{figure}
\centering
\includegraphics[width=.5\linewidth]{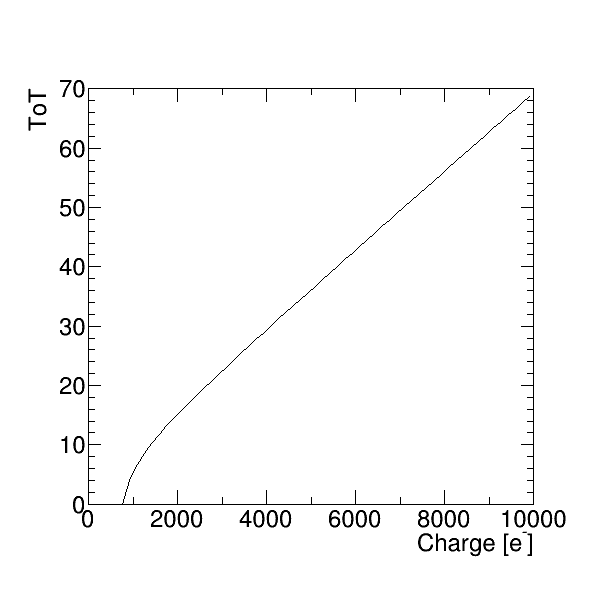}
\captionof{figure}{The measured ToT versus injected charge, using test pulses for a single pixel. }
\label{fig:Timepix3/chargeCal}
\end{figure}

With the injection of a known amount of charge carriers via test pulses, a charge calibration can be performed. This charge calibration is needed to convert the measured ToT into a charge measurement, making it independent of the settings and variations over the chip and solely relying on interaction of the charged particles in the silicon. The ToT response of a pixel within the chip can be modelled by \citep{pitters2018time}
\begin{equation}
	\text{ToT}(Q)=\frac{a_{tot}}{Q-b_{tot}}+c_{tot}+d_{tot}Q.
\end{equation}
Where $Q$ is the charge injected, and $a_{tot}$, $b_{tot}$, $c_{tot}$, and $d_{tot}$ are free parameters. The inverse function can later be used to convert the measured ToT to the total charge $Q$:
\begin{equation}
	Q=\frac{\text{ToT} +b_{tot}d_{tot} - c_{tot}+\sqrt{ -4a_{tot} d_{tot} + \left( c_{tot} + b_{tot}d_{tot} - \text{ToT} \right)^2 }}{2d_{tot}}
\end{equation}
An example of this curve for a single pixel of a Timepix3 is shown in \autoref{fig:Timepix3/chargeCal}. By calculating these parameters for each pixel, a per-pixel charge calibration can be performed to exclude chip effects in the measured charge.

\chapter{Design and properties of a pulsed focussed laser setup} 
\label{sec:LaserSetup}

The use of optical photon in order to generate pixel hits on the Timepix3 is chosen as an alternative to the use of particle accelerators and test pulses to investigate the per-pixel behaviour of the Timepix3. In the next Sections, the implementation of this method will be discussed as well as the final laser system and its properties that we have build for this work. This laser system partly consists of components that were used previously in a similar laser setup build at Nikhef \citep{bosma2012cutting}.

\section{General setup}
\label{sec:LaserSetup/Setup}
The idea behind the setup is, besides the focusing of the photons, fairly simple. A laser is used to liberate charge carriers in the silicon that is acts as the detection material connected to the Timepix3 ASIC. These charge carriers will be detected by the Timepix3 ASIC, and the raw data collected from the Timepix3 can be analysed offline to investigate the Timepix3. 

To achieve a constant source of optical photons, a Fibre Bragg Grating (FBG, see \citep{hill1997fiber}) stabilized pulsed laser diode with a wavelength of \SI{680}{\nano\meter} is used. The advantage of such a laser is the constant wavelength, and thus energy, per photon due to the FBG. This FBG ensures by satisfying the Bragg condition only for a specific wavelength. The laser diode is driven by a pulse generator with a quick rise-time, and is used to achieve a constant period between the individual laser pulses. It is important to have a quick rise-time on the signal that drives the laser to prevent any fluctuations in the time when the laser will reach population inversion. Otherwise the onset of the laser pulse will vary in time, as will the measured fToA associated to the laser pulse. 

Without any other reference signal, it is not possible to determine the absolute delay. Therefore, a second signal should be supplied by the pulse generator that is used to drive the laser diode. This second signal will be provided as an external trigger to the SPIDR to compare the time of the trigger signal to the measured time of the induced charge carriers. This way the absolute delay between the measurement of the charge carriers and the creation of the photons can be calculated. The absolute delay will depend on the length of the cables and the optical fibre. However, as long as these lengths are kept constant, this will only produce a constant shift in the absolute delay and will thus not influence the measurements.

One of the advantages of using a laser, is that time of the injection of the charge within the sensor can be tuned. For randomly generated particles this is normally not possible, while such a freedom can be of significant use to investigate timing aspects. By introducing a delay with respect to the \SI{40}{\mega \hertz} clock in the signal that drives the laser diode, the signal from the pulse generator that is used to drive the laser can be delayed with a precision of up to \SI{5}{\pico \second}, depending on the pulse generator that is used. This delay will thus determine the delay of the arrival of charge carriers in the Timepix3, and gives the possibility to investigate aspects of the Timepix3 that otherwise could only have been investigated using test pulses. 

The Timepix3's \SI{40}{\mega\hertz} clock should be synchronized to the external source of charge carriers (to prevent clock drift between the two systems and keep the phase constant). To achieve synchronization between the Timepix3 and the pulse generator, an external \SI{40}{\mega\hertz} signal is supplied to the Timepix3; which can be used by the Timepix3 instead of the internal clock. This same external clock is also supplied to the pulse generator which can accept this as a reference clock to generate the pulse that drives the laser. 

On the other hand, one can also decide to not synchronize the laser with respect to the Timepix3. This will mimic the normal statistical time of arrival of charged particles within the \SI{25}{\nano\second} period. This is due to the slight mismatch in frequency between the internal clock of the Timepix3 and that of the pulse generator, which is inherent to different equipment that are not synchronized. 

\begin{figure}
\centering
\includegraphics[width=.75\linewidth]{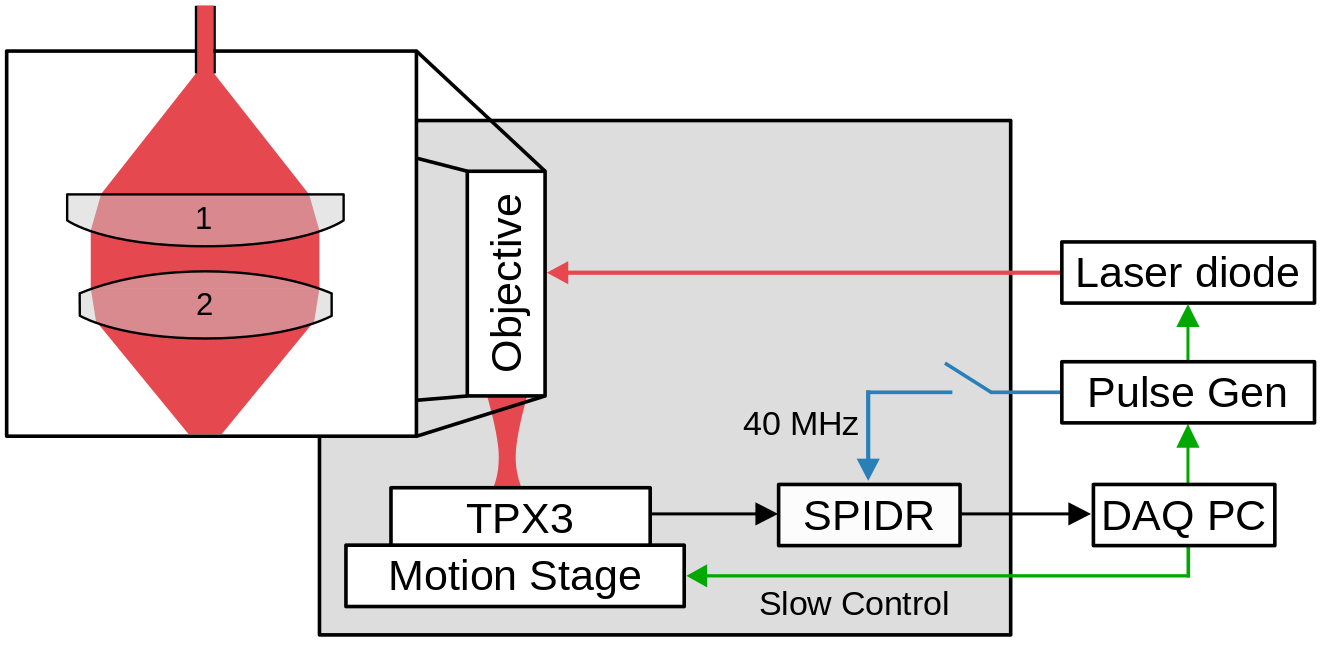}
\captionof{figure}{A diagram illustrating the setup of the laser system at Nikhef. The black arrows indicate the direction of data flow, the green arrows indicate the direction of the control signals, the blue arrow indicates the \SI{40}{\mega\hertz} clock generated externally generated by the pulse generator and the red arrow indicates the direction of the light. }
\label{fig:LaserSetup/Setup}
\end{figure}

\begin{figure}
\centering
\includegraphics[width=.75\linewidth]{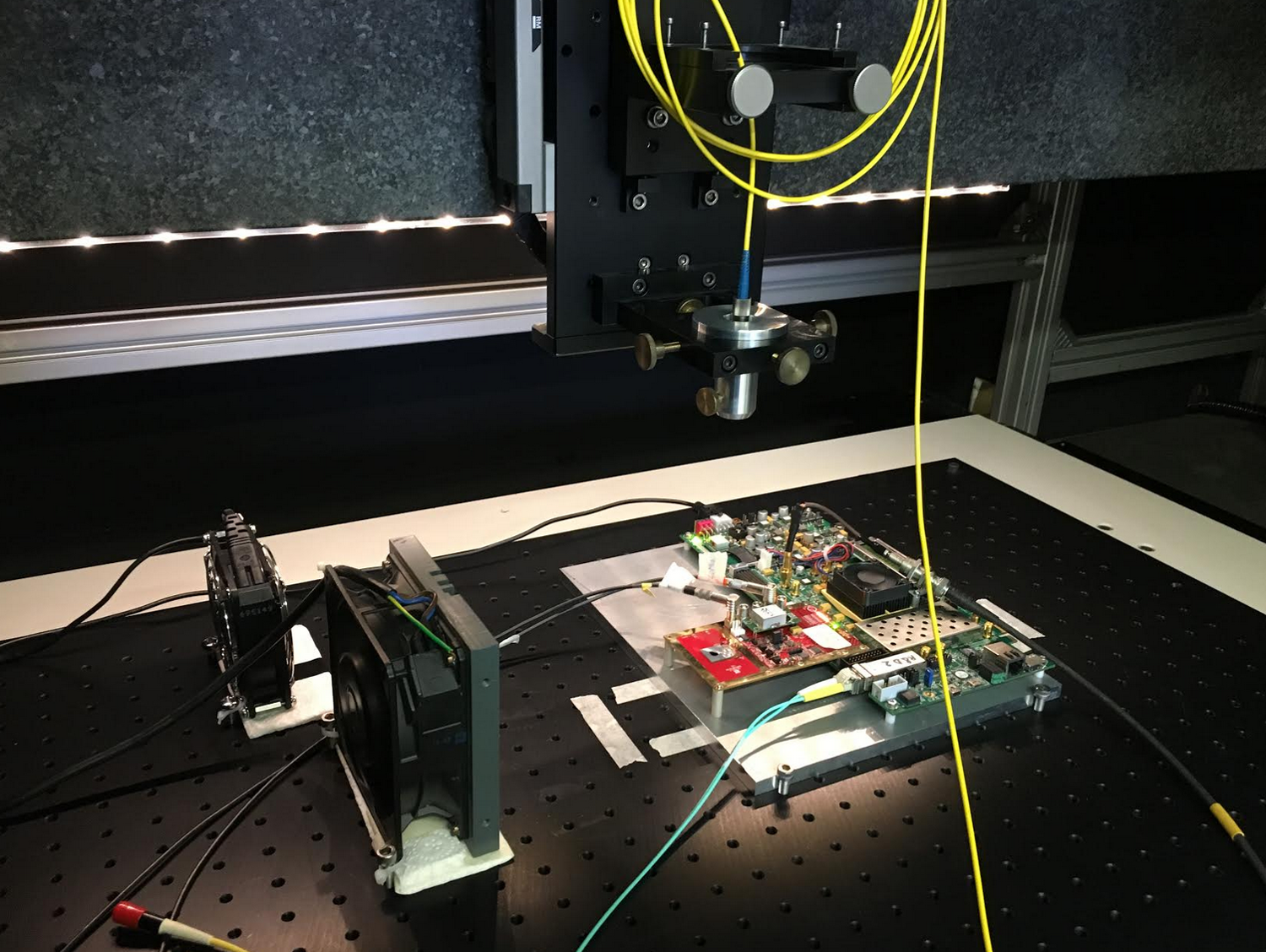}
\captionof{figure}{A picture of the laser setup that we build for this work. The Timepix3, the SPIDR and the mounting of the lens system are visible in this picture. Both the pulse generator and the laser diode are located outside of this picture.}
\label{fig:LaserSetup/SetupPicture}
\end{figure}

A diagram of the setup at Nikhef that we constructed for this thesis is shown in \autoref{fig:LaserSetup/Setup} as well as a picture of the setup in \autoref{fig:LaserSetup/SetupPicture}. A computer (DAQ PC) is used to control the SPIDR, a motion stage and a pulse generator. As mentioned before, the laser diode is driven by the pulse generator, while the laser diode itself is coupled to an optical fibre which is connected to a lens system (objective). The Timepix3 is placed on a xy-motion stage with a resolution of \SI{500}{\nano \meter}. This allows us to scan the pixel matrix, while the objective is attached to a motion stage that is able to move in the z-direction to be able to find the focal spot of the laser. 

\section{Focusing of the beam spot}
Besides the generation of photons, it is also important to focus the photons such that the beam spot is smaller than the pixel dimensions. The photons that are created in the laser diode are coupled to an optical fibre, which will guide the light to just above the sensor. At this point the pulse is coupled to air. Two lenses are used to first expand and then focus the pulse to achieve a spot size smaller than the pitch of the pixels. This theory behind this process is first discussed below, after which the system that is made for this work is discussed.

\subsection*{Theory of focusing a Gaussian beam}

After focusing a parallel beam of light using a single lens, in general the focusing of the light will not be in a perfect straight cone, but will form a parabolic shape. For lenses with a big diameter and large focus spots, this cone is well approximated by a normal straight cone, however when a relatively small spot size is desired, the shape of the beam around the focus point starts to depend on the wavelength, lens diameter, and spot size. To calculate the beam shape around the focus point, a Gaussian beam profile is assumed. The transverse profile of the optical density $I$ of a Gaussian beam with power $P$ can be described with the Gaussian function \citep{varga1998gaussian}:
\begin{equation}
    I(r,z)=\frac{2P}{\pi w^2\left(z\right)}\exp(-2\frac{r^2}{w^2\left( z \right)}),
\end{equation}
where $w\left( z \right)$ is the beam radius at $z$. The beam radius is defined as the radius at which the intensity has dropped to a factor $1/e^2$ of the maximum intensity. Gaussian beams are generally only considered when the beam divergence is small. Because of the small divergence, the Fresnel approximation \citep{Steane:89} (also known as the paraxial approximation) can be applied. This approximation assumes that normal vector of the wave front at any position within the beam is only slightly deviating from the beam axis were the angle between them is given by $\theta$. This condition then implies $\sin(\theta)\approx\theta,\tan(\theta)\approx\theta, \ \text{and}\ \cos(\theta)\approx1$. By applying this approximation, the second-order derivative in the propagation equation can be omitted. This propagation equation can be derived from Maxwell's equation, and becomes a first order differential equation after omitting the second-order derivative. For a Gaussian beam, the time and space dependence is given by \citep{varga1998gaussian}
\begin{equation}
    E(r,z;w_0)=E_0\frac{w_0}{w\left( z \right)}\exp[-r^2\left( \frac{1}{w^2\left( z \right)} -\frac{ik_0}{2 R\left( z \right)} \right)-i\Phi\left( z \right)],
    \label{eq:Theory/EField}
\end{equation}
where 
\begin{align}
  \phantom{w^2\left( z \right)}
  &\begin{aligned}
    \mathllap{w^2\left( z \right)} &= w_0^2\left( 1+\frac{z^2}{z_0^2} \right),
  \end{aligned}\\
  \phantom{w^2\left( z \right)}
  &\begin{aligned}
    \mathllap{R\left( z \right)} &= z\left( 1+\frac{z^2}{z_0^2} \right),
  \end{aligned}\\
  \phantom{w^2\left( z \right)}
    &\begin{aligned}
    \mathllap{\Phi\left( z \right)} &= \arctan(\frac{z}{z_0}),
  \end{aligned}\\
    &\begin{aligned}
    \mathllap{z_r} &= \frac{\pi w_0^2}{\lambda}.
    \label{eq:Theory/Laser/Rayleigh}
  \end{aligned}
\end{align}
Here, $k$ is the wavenumber of the light in the beam, and is given by $k=2\pi/\lambda$. $z_r$ is known as the Rayleigh length, and gives an indication over which length the beam can propagate without deviating significantly in radius. The beam waist $w_0$ can be expressed in terms of the wavelength and the angular width $\theta$ of the beam \citep{varga1998gaussian}:
\begin{equation}
	w_0=\frac{2 \lambda}{\pi \theta}.
    \label{eq:Theory/beamWaist}
\end{equation}

\begin{figure}
\centering
\includegraphics[width=.45\linewidth]{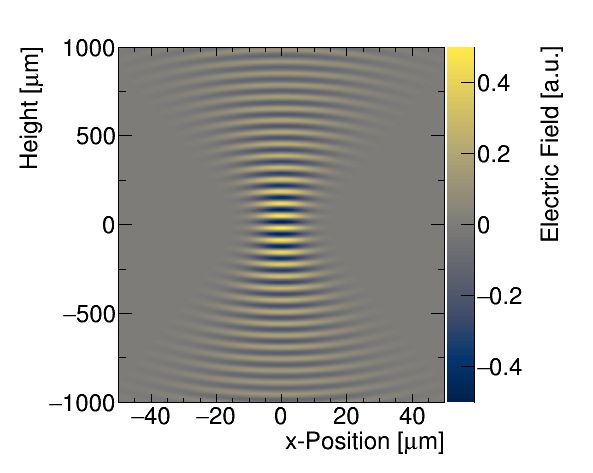}
\captionof{figure}{The calculated electric field for a Gaussian beam that is focused to a radius of \SI{7}{\micro\meter}. The Gaussian beam is moving towards the negative height direction.}
\label{fig:Theory/EFieldSpot}
\end{figure}

The real part of the electric field can be obtained by multiplying \autoref{eq:Theory/EField} with $\exp(-2i\pi c t / \lambda)$, where $c$ is the speed of light. The real part of the electric field for a wavelength of 680 \si{nm} with a beam waist radius of $\SI{7}{\micro\meter}$ is shown in \autoref{fig:Theory/EFieldSpot}. For a complete derivation of the solution for a Gaussian beam see \citep{varga1998gaussian}.

\subsection*{Minimum spot size due to magnification and theory}
\label{sec:Theory/FocusLaser}
To be able to focus a laser beam to a few \si{\micro\meter} it is important to choose the correct lens or lenses. In the case of a Gaussian profile, two equations govern the spot size that can fundamentally be achieved. The first one is due to the imaging of the fibre to the position of the focal point. A lens system images an object before the lens, to an image after the lens. By choosing the focal length and position of the object and image, a specific magnification can be achieved. Depending on the application of the lens, this magnification can be greater or smaller than one. 

For the laser setup that is used later on (see \autoref{sec:LaserSetup}), the output of the fibre will be imaged to the focal point. Therefore, a fundamental limit by this imaging on the spot size can be determined. For a two lens system, in which the first one collimates the light ($f_{in}$), and the second one focuses it ($f_{out}$), the magnification is given by the ratio of the second lens over the first lens \citep{Goodman:85}. Therefore, the minimum spot size due to imaging is
\begin{equation}
	d_{spot}=\frac{f_{out}}{f_{in}}d_{fiber}.
\end{equation}

The second constraint on the size of the focal spot is due to the beam waist approaching, but never reaching zero. To determine this size, \autoref{eq:Theory/beamWaist} can be rewritten to find an expression for the angular width of the beam. This yields
\begin{equation}
	\theta = \frac{4\lambda}{2 \pi w_0}.
    \label{eq:Theory/angularWidth1}
\end{equation}
The angular width can also be geometrically be determined assuming that the beam covers the complete lens, which has a diameter of $D$. The angular width is given by
\begin{equation}
	\theta=2\tan{\left(\frac{D}{2f_{out}}\right)} \approx \frac{D}{f_{out}}.
    \label{eq:Theory/angularWidth2}
\end{equation}
The paraxial approximation is again used to approximate $\tan(x)$ as $x$. 

By equating \autoref{eq:Theory/angularWidth1} to \autoref{eq:Theory/angularWidth2}, an expression for the beam waist can be found as a function of the properties of the second lens. The beam waist is given by
\begin{equation}
	w_0 = \frac{4\lambda}{\pi}\frac{f_{out}}{D}.
\end{equation}

However, this beam waist can never be reached if the image size of the initial object is bigger than this beam waist. Therefore the diameter of the minimal spot size is given by
\begin{equation}
	d_{min,spot}=\text{min} \left\{\frac{f_{out}}{f_{in}}d_{fiber}, \ \frac{4\lambda}{\pi}\frac{f_{out}}{D} \right\}.
    \label{eq:Theory/minSpotsize}
\end{equation}

\subsection*{Design of lens system}
For the setup we developed for this work, a laser is used with a wavelength of \SI{680}{\nano\meter} is used. The first lens of the lens system that is build is an aspherical lens (lens 1) with a focal length of \SI{18.4}{\milli \meter} and a diameter of \SI{5.52}{\milli \meter} in order to couple the pulse from the fibre to air. This lens is placed at such a distance that the focal point coincides with the opening of the fibre, and produces a parallel beam after the lens. In this parallel beam another aspherical lens (lens 2) is used to focus at the Timepix3. This lens has a focal length of \SI{2.75}{\milli \meter} and a diameter of \SI{3.6}{\milli \meter}. Using \autoref{eq:Theory/minSpotsize}, the minimum spot size that this setup is able to achieve can be calculated; and is \SI{1.35}{\micro \meter}. This spot size however can only be achieved if all of the photons are absorbed within the Rayleigh length (\autoref{eq:Theory/Laser/Rayleigh}), otherwise the beam will already have diverged significantly before all photons are absorbed. 

The setup is first simulated in Zemax OpticStudio to determine the minimum size and profile of the focal spot, in order to compare with the theoretical spot size. An optical fibre is modelled by two cylinders representing the core and the cladding of the fibre. The refractive indices were chosen such that the numerical aperture (NA) of the simulated fibre matched the fibre used in the setup. To determine the relation between the refractive index of the core and that of the cladding, the following relation for the NA is used
\begin{equation}
    \text{NA}=\sin(\alpha)=\sqrt{n_1^2-n_2^2}.
\end{equation}
Here $2\alpha$ is the complete acceptance angle of the fibre, $n_1$ is the refractive index of the core, and $n_2$ is the refractive index of the cladding. To couple light into this fibre, a lens is designed that matches the NA of the fibre. A plane emitting light perpendicular to its surface with a Gaussian profile is used as a source for the photons. At the other end of the fibre, a model of lens 1 is placed. The position is chosen such that the focal point of this lens coincided with the end of the fibre as described earlier. For multiple positions behind the lens it is confirmed that the beam did not diverge significantly. A model of lens 2 was then added to simulate the spot size and shape. Due to the parallel beam, the position of this lens does not influence the final spot size as the lens is well aligned with the previous one. To simplify the simulation, and thus make the simulation quicker, the fibre is replaced by a Gaussian source that matched the opening angle, NA, and the transverse mode of the fibre. 

\begin{figure}
\centering
\begin{minipage}{.47\textwidth}
\centering
\includegraphics[width=1\linewidth]{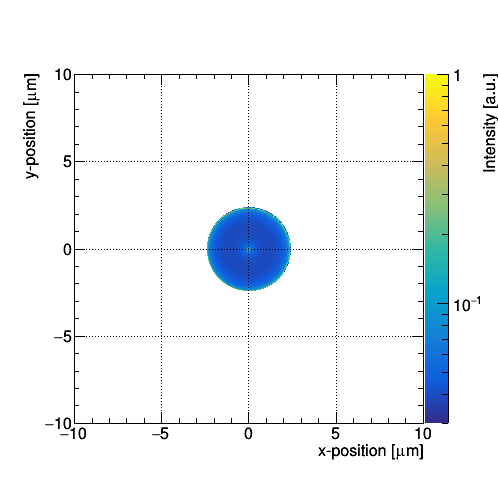}
\end{minipage}\qquad
\begin{minipage}{.47\textwidth}
\centering
\includegraphics[width=1\linewidth]{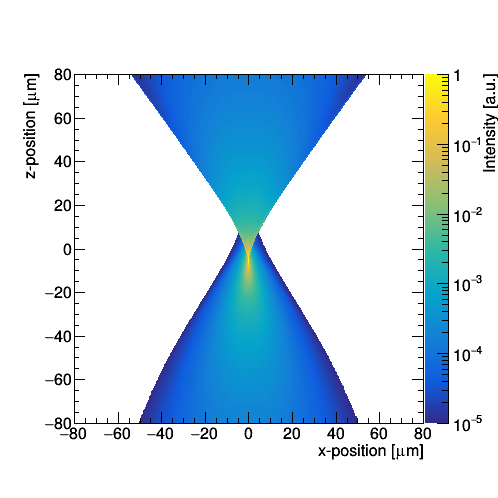}
\end{minipage}


\begin{minipage}[t]{.47\textwidth}
\centering
\captionsetup{width=.9\linewidth}
\captionof{figure}{The simulated intensity profile perpendicular to the beam direction at the focal point. The spot has a diameter of \SI{4.7}{\micro \meter}.}
\label{fig:LaserSetup/profileAtFocus}
\end{minipage}\qquad
\begin{minipage}[t]{.47\textwidth}
\centering
\captionsetup{width=.9\linewidth}
\caption{The simulated intensity profile in the centre of the beam parallel to the beam for a range of \SI{80}{\micro\meter} around the focal point.}
\label{fig:LaserSetup/CrossectionBeam}
\end{minipage}
\end{figure}

The parameters that are used for the simulation are shown in \autoref{tab:LaserSetup/SimulationTable}. Both the profile of the beam at the focal point and the transverse profile of the beam are calculated using the simulation described above. The simulated intensity profile perpendicular to the beam direction is shown in \autoref{fig:LaserSetup/profileAtFocus}. From this result, the minimum spot size is determined to be \SI{4.7}{\micro \meter} in diameter, and is thus bigger than the spot size that is expected from the calculation. This difference is most likely due to the physical imperfections of the lenses, as well as the placement of the lenses, which is never perfect. \autoref{fig:LaserSetup/CrossectionBeam} shows the beam profile parallel to the beam direction, and provides information about the Rayleigh length, and thus determines the length over which the light should be absorbed. For \SI{680}{\nano\meter}, the \SI{30}{\micro\meter} over which the beam does not diverge significantly is enough to absorb all the light. Therefore, the simulation confirmed that the two lenses that were chosen, are sufficient to achieve at least a sub-pixel spot size.

\begin{table}
   \caption{ Summary of the settings used during the simulation of the beam profile.} 
   \label{tab:LaserSetup/SimulationTable}
   \small 
   \centering 
   \begin{tabular}{lcr} 
   \toprule[\heavyrulewidth]\toprule[\heavyrulewidth]
   \textbf{Setting} & \textbf{Magnitude} & \textbf{Unit}\\ 
   \midrule
   Wavelength & 680 & \si{\nano\meter} \\
   Number of analysis rays & $5\cdot 10^{7}$&  \\
   Diameter of core & 9 & \si{\micro \meter} \\
   Refractive index core & 1.44 &  \\
   Diameter of cladding & 1.25 & \si{\milli\meter} \\
   Refractive index cladding & 1.43652 &  \\
   Length of fibre & 200 & \si{\milli \meter} \\
   Material of lens 1 & D-ZK3M &  \\
   Material of lens 2 & D-ZLaF52LAM &  \\
   \bottomrule[\heavyrulewidth] 
   \end{tabular}
\end{table}

The lens setup has been purchased at Thorlabs and placed in the setup after minor mechanical adjustments to fit in the laser setup.

\subsection{Achieved beam spot}
The size of the focus spot with the lens system installed has been experimentally determined. However, the size of the spot that can be measured with the Timepix3 is convoluted with the diffusion of the charge cloud during the drift to the analogue front-end. Though, this can still give an indication of the spot size. To calculate the spot size with the Timepix3, the spot should first be both vertically and horizontally aligned to the middle of a single pixel. The horizontal middle can be found by moving the laser horizontally until charge is collected in the neighbouring pixel. This way all four sides of the pixel can be found, after which the centre can be calculated. By moving the spot outside the centre pixel along one axis, it is possible to deposit zero charge inside the centre pixel. When this point is reached, the spot is slowly moved back towards the centre pixel in increments of \SI{2}{\micro\meter} until the spot crosses the complete pixel and no charge is injected again in this centre pixel. At each increment the average ToT is measured, which is proportional to the fraction of the spot that lies within the pixel. This way a charge versus position scan can be made. Such a scan should show no charge on both ends, because at those positions the spot is not positioned on the centre pixel, but on a neighbouring pixel. However, when the pixel is approached, the pixel collects gradually more charge until it collects all charge carriers liberated by the laser pulse. At this point the collected charge should stay constant until the spot reaches the neighbouring pixel on the other side. By fitting this profile with two error functions, the standard deviation $\sigma_{erf}$ is calculated. This width arises due to the intensity profile of the spot convoluted with the normal distribution of the charge cloud in the sensor ($\sigma_{diffusion}$). Because the spread of the charge cloud can be calculated for the sensor, the standard deviation $\sigma_{spot}$ of the laser spot can be found by using
\begin{equation}
	\sigma_{spot}=\sqrt{\sigma_{erf}^2-\sigma_{diffusion}^2}\ .
\end{equation}
\begin{figure}
\centering
\includegraphics[width=.85\linewidth]{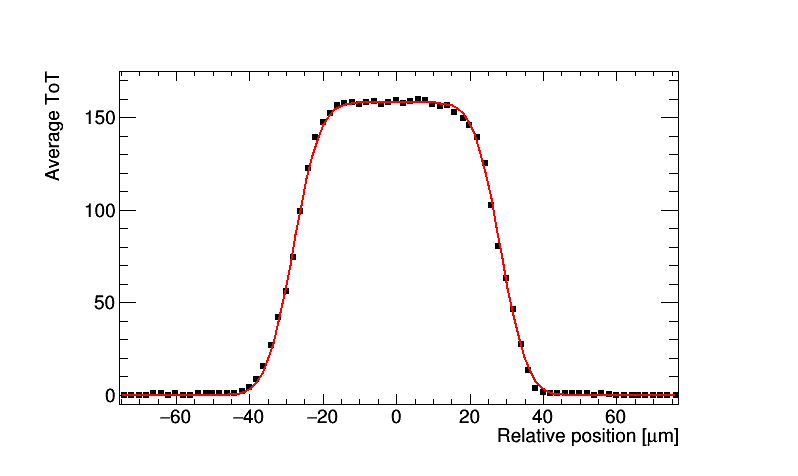}
\captionof{figure}{A ToT scan for the relative position of a single pixel. The black points indicate the measurement, and the red line indicates a fit to \autoref{eq:Laser/ToTScan}.}
\label{fig:LaserSetup/ChargeProfile}
\end{figure}

The measured charge profile made with the Timepix3 is shown in \autoref{fig:LaserSetup/ChargeProfile}, along with a fit to the combination of two error functions describing the integral of the Gaussian profile. The formula that is used to fit the scan is
\begin{equation}
	\text{ToT}(x)=a\left[ \erf\left(\frac{x-\mu_1}{\sigma_1}\right) - \erf\left(\frac{x-\mu_2}{\sigma_2}\right) \right].
	\label{eq:Laser/ToTScan}
\end{equation}
For a \SI{200}{\micro\meter} sensor biased at \SI{200}{\volt}, the average standard deviation of the two error functions is \SI{8.19}{\micro \meter} (see \autoref{fig:LaserSetup/ChargeProfile}). The standard deviation of due diffusion by thermal effects, assuming the voltage to reach full depletion is \SI{115}{\volt} and using \autoref{eq:Theory/ThermalDiffusion}, is \SI{2.8}{\micro \meter}, and the radius of the charge cloud due to diffusion at $26\cdot10^3$ electrons (ToT of 159) is \SI{10.86}{\micro \meter} (\autoref{eq:Theory/ElectrostaticDiffusion}). Assuming this radius is reached at a width of $3\sigma$, this gives a standard deviation for the laser spot of $6.79\pm0.09$ \si{\micro\meter}, corresponding to a FWHM of $16.0\pm0.2$ \si{\micro\meter}. The error on this standard deviation corresponds to the error on the value of the spot size determined by the fit. The standard deviation is relatively large compared to the calculated optimal focus spot radius of \SI{1.35}{\micro \meter}, however in this calculation it is assumed that all the components are aligned perfectly and the lenses do not suffer from spherical aberration, which is an optimistic assumption for lenses of this size. Even more, a small difference in the full depletion voltage changes the spread due to diffusion quite a lot, and thus influences $\sigma_{diffusion}$. This could explain the difference in the calculated and simulated spot size compared to the measured spot size.

After the installation of a lens system that was selected using an optical simulation, we have been able to achieve a focused beam spot FWHM of $16.0\pm0.2$ \si{\micro\meter}. This narrow beam spot is sufficient to perform our intended studies with the Timepix3 as it is clearly more narrow than the pixel size dimensions.

\section{Timing properties of the laser setup}
\begin{figure}
\centering
\begin{minipage}{.47\textwidth}
\centering
\includegraphics[width=1\linewidth]{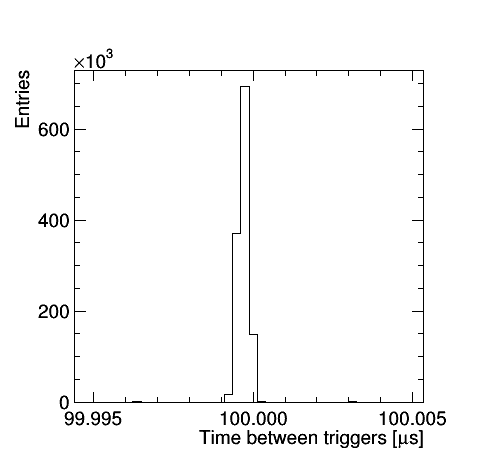}
\end{minipage}\qquad
\begin{minipage}{.47\textwidth}
\centering
\includegraphics[width=1\linewidth]{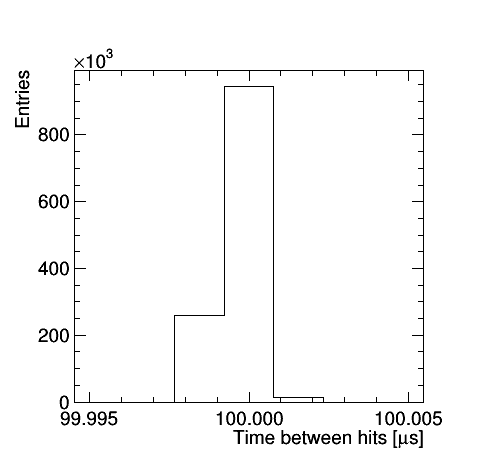}
\end{minipage}


\begin{minipage}[t]{.47\textwidth}
\centering
\captionsetup{width=.9\linewidth}
\captionof{figure}{The time between two subsequent triggers that were generated on the pulse generator and measured by the TDC on the Timepix3. The pulse generator is set to \SI{10}{\kilo\hertz}, resulting in a period of \SI{100}{\micro\second}.}
\label{fig:LaserSetup/TriggerPeriod}
\end{minipage}\qquad
\begin{minipage}[t]{.47\textwidth}
\centering
\captionsetup{width=.9\linewidth}
\caption{The time between two subsequent hits on the Timepix3. The laser is driven by a pulse with a frequency of \SI{10}{\kilo\hertz}, resulting in a period of \SI{100}{\micro\second}.}
\label{fig:LaserSetup/HitPeriod}
\end{minipage}
\end{figure}

The laser diode is driven by a pulse generator, which can be set to different repetition frequencies that define the time between subsequent laser pulses. This frequency should not be too high to ensure that the Timepix3 will not reach its maximum read-out speed of \SI{50}{\kilo\hertz} per pixel \citep{poikela2015readout}. To be well below this limit, a frequency of \SI{10}{\kilo\hertz} is used during the measurements. The period of the trigger signal at this frequency for $12\cdot10^4$ trigger pulses is shown in \autoref{fig:LaserSetup/TriggerPeriod}. This Figure shows the time between subsequent triggers measured with the TDC on the SPIDR. For the same measurement, the time between subsequent hits induced by the laser pulse and measured with the Timepix3 is shown in \autoref{fig:LaserSetup/HitPeriod}. From this it is concluded that there is a small deviation on the period of the laser pulses and the triggers. This is most likely induced by the electrical pulse propagation though the cables, which gives rise to a slight jitter on the rise time of the pulse.

\section{Optimizing laser performance}

\begin{figure}[h]
\centering
\begin{minipage}{.47\textwidth}
\centering
\includegraphics[width=1\linewidth]{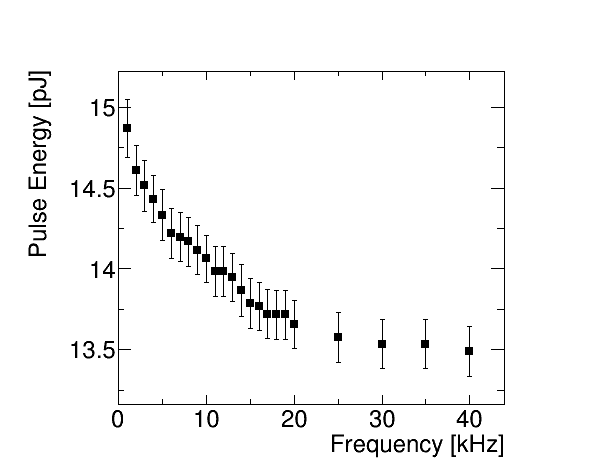}
\end{minipage}\qquad
\begin{minipage}{.47\textwidth}
\centering
\includegraphics[width=1\linewidth]{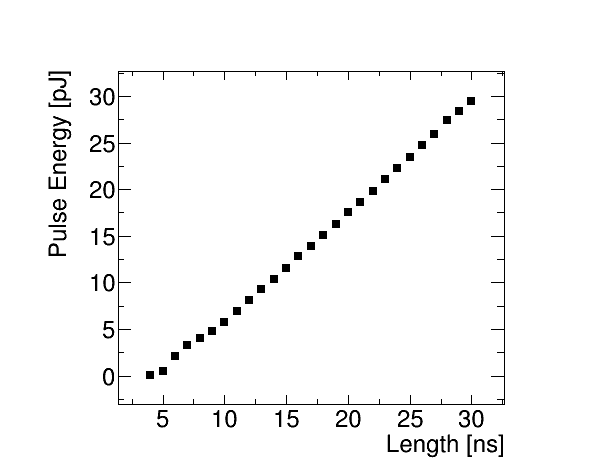}
\end{minipage}


\begin{minipage}[t]{.47\textwidth}
\centering
\captionsetup{width=.9\linewidth}
\captionof{figure}{The average energy per laser pulse for a range of repetition frequencies. The length of the pulse was fixed at \SI{16.6}{\nano\second} with a magnitude of \SI{3.0}{\volt}.}
\label{fig:LaserSetup/LaserFrequency}
\end{minipage}\qquad
\begin{minipage}[t]{.47\textwidth}
\centering
\captionsetup{width=.9\linewidth}
\caption{The average energy per laser pulse for different lengths of the pulse that drives the laser. The repetition frequency was fixed at \SI{10}{\kilo\hertz} with a pulse magnitude of \SI{3.0}{\volt}.}
\label{fig:LaserSetup/LaserLength}
\end{minipage}
\end{figure}

To quantify the behaviour of the laser diode that is used, three different scans were performed. \autoref{fig:LaserSetup/LaserFrequency} displays the energy of the laser pulse in picojoule for a range of repetition frequencies. From this curve it can be concluded that the energy per pulse slightly decreases when the frequency increases. However, at \SI{50}{\kilo\hertz} the maximum hit rate of a single pixel is reached, therefore the repetition frequency of the laser should not be increased beyond this value. To ensure that each hit is measured, a repetition frequency of \SI{10}{\kilo\hertz} is chosen. 

\begin{figure}[h]
\centering
\includegraphics[width=.47\linewidth]{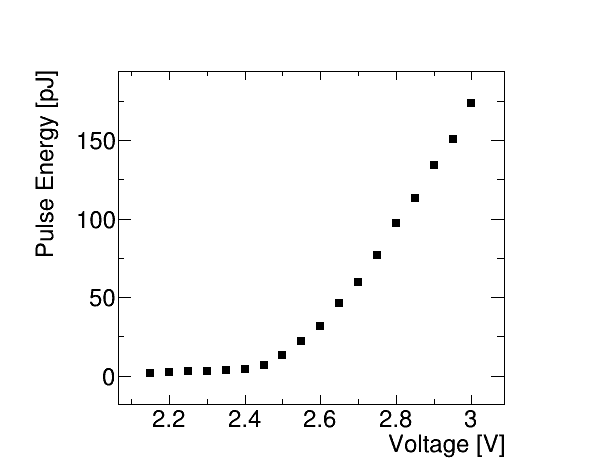}
\captionsetup{width=.423\linewidth}
\captionof{figure}{The average energy per laser pulse for different magnitudes of the pulse that drives the laser. The repetition frequency was fixed at \SI{10}{\kilo\hertz} with a pulse length of \SI{10.0}{\nano\second}.}
\label{fig:LaserSetup/LaserVoltage}
\end{figure}

Another variable that influences the energy per laser pulse is the length of the signal that drives the laser diode. The minimum length that can be achieved with the pulse generator used is \SI{4}{\nano\second}. The average pulse energy per laser pulse for a length of \SI{4}{\nano\second} to \SI{30}{\nano\second} is shown in \autoref{fig:LaserSetup/LaserLength}. In addition to the length of the pulse, the magnitude can also be changed. \autoref{fig:LaserSetup/LaserVoltage} illustrates the average pulse energy per laser pulse for a range of magnitudes. By choosing appropriate values for the length as well as the magnitude of the pulse, the intensity of the pulse can be tuned to the desired pulse energy. One should note that \SI{34}{\percent} of \SI{680}{\nano\meter} light is reflected at the air-silicon interface (see \citep{hecht2016optics} for the Fresnel equations). Therefore the intensity of the pulse within the silicon is just \SI{66}{\percent} of the total intensity.

\newpage

\chapter{Testbeam experiment}
\label{sec:Testbeam}

We have performed a beam test experiment in October of 2018 at the Super Proton Synchrotron (SPS) at CERN. In this experiment we have taken measurements using the Timepix3 telescope \citep{akiba2019arxiv} and studied several Timepix3 assemblies (which are summarised in \autoref{sec:Results}).

In this Chapter the various components of the Timepix3 telescope, as well as the offline analysis software used to convert the data from the Timepix3 telescope to a ROOT file is discussed in detail.

The Timepix3 telescope was located in H8A in the North Area at CERN. During the measurements the H8 beam line was used and consisted mostly of positively charged pions with a momentum of around \SI{180}{\giga\eV}. The beam that is supplied to the North Area is not a constant beam, but consists of spills. These spills are around \SI{4.7}{\second}, each containing around $10^6$ particles, and are followed by a period in which no particles are traversing the beam line. This period depends on the operating conditions of the SPS, and typically lasts around \SI{33}{\second}, though it can be as low as \SI{10}{\second}.

\section{Timepix3 telescope}
\label{sec:Timepix3/Telesope}

\begin{figure}
\centering
\captionsetup{width=.9\linewidth}
\includegraphics[width=.95\linewidth]{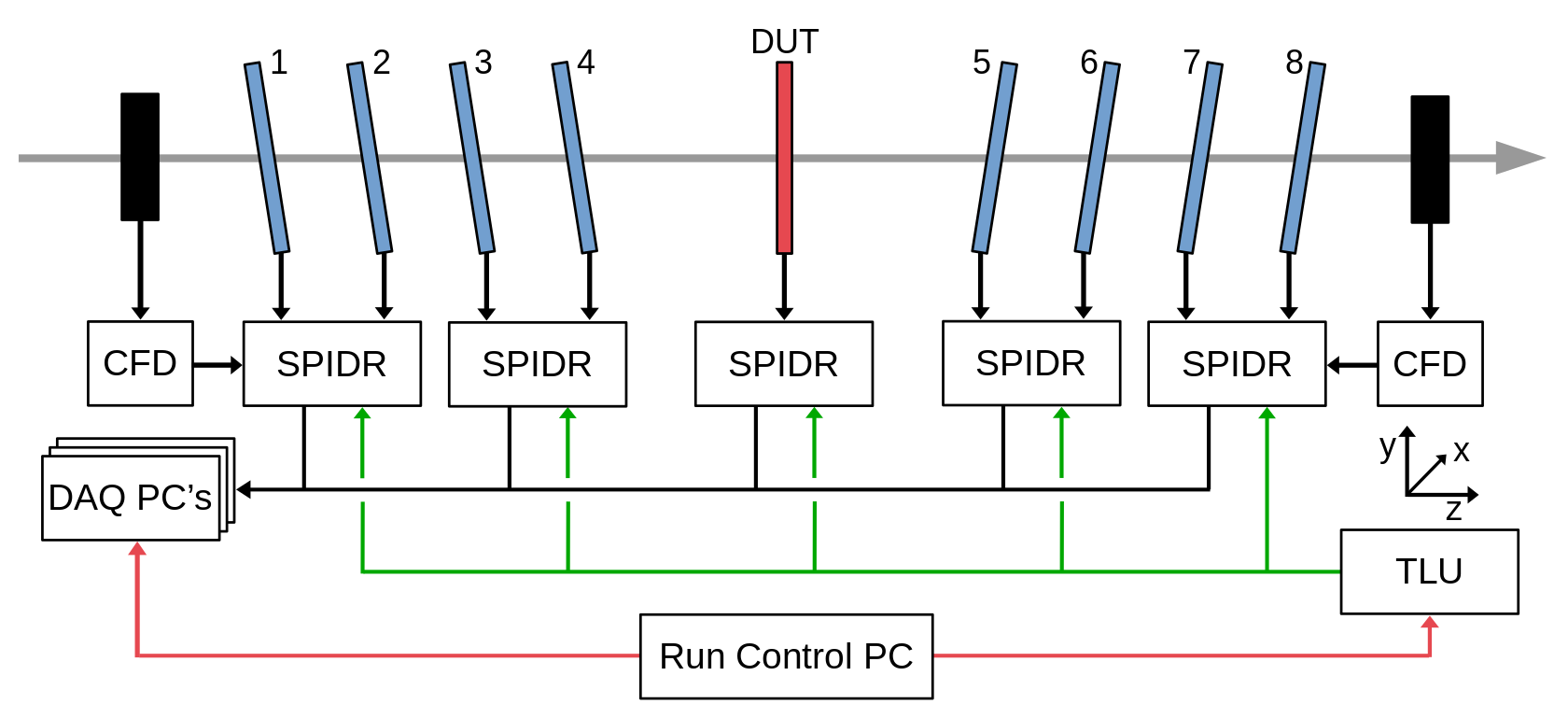}
\captionof{figure}{A schematic overview of the Timepix3 telescope; the eight blue planes represent the telescope's Timepix3s. The black squares represent the two scintillators, while the red plane represents the device-under-test. The black arrows indicate the direction of the data flow, and the green arrows indicate the \SI{40}{\mega\hertz} clock and the shutter signal. The beam is indicated by the grey arrow.}
\label{fig:TestbeamSetup/TelescopeSetup}
\end{figure}

\begin{figure}
\centering
\captionsetup{width=.9\linewidth}
\includegraphics[width=.95\linewidth]{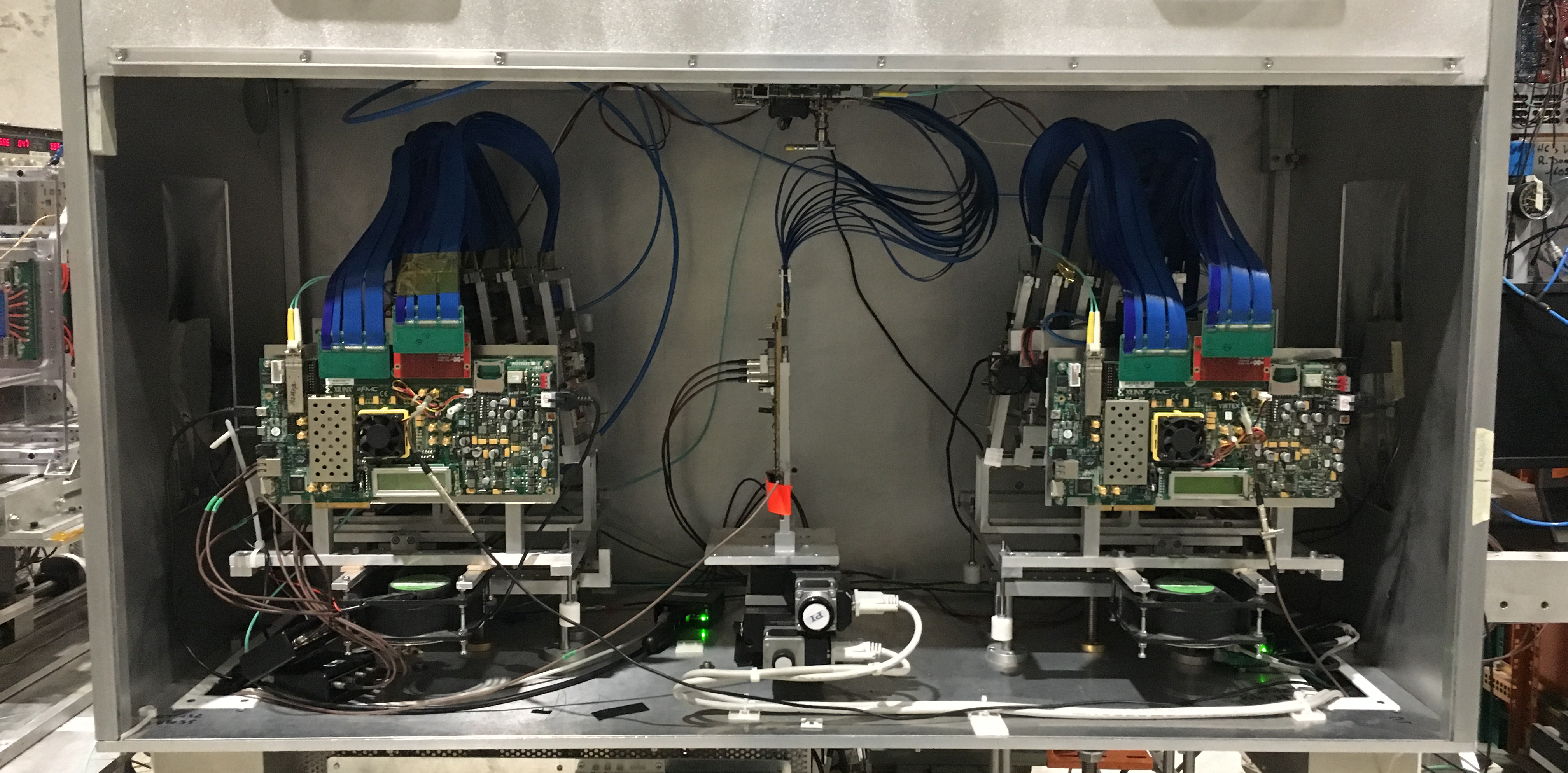}
\captionof{figure}{The Timepix3 telescope at the testbeam facilities at the North Area at CERN.}
\label{fig:TestbeamSetup/TelescopeSetupPicture}
\end{figure}

The Timepix3 telescope consists of eight Timepix3 planes with a \SI{300}{\micro\meter} thick silicon sensor attached to each Timepix3. A diagram of the telescope is shown in \autoref{fig:TestbeamSetup/TelescopeSetup}, as well as a picture of the telescope in \autoref{fig:TestbeamSetup/TelescopeSetupPicture}. The Timepix3s are connected to four SPIDRs, divided over two sliding arms. Using the position information from these eight Timepix3s, it is possible to reconstruct the track of a single particle. The telescope planes are tilted \SI{9}{\degree} in both the column and the row direction to ensure that most hits are spread over at least two pixels in both directions. By doing so, the resolution on the measured position of the particle improves \citep{akiba2019arxiv}. On both ends of the two arms, a scintillator is placed that is connected to a constant fraction discriminator (CFD). The coincidence signal from these two scintillators provides a trigger to the TDC input of one of the four SPIDRs, which will timestamp the track with a precision of better than \SI{300}{\pico\second} \citep{akiba2019arxiv}. The timestamp of the coincidence unit can be used in the offline analysis to determine the absolute delay between the recorded timestamp of the scintillator and the recorded timestamp of the Timepix3. The individual outputs of the CFDs are also used as second and third triggers, in order to have the option to use a single scintillator in the offline analysis. These triggers are solely used to timestamp the tracks, and are not used as a trigger to readout the data because the Timepix3 telescope uses is a data-driven readout (enabled by the large bandwidth of both the Timepix3 and SPIDR as well as the low noise performance of the Timepix3).

\begin{figure}
\centering
\includegraphics[width=\linewidth]{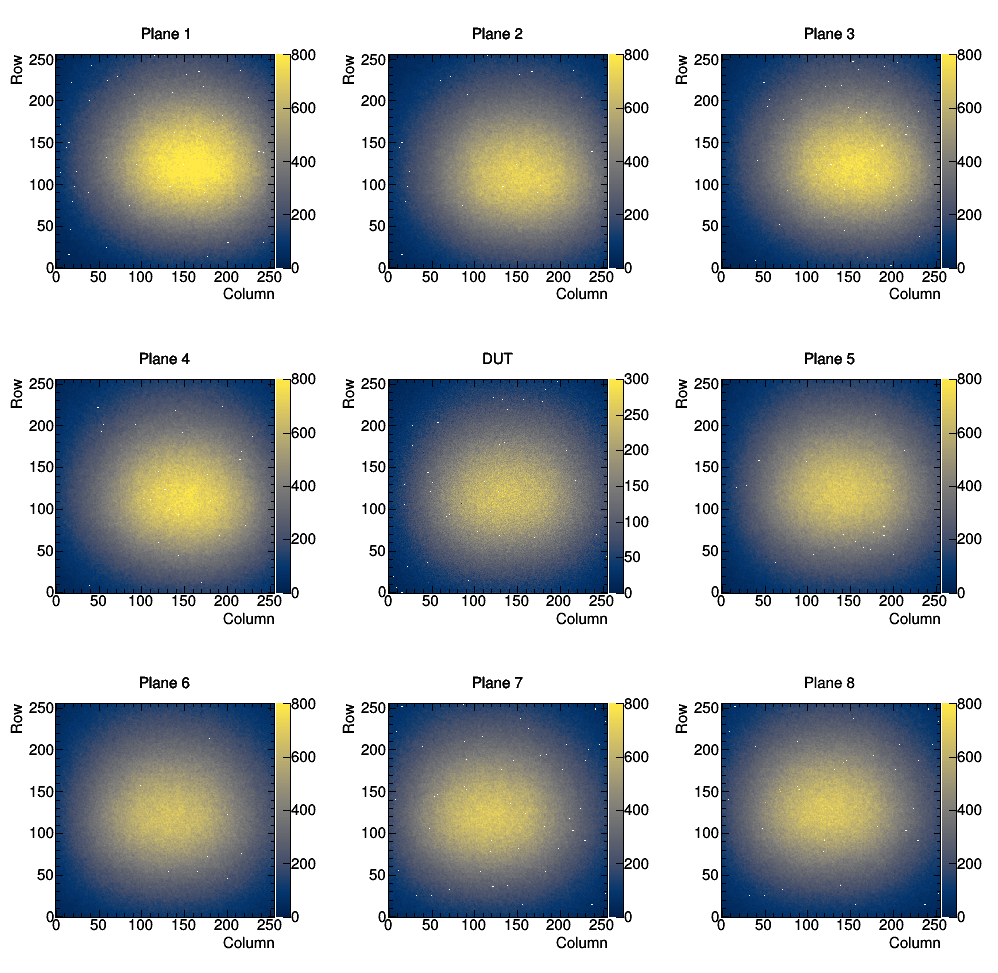}
\captionof{figure}{Hitmaps showing the location and number of hits for all nine planes, including the DUT, for a single run of the telescope. The beam spot is clearly visible on all planes.}
\label{fig:TestbeamSetup/hitmapTestbeam}
\end{figure}

In the middle of the two sliding arms, room is reserved for a separate device to be tested. This device is usually called the device-under-test (DUT), and is the detector that will be investigated. At the position of the DUT, the pointing resolution of the telescope is the best, with a resolution in the x-direction of $1.69\pm 0.16$ \si{\micro \meter}, and $1.55\pm 0.16$ \si{\micro \meter} in the y-direction \citep{akiba2019arxiv}. The DUT is placed on a motion stage that can move the DUT within the beam, while also having the ability to rotate \SI{360}{\degree} around the y-axis. The position of the beam within the eight telescope planes and the DUT for a real measurement is shown in \autoref{fig:TestbeamSetup/hitmapTestbeam}.

All five SPIDRs are connected to a Telescope Logic Unit (TLU), which ensures the synchronisation between the different telescope planes. The TLU provides the system clock to all SPIDRs such that there is no difference in the frequency and phase between the different detector planes. In addition to the clock, the TLU also supplies a signal to synchronise all time counters (indicated as green arrows in \autoref{fig:TestbeamSetup/TelescopeSetup}), as well as providing a shutter signal to control the start and stop of measurements. The TLU itself is controlled via a designated run control PC. This PC is used to control the complete telescope, as well as the motion stage on which the DUT is placed. 

When the SPIDRs receive the signal from the TLU to start a measurement, the data is not stored locally. The data is immediately transmitted to several Data Acquisition (DAQ) PC's. These DAQ PC's can be controlled by the run control PC as well, and can be accessed remotely. In this way, data can quickly be analysed after a measurement to check the position of the telescope and the DUT with respect to the beam. The data can later be copied from the DAQ PC's to other machines to perform both the alignment of the telescope and DUT, and to fully analyse the data. Both these actions are performed by the software application Kepler, described below.

\section{Offline analysis software}
\label{sec:Testbeam/Kepler}

The software application Kepler has been created to process the data from the Timepix3 telescope. Kepler is based on the Gaudi event-processing framework \citep{clemencic2010recent}, which is used in the CMS and LHCb collaborations. Kepler has been created to process the measurements of the telescope, as well as to perform the offline alignment for the telescope. It takes the raw data files produced by the five SPIDRs, and produces a single ROOT \citep{rootCern} file that can be analysed by the user.

The clustering algorithm that is used by Kepler is discussed first in \autoref{sec:Testbeam/Clustering}. Afterwards the algorithm that is used to construct tracks through the different planes is discussed (\autoref{sec:Testbeam/Tracking}), and finally the alignment of the telescope as well as the alignment of the DUT is discussed in \autoref{sec:Testbeam/Alignment} and in \autoref{sec:Testbeam/DUTTracks}.

\subsection{Pixel hit collection and clustering}
\label{sec:Testbeam/Clustering}

The collected pixel hits during a measurement are time sorted according to their timestamp after which they are assigned to events covering certain time windows. This timestamp, is the time the hit arrived after the start of the measurement. The length of the time windows is chosen such that the processing of a single events can be performed quickly by having a limited number of pixel hits that need to be assigned to clusters. Each of these events is processed separately. During the processing the both the clustering and tracking is performed. The clustering algorithm is discussed first below.

The clustering algorithm starts by taking the first hit in time, which is referred to as the seed pixel. The clustering algorithm loops over neighbouring pixels of the seed pixel and adds them to the cluster if these hits lie within a \SI{100}{\nano\second} time window. After all hits are assigned to the cluster, the various properties of the cluster are determined. The timestamp associated to the cluster is the earliest timestamp of the pixel hits within the cluster; the charge associated is the sum of the individual ToT values (converted to charge using the method described in \autoref{sec:Timepix3/ChargeCalibration}); the size of the cluster is defined as the number of pixels that are part of the cluster, and the position of the cluster is determined by the charged-weighted centre of gravity of the pixels. This position is calculated by
\begin{equation}
	p_{i,cluster} = \frac{\sum_j p_{i,j} Q_{j}}{\sum_j{Q_j}},
\end{equation}
where $j$ denotes the pixels within the cluster, $i$ denotes either the column position or the row position, and $Q_j$ is the charge of the pixel. 

\begin{figure}
\centering
\begin{minipage}{.47\textwidth}
\centering
\includegraphics[width=1\linewidth]{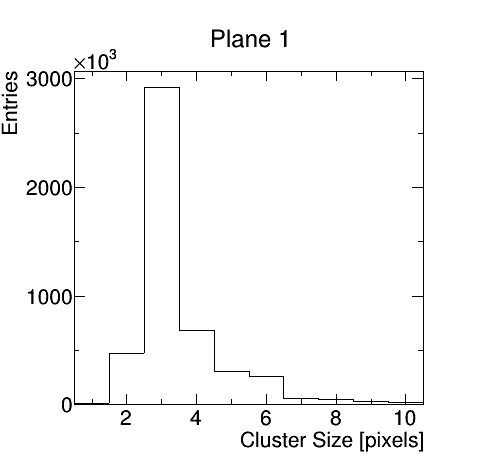}
\end{minipage}\qquad
\begin{minipage}{.47\textwidth}
\centering
\includegraphics[width=1\linewidth]{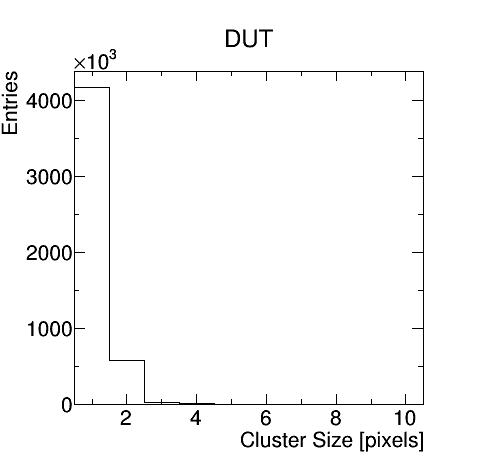}
\end{minipage}


\begin{minipage}[t]{.47\textwidth}
\centering
\captionsetup{width=.9\linewidth}
\captionof{figure}{The distribution of the number of hits that are assigned to a cluster for the first plane of the telescope. Because of the \SI{9}{\degree} tilt in both directions of the detector the average cluster size is three.}
\label{fig:TestbeamSetup/clusterSizePlane0}
\end{minipage}\qquad
\begin{minipage}[t]{.47\textwidth}
\centering
\captionsetup{width=.9\linewidth}
\caption{The distribution of the number of hits that are assigned to a cluster for a \SI{50}{\micro\meter} DUT.}
\label{fig:TestbeamSetup/clusterSizeDUT}
\end{minipage}
\end{figure}

Due to the angle of the telescope planes with respect to the beam, the average cluster size is around three, while for the DUT, which is placed perpendicular to the beam, the average size is around one. This angle of the telescope planes with respect to the beam is chosen such that the created charge cloud is spread over at least \SI{55}{\micro\meter} in both lateral direction before having traversed the complete depth of the telescope planes. Depending on the position of the hit within a pixel, the charge can spread to neighbouring pixels (giving rise to larger cluster sizes). This effect can be observed in the size distribution of both the telescope planes as well as in the DUT. The cluster size distribution for the first telescope plane is shown in \autoref{fig:TestbeamSetup/clusterSizePlane0}, and the cluster size distribution for the DUT is shown in \autoref{fig:TestbeamSetup/clusterSizeDUT}. 

\begin{figure}

\begin{minipage}{.5\linewidth}
\centering
\subfloat[]{\label{fig:TestbeamSetup/oneHitClusters}\includegraphics[width=1\linewidth]{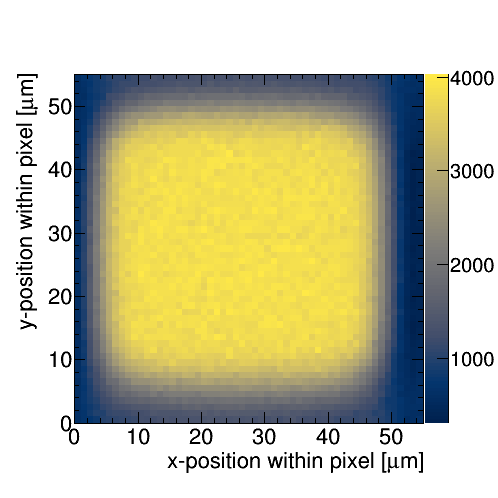}}
\end{minipage}%
\begin{minipage}{.5\linewidth}
\centering
\subfloat[]{\label{fig:TestbeamSetup/twoHitClusters}\includegraphics[width=1\linewidth]{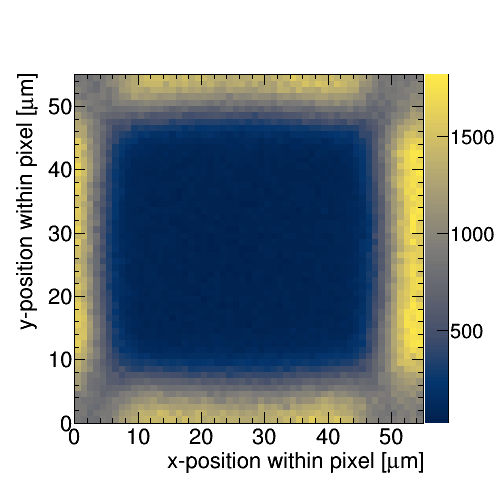}}
\end{minipage}\par\medskip
\begin{minipage}{.5\linewidth}
\centering
\subfloat[]{\label{fig:TestbeamSetup/threeHitClusters}\includegraphics[width=1\linewidth]{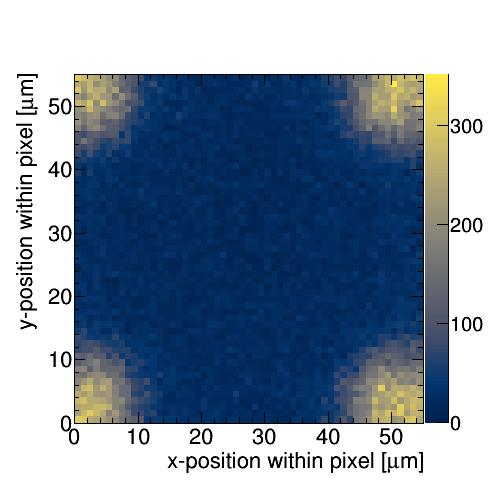}}
\end{minipage}%
\begin{minipage}{.5\linewidth}
\centering
\subfloat[]{\label{fig:TestbeamSetup/fourHitClusters}\includegraphics[width=1\linewidth]{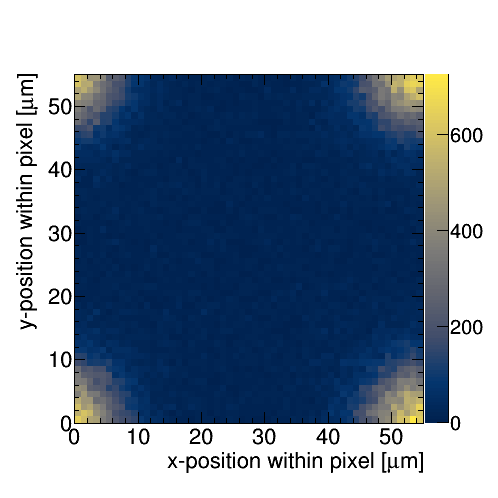}}
\end{minipage}

\caption{The position of the tracks within the pixels for different cluster sizes on the DUT. Clusters with size one, two, three, and four are shown in (A), (B), (C), and (D) respectively.}
\label{fig:mainClusterSizes}
\end{figure}

The size of the cluster on the DUT also gives an indication of the position of the track within the pixel. When a particle traverses the DUT in the middle of a pixel, it is less likely that the cluster will be more than one pixel in size. Closer to the sides of the pixel the average cluster size will become two, due to the spread of charge within the sensor (as described in \autoref{sec:Theory/DiffusionChargeCloud}). Following this, it is most likely to have a three or four pixel cluster in the corners of a pixel. This can also be observed in the position of the hits within a pixel. The positions of the tracks of which the associated cluster has either one, two, three, or four pixels, are shown respectively in \autoref{fig:TestbeamSetup/oneHitClusters}, \autoref{fig:TestbeamSetup/twoHitClusters}, \autoref{fig:TestbeamSetup/threeHitClusters}, and \autoref{fig:TestbeamSetup/fourHitClusters}. The slight offset in the x-profile, is due to a small offset in the rotation around the y-axis for the DUT. This indicates that the detector is not completely perpendicular to the beam.

When the clustering algorithm encounters a pixel that is either dead or masked, it would normally stop at this pixel. This could result in a cluster being smaller than they actually are. To counteract this problem, the cluster algorithm is allowed to skip over one pixel, and continue the clustering on the neighbouring pixels of the dead or masked pixel. 

After all hits are assigned to a cluster, the tracking procedure can start. This tracking procedure uses the clusters to construct tracks through the telescope.

\subsection{Time based tracking}
\label{sec:Testbeam/Tracking}

The tracking algorithm uses the precise timestamp of the clusters to efficiently reconstruct particle trajectories throughout the telescope.. During the tracking procedure the DUT is excluded to prevent a bias on the track positions due to the clusters on the DUT. The tracking algorithm starts with a seed cluster on the first telescope plane that is not yet associated to a track. It then searches for a cluster on the second telescope plane that is within \SI{10}{\nano\second} of this seed cluster; if a cluster is found, these two clusters are combined to construct a seed track. A straight line is fitted through these two cluster positions and extrapolated to the third plane, where a search is performed for a cluster within the \SI{10}{\nano\second} window as well as within a spatial window. The spatial window is determined by a maximum opening angle set at \SI{0.01}{\radian}, which translate to a spatial window of \SI{0.01}{\radian}$\cdot dz$, where $dz$ is the distance between the second and third plane. This spatial window allows for possible scattering of particles in the detectors. This process continues for the next plane, and is repeated until a track is reconstructed throughout all eight planes. The track is disregarded if an associated cluster can not be found on one or more planes. This tracking sequence is repeated until all clusters are either associated to a track or have been determined to not be part of a track at all.

\begin{figure}
\centering
\includegraphics[width=0.75\linewidth]{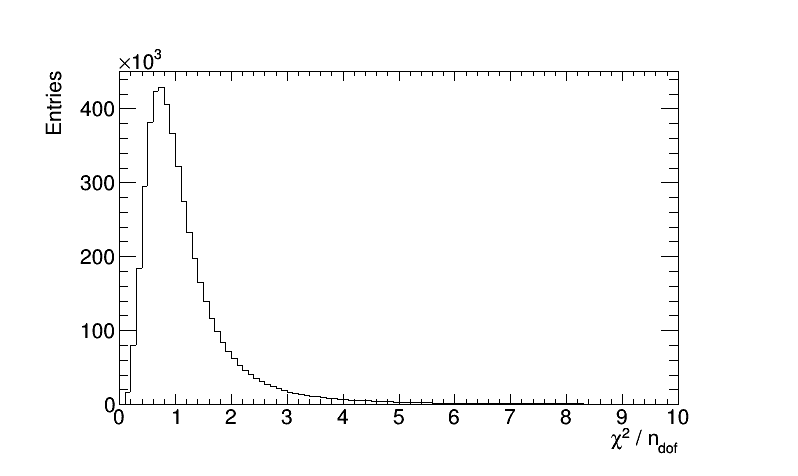}
\captionof{figure}{The distribution of $\chi^2$ per degree of freedom for all the tracks in a single run.}
\label{fig:TestbeamSetup/chi2Tracks}
\end{figure}

Once tracks have been reconstructed, they are fitted with a straight line to reconstruct the trajectory of the particle. Tracks with a $\chi^2$ per degree of freedom that exceeds ten are not considered as tracks, and are thus neglected from this point onward. The distribution of the $\chi^2$ per degree of freedom for these tracks is shown in \autoref{fig:TestbeamSetup/chi2Tracks}. Most tracks have a $\chi^2$ per degree of freedom of less than two, and the distribution peaks at one, indicating that a straight line through the individual clusters is a good approximation for the track.

The timestamp of the track is reconstructed from the average timestamp of the eight clusters that make up the track. However, during the testbeam, scintillators are also used to provide a trigger to the SPIDR's TDC. These triggers can also be used to timestamp the track. To do so, all triggers that fall within \SI{100}{\nano\second} of the track timestamp are associated to the track. 

A problem which may occur frequently in the almost continuous flux of SPS spills, is when two tracks are too close together in time; in this case, multiple triggers are associated to a single track. To avoid this, the track is only selected in the offline processing when a single trigger is associated, otherwise it is disregarded. Due to the physical layout of the scintillators, not all tracks will be assigned a timestamp from the scintillators. This results in a lower tracking efficiency of the telescope compared to timestamping based on the average timestamp of the associated clusters.

\subsection{Alignment procedure}
\label{sec:Testbeam/Alignment}

The alignment of the telescope with respect to itself and the DUT is an essential part of the analysis process. Without the alignment, the telescope cannot be used as a tool to determine the track position at the DUT. This is performed in two stages using Kepler; first the eight planes of the telescope are aligned, following which the relative position and rotation of the DUT is determined.

Around $10^4$ tracks are used to align the telescope. To begin with, the algorithm loads the $10^4$ tracks from the raw data files which where produced during the measurement. The Millepede algorithm \citep{blobel2006software} is used to align the telescope planes in space. This will determine six parameters for each telescope plane: $X,Y,Z,\phi,\theta,$ and $\psi$. These parameters are the x-, y-, z-position, rotation around the x-,  y-, and z-axis respectively. Initially the alignment sequence tries to align the planes with respect to the first plane by minimizing the x- and y-cluster residuals of the track on each plane, allowing only to vary $X,Y,$ and $\psi$. After this partial alignment, the Millepede algorithm is used to minimize the $\chi^2$ of tracks that have a $\chi^2$ of less than 1000, allowing to change all variables. This minimization is now repeated a second time, but with a tighter restriction on the $\chi^2$ of the tracks, allowing this to be at maximum 100. 

\begin{figure}
\centering
\includegraphics[width=\linewidth]{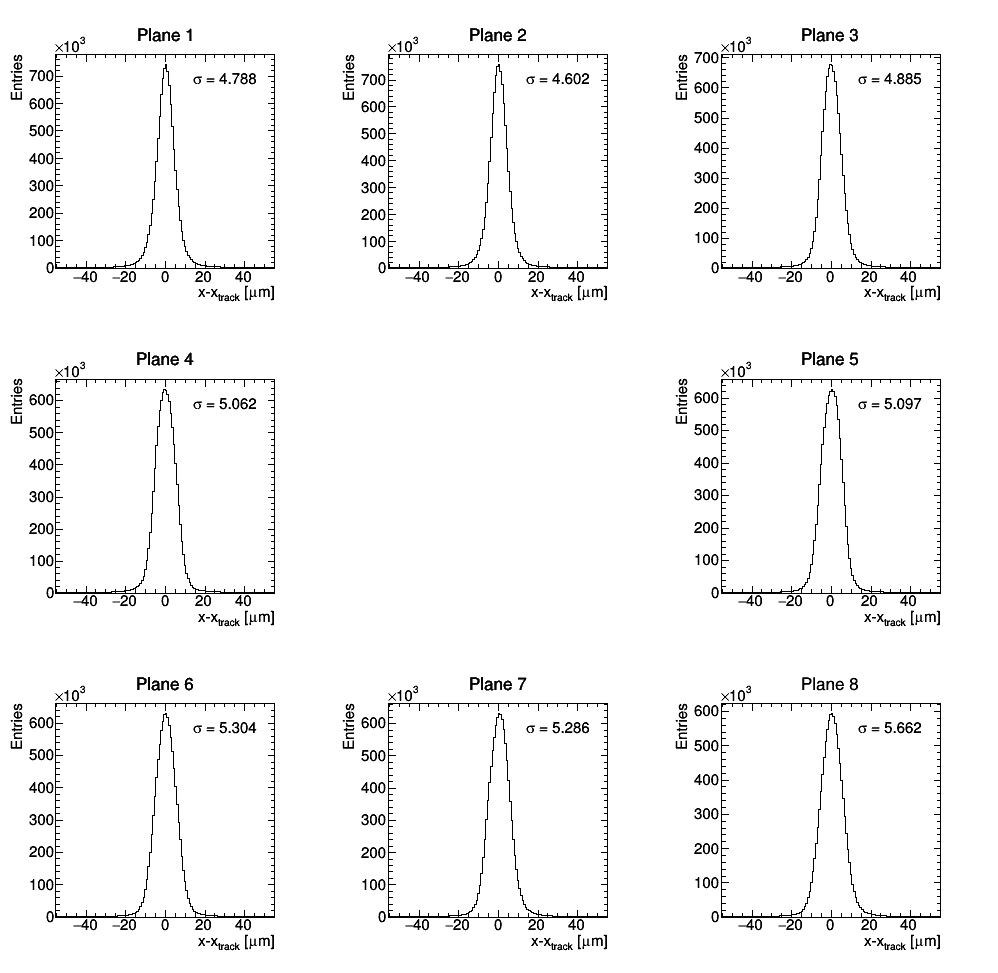}
\captionof{figure}{The unbiased residuals for all telescope planes. In the top right of the Figures, the standard deviation in \si{\micro\meter} of the unbiased residuals is displayed. This standard deviation is determined via a fit to a normal distribution.}
\label{fig:TestbeamSetup/UnbiasedResTelescopePlanes}
\end{figure}

By looking at the difference between the calculated track position and the position of the hit on the planes of the telescope, the alignment can be checked manually after the alignment sequence has finished. These differences should follow a normal distribution with a standard deviation of around \SI{5}{\micro\meter}-\SI{6}{\micro\meter}. These distributions are shown in \autoref{fig:TestbeamSetup/UnbiasedResTelescopePlanes} for all telescope planes. The standard deviation of the normal distribution is shown in the top right of the individual plots in this Figure.

Now that the telescope is aligned, the same set of tracks that is used for the alignment of the telescope is used to align the DUT with respect to the telescope. For this part of the alignment the Minuit algorithm \cite{james2004minuit} is used. Using this algorithm, the six free parameters for the DUT are determined by minimizing the residual distribution. This alignment can then be used to reconstruct measurements that are taken in the same configuration.

\subsection{Cluster association on DUT}
\label{sec:Testbeam/DUTTracks}
As mentioned before, the DUT is excluded from the track fitting algorithm. By excluding the DUT, the telescope provides unbiased track positions at the z-position of the detector. This z-position is known from the alignment of the DUT, and provides the conversion from the global coordinates, in which tracks are defined, to the local coordinates of the DUT. The residuals of the x- and y-position of the clusters on the DUT can be used to check the quality of the alignment and allows the measurement of the spatial resolution of the DUT. If the centre of the distribution is non-zero, the alignment of the DUT has not converged and should be checked and re-performed. 

\begin{figure}
\centering
\begin{minipage}{.47\textwidth}
\centering
\includegraphics[width=1\linewidth]{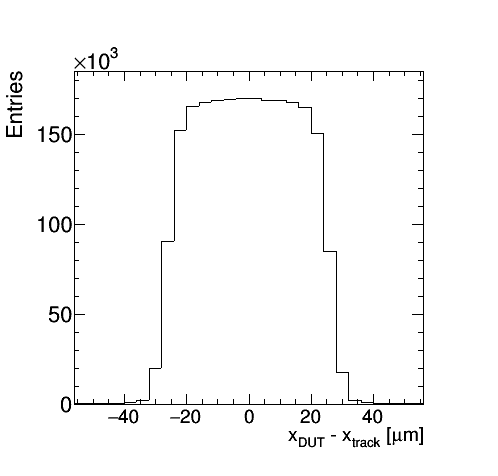}
\end{minipage}\qquad
\begin{minipage}{.47\textwidth}
\centering
\includegraphics[width=1\linewidth]{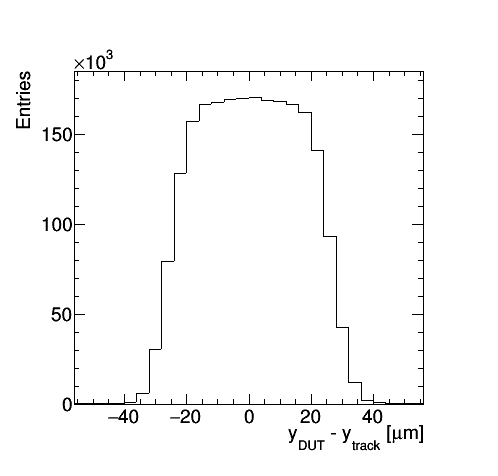}
\end{minipage}


\begin{minipage}[t]{.47\textwidth}
\centering
\captionsetup{width=.9\linewidth}
\captionof{figure}{The difference between x-position of the cluster on the DUT and the x-position of the track that is reconstructed with the telescope.}
\label{fig:TestbeamSetup/XResiduals}
\end{minipage}\qquad
\begin{minipage}[t]{.47\textwidth}
\centering
\captionsetup{width=.9\linewidth}
\caption{The difference between x-position of the cluster on the DUT and the x-position of the track that is reconstructed with the telescope.}
\label{fig:TestbeamSetup/YResiduals}
\end{minipage}
\end{figure}

The unbiased x- and y-residuals of the DUT for a single run of the testbeam are shown in \autoref{fig:TestbeamSetup/XResiduals} and \autoref{fig:TestbeamSetup/YResiduals}. The RMS of the x-residuals is \SI{15.6}{\micro\meter}, and the RMS of the y-residuals is \SI{16.0}{\micro\meter}. This RMS is the around the expected resolution of \SI{55}{\micro\meter}/$\sqrt{12}=$ \SI{15.9}{\micro\meter}, assuming a spatial bin with a width of \SI{55}{\micro\meter} corresponding to the pixel pitch. Indicating that the resolution of the DUT is dominated by single pixel hits, and not by charge shared hits (which would improve the resolution beyond \SI{15.9}{\micro\meter}).

\subsection{Data processing}
To construct an entire measurement, Kepler needs to iterate through a few steps. The total data that is taken during a single measurement could be up to gigabytes in size, therefore Kepler will read through and process the data in smaller parts. For each part, it reads the raw data and locally stores the pixel hits within that data. For each plane it clusters the hits according to the process described above. After the construction of the clusters, it will construct tracks through the cluster as described before. The clusters on the DUT are excluded from the fitting of the track to not bias the position of the track. After all tracks are processed, the desired data is written to a ROOT file for further analysis by the user. This data includes values such as the global position of the tracks, the clusters and their timestamp associated to the track, the intersection point of the track with the DUT, and the associated triggers and their corresponding timestamps. The parameter windows for both track selection and fitting as well as for the clustering of the hits are summarized in \autoref{tab:TestbeamSetup/KeplerTable}.

\begin{table}
   \caption{The constraints on the various variables that are used by Kepler during the analysis of the testbeam data.}
   \label{tab:TestbeamSetup/KeplerTable}
   \small 
   \centering 
   \begin{tabular}{lc} 
   \toprule[\heavyrulewidth]\toprule[\heavyrulewidth]
   \textbf{Requirement} & \textbf{Constraint}\\ 
   \midrule
   Time between hits in a cluster & $\leq$ \SI{100}{\nano\second} \\
   Track time window & $\leq$ \SI{10}{\nano \second} \\
   Number of clusters per track & 8 \\
   Number of clusters per track per plane & 1 \\
   Trigger time window & $\leq$ \SI{100}{\nano\second} \\
   Number of triggers within time window & 1 \\
   \bottomrule[\heavyrulewidth] 
   \end{tabular}
\end{table}

\section{The telescope scintillators}
The scintillators mounted on either sides of the telescope are an important part of determining the track timing resolution. It is therefore important to understand the signal that is created by the coincidence setup as well as the separate CFDs; these signals are eventually coupled to the trigger input of the SPIDR. The scintillators both have a dead time after the registration of a hit, due to the length of the pulse that is generated by the scintillator as well as the length of the pulse generated by the NIM logic that is used. 

\begin{figure}
\centering
\begin{minipage}{.47\textwidth}
\centering
\includegraphics[width=1\linewidth]{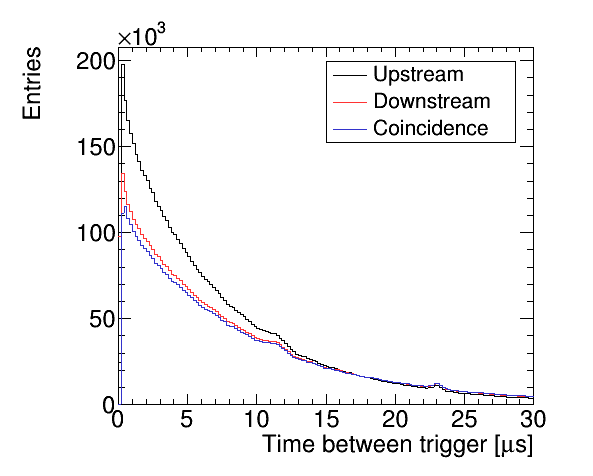}
\end{minipage}\qquad
\begin{minipage}{.47\textwidth}
\centering
\includegraphics[width=1\linewidth]{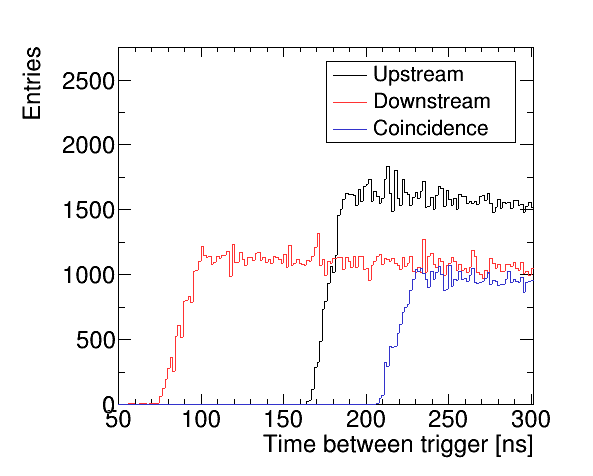}
\end{minipage}


\begin{minipage}[t]{.47\textwidth}
\centering
\captionsetup{width=.9\linewidth}
\captionof{figure}{The distribution of time differences between subsequent triggers of the two scintillators and the coincidence unit. The top curve indicates the upstream scintillator, the middle curve indicates the downstream scintillator, and the bottom curve indicates the coincidence setup.}
\label{fig:TestbeamSetup/TriggerPeriod}
\end{minipage}\qquad
\begin{minipage}[t]{.47\textwidth}
\centering
\captionsetup{width=.9\linewidth}
\caption{ The same measurement as \autoref{fig:TestbeamSetup/TriggerPeriod}, but now only showing the first \SI{300}{\nano\second}. The dead time of the scintillators and the NIM logic can be derived from this Figure. }
\label{fig:TestbeamSetup/TriggerPeriodZoomed}
\end{minipage}
\end{figure}

This dead time is determined by looking at the time between triggers from the scintillators. The distribution of times between sequential triggers is shown in \autoref{fig:TestbeamSetup/TriggerPeriod} for the two scintillators individually, and for the coincidence unit. \autoref{fig:TestbeamSetup/TriggerPeriodZoomed} shows a zoom-in of the range from \SI{50}{\nano \second} to \SI{300}{\nano\second}, to illustrate the different dead times of the scintillators and the NIM logic. The difference in the number of entries is due to the different positions of the scintillators with respect to the beam; while smaller number of entries of the coincidence trigger with respect to the downstream scintillator indicates that the two scintillators are not perfectly aligned with respect to each other and the beam. The dead time differs between both scintillators and NIM logic, and is \SI{164}{\nano\second} for the upstream scintillator and \SI{75}{\nano\second} for the downstream scintillator, while the dead time for the coincidence unit is \SI{206}{\nano\second}. This dead time implies that there is a loss in efficiency of the tracked particles due to sequential particles sometimes being less than \SI{206}{\nano\second} apart.

The exponential decay of the time between triggers is expected, due to the Poisson statistics of the beam. Poisson statistics dictate that the probability of two tracks occurring with a time interval $\delta_t$ is
\begin{equation}
	P(\delta_t) = 1-\exp(-\lambda \delta_t).
\end{equation}
The particle rate $\lambda$ that traverses the telescope can be determined by fitting this probability equation to the integrated distribution of the time between triggers. In this way the particle rate is determined to be $940\pm 0.8$ \si{\kilo\hertz}. The peak in triggers with a multiple time delays of \SI{11.5}{\micro\second} is an artifact of how particle bunches are distributed throughout the SPS. This time corresponds to exactly half the length of the SPS, and thus explains the bump in the distribution.

\chapter{Timing performance studies of the Timepix3}
\label{sec:Results}

We have studied in total three different sensors that are attached to Timepix3 ASICs using data from test pulses, the October 2018 testbeam, and the laser setup. The main difference between these sensors is the thickness and design. Two planar silicon sensors are studied, along with a 3d silicon sensor \citep{pellegrini2008first}. In \autoref{tab:Results/DataTable} you find a detailed summary of the different sensors as well as the measurements performed with these sensor and the settings used during these measurements. A detailed run list of the different runs and sensors from the testbeam is shown in \autoref{tab:Results/RunTable}. Several analysis with all three methods have been performed and will be presented in this Chapter.

\begin{table}
   \caption{The three different sensors that are investigated during this work. The sensor properties are shown, as well as the corresponding methods and operating conditions that are used to measure the sensors.}
   \label{tab:Results/DataTable}
   \scriptsize 
   \centering 
   \begin{tabular}{c|c|c|c} 
   \toprule[\heavyrulewidth]\toprule[\heavyrulewidth]
   \textbf{Device} & \textbf{Sensor} & \textbf{Laser} & \textbf{Testbeam} \\ 
   \midrule
   \makecell{W0011I05 } & \makecell{\SI{285}{\micro\meter} 3D silicon sensor \\ Height of pillars: \SI{250}{\micro\meter} \\ Radius of pillars: \SI{5}{\micro\meter}\\Produced by: CNM-IMB \\ Wafer number: 11 \\ Position on wafer: I05 \\ Depletion voltage: \SI{20}{\volt}} & No & \makecell{ Threshold: 800 e$^-$ \\ Ikrum: 5 \\ Bias voltage telescope: \SI{200}{\volt}} \\
   \midrule
   
   W0020J07 & \makecell{\SI{200}{\micro\meter} planar silicon sensor \\ n-on-p \\ Produced by: Hamamatsu \\ Implant size: unknown\\Guard ring: none \\ Wafer number: 20 \\ Position on wafer: J07 \\ Depletion voltage: \SI{115}{\volt}} & \makecell{Threshold: 700 e$^-$ \\ Ikrum: 10 \\Bias voltage: \SI{200}{\volt}} & \makecell{Threshold: 700 e$^-$ \\ Ikrum: 5 \\ Bias voltage telescope: \SI{200}{\volt}} \\
   \midrule
   
   W0039I11 & \makecell{\SI{50}{\micro\meter} planar silicon sensor \\ n-in-p \\Implant size: \SI{30}{\micro\meter} (circular)\\Guard ring: floating \\Produced by: Advacam \\ Wafer number: 39 \\ Position on wafer: I11 \\ Depletion voltage: \SI{20}{\volt}} & No & \makecell{Threshold: 700 e$^-$ \\ Ikrum: 5 \\ Bias voltage telescope: \SI{200}{\volt}} \\
   
   \bottomrule[\heavyrulewidth] 
   \end{tabular}
\end{table}

\begin{table}
   \caption{The bias voltage that is used for the different sensors and the number of spills (1 spill $\approx$ $10^6$ particles) that are measured, as well as the corresponding run numbers.}
   \label{tab:Results/RunTable}
   \scriptsize 
   \centering 
   \begin{tabular}{c|c} 
   \toprule[\heavyrulewidth]\toprule[\heavyrulewidth]
   \textbf{Device} & \textbf{Runs} \\ 
   \midrule
   W0011I05 & \begin{tabular}{c|c|c}
   Bias voltage (V) & \# of spills & Run number(s) \\
   \midrule
   50 & 20 & 31192 \\
   60 & 20 & 31193 \\
   \end{tabular} \\
   \midrule
   
   W0020J07 &  \begin{tabular}{c|c|c}
   Bias voltage (V) & \# of spills & Run number(s) \\
   \midrule
   25 & 20 & 31513 \\
   50 & 20 & 31512 \\
   75 & 20 & 31511 \\
   100 & 20 & 31510 \\
   125 & 20 & 31509 \\
   150 & 20 & 31506 \\
   175 & 20 & 31505 \\
   200 & 20 & 31504 \\
   225 & 20 & 31503 \\
   250 & 20 & 31502 \\
   275 & 20 & 31501 \\
   300 & 20 & 31500 \\

   \end{tabular} \\
   \midrule
   
   W0039I11 &  \begin{tabular}{c|c|c}
   Bias voltage (V) & \# of spills & Run number(s) \\
   \midrule
   10 & 20 & 31273 \\
   15 & 14 & 31270 \\
   15 & 7 & 31268 \\
   20 & 20 & 31267 \\
   30 & 20 & 31263 \\
   40 & 20 & 31262 \\
   50 & 20 & 31261 \\
   60 & 20 & 31260 \\
   70 & 20 & 31259 \\
   80 & 20 & 31258 \\
   90 & 20 & \makecell{31257, 31439, 31440 \\ 31441, 31442, 31443} \\
   \end{tabular} \\

   \bottomrule[\heavyrulewidth] 
   \end{tabular}
\end{table}

First we discuss the timewalk correction and some observed features in the delay spectrum. After this the timing structure observed over the Timepix3 ASIC is discussed, as well as the fToA bin sizes for different pixels. Finally, using the measured timing structure of the Timepix3, an improvement of the time resolution is presented.

\section{Performance before and after timewalk correction}

\begin{figure}
\centering
\begin{minipage}{.47\textwidth}
\centering
\includegraphics[width=1\linewidth]{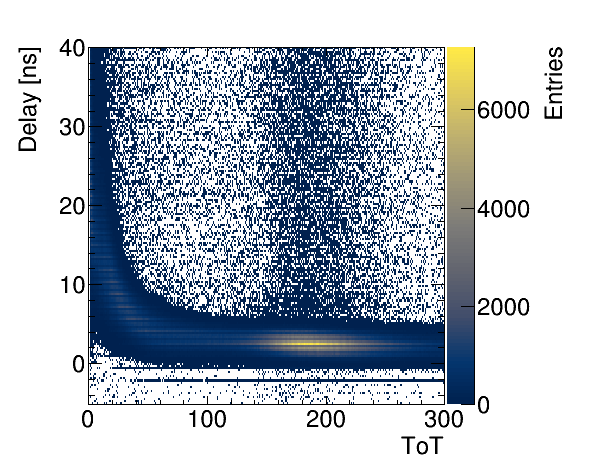}
\end{minipage}\qquad
\begin{minipage}{.47\textwidth}
\centering
\includegraphics[width=1\linewidth]{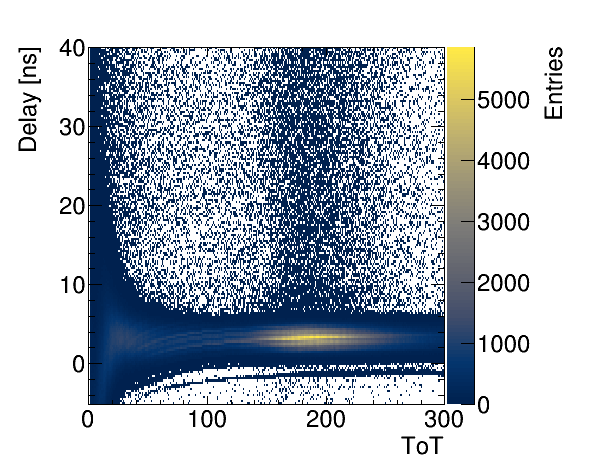}
\end{minipage}


\begin{minipage}[t]{.47\textwidth}
\centering
\captionsetup{width=.9\linewidth}
\captionof{figure}{The ToT of each hit plotted against the corresponding delay of the hit for W0020J07 for run 31500. The horizontal lines are due to the non-uniformity of the TDC.}
\label{fig:Results/TimewalkCurve50um}
\end{minipage}\qquad
\begin{minipage}[t]{.47\textwidth}
\centering
\captionsetup{width=.9\linewidth}
\caption{The ToT of each hit plotted against the corresponding timewalk corrected delay of the hit for W0020J07 for run 31500.}
\label{fig:Results/TimewalkCurve50umCorrected}
\end{minipage}
\end{figure}

The timing performance of the Timepix3 suffers from timewalk \citep{gromov2012development} as discussed in \autoref{sec:Timepix3/Timewalk}. This can clearly be seen in the data taken at the testbeam as shown in \autoref{fig:Results/TimewalkCurve50um}. This Figure illustrates the delay of the measurement versus the charge of the measurement. Because timewalk depends on the charge per pixel, and not on the total charge of the cluster, only the delay and charge of the hit with the earliest timestamp within the cluster are considered in this case. In this Figure, the larger delay of hits with a low charge (and thus low ToT) can clearly be observed.

The effect on the timing resolution of the Timepix3 due to timewalk (up to tenths of \si{\nano\second}) is significantly larger than the systematic per-pixel variation on the timing precision of the Timepix3 (expected to be less than \SI{1.56}{\nano\second} based on the constraints during the design of Timepix3). Therefore, either a timewalk correction should be performed on the data, or a selection of events with the same charge should be made, to be able to further investigate the timing aspects of the Timepix3 chip. By selecting events on their charge, a large number of events is not used. Therefore, it is more efficient to correct for timewalk instead of selecting events on their charge. 

The method to correct the for the timewalk effect, is already described in \autoref{sec:Timepix3/Timewalk}. This correction is applied to the testbeam data, 
and an example of the delay distribution of a measurement before and after the correction are presented in \autoref{fig:Results/TimewalkCurve50um} and \autoref{fig:Results/TimewalkCurve50umCorrected} subsequently. After correction, the delay is reduced to a straight line for all charge deposits, and therefore no longer dependent on the charge. However, the spread in the delay at low charge is still significant. As mentioned before, a further improvement of the timewalk correction can be applied by extending the ToT value from a measurement with the corresponding fToA value. By extending the ToT value, the timewalk correction can still be improved further by the increased precision of the charge measurement. This increase in the precision of the charge measurement will result in the ability to determine the average delay for smaller charge intervals.

\subsection*{Measured SPIDR's TDC bin structures}

\begin{figure}
\centering
\includegraphics[width=.75\linewidth]{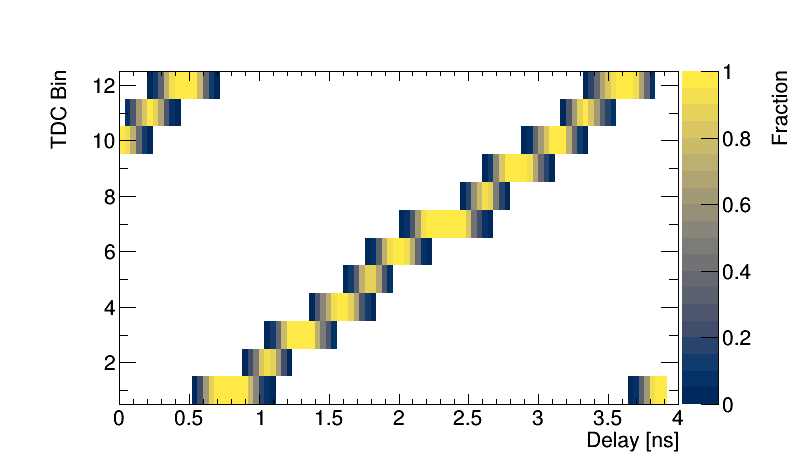}
\captionof{figure}{The measured relative position and population of the twelve different trigger bins from the SPIDR TDC. The TDC bins have significantly different widths.}
\label{fig:Results/TriggerBinScan}
\end{figure}

In \autoref{fig:Results/TimewalkCurve50um}, it is evident that there are horizontal lines visible in the timewalk curve. These lines are due to a non-uniformity of the SPIDR's TDC bins which are used to measure the timestamp of the trigger, and are caused by more triggers being measured in specific time bins of the TDC. Assuming that the triggers from the coincidence setup arrive uniformly within a period of the \SI{320}{\mega\hertz} clock, one would expect a flat distribution instead of the peaks that are visible. This effect is caused by the non-uniformity of the TDC bins. To quantify this, a delay scan of the time bins of the TDC is performed. The results of this delay scan is shown in \autoref{fig:Results/TriggerBinScan}. In this Figure it is evident that the size of the twelve different time bins is not equal. This non-uniformity is due to a non-perfect phase shift of the \SI{320}{\mega\hertz} clocks, which is used to create the twelve bins. A further indication for this phase shift is the symmetry present within these bins. 
After half a period, the clocks are in precise opposite polarity and thus explaining the same width of bin one and bin seven. If the phase shift between the individual copies of the clock is not precisely $\pi / 3$, but a slight deviation is present in the different phase shifts, the time bins that are created are not equal in width. The average width however, will still be one-twelfth of the period of the \SI{320}{\mega\hertz} clock because the frequency of the six clocks is still \SI{320}{\mega\hertz}. The standard deviation of the bins is still $\delta_{t}/\sqrt{12}$, where $\delta_t$ is the width of the bin. However, the average standard deviation of the twelve trigger bins is bigger than if the bins were equal in width. The difference in width, causes the difference in the number of triggers measured assuming that the triggers arrive uniformly throughout the \SI{320}{\mega\hertz} period. This is precisely the process that causes the horizontal lines in the timewalk curve. 

By selecting only events that fall within time bins two, four, seven, and nine, the precision of the trigger signal can be increased due to the smaller size of these bins. However, by selecting just one third of the data, the total number of events that can be used is smaller, and thus does not provide enough statistics to process the data. Therefore, such a selection of the data could not yet be applied during this analysis. However, it might proof useful in further studies on the Timepix3 in order to increase the timing precision of the reference signal when a lower efficiency is acceptable.

\section{Timing structures over the pixel matrix}
\label{sec:Results/DelayPixelMatrix}


After the timewalk correction has been applied, a global structure is observed in the time delay over the pixel matrix. This structure has been associated with the propagation of the electric signal of the clocks over the matrix structure. The time difference expresses itself as a difference in the delay between pixels located in different parts of the pixel matrix. This variation was already investigated during the design phase of the Timepix3, however this investigation only included simulations of the electronics, and did not include measurements on the real chip. During this design phase, the time variations during normal operating conditions were designed in such a way that the variation did not exceed one fToA bin of the Timepix3 (\SI{1.56}{\nano\second}). Using the timing information from the Timepix3 telescope, the average delay per pixel has been determined. Besides the use of the Timepix3 telescope, the laser setup that is described in \autoref{sec:LaserSetup}, can also be used to measure the average delay per pixel.

In the first Section the measurements taken during the testbeam with the Time\-pix3 telescope will be discussed. After this the same set of measurement taken with the laser setup will be discussed and compared to the testbeam results.

\subsection{Testbeam measurements}

During the testbeam three different detectors have been tested. The properties and operating conditions of the detectors are shown in \autoref{tab:Results/DataTable}. The results in this Section will mostly focus on W0039I11 because this is the only detector that was displaced in the xy-plane in between different runs in order to scan the complete pixel matrix.

A coincidence unit is used during the testbeam to generate a timestamp for the tracks traversing the DUT. By using the timestamp from the tracks and comparing it to the timestamp of the hit on the DUT, the average delay for each pixel of a Timepix3 chip can be determined. In the first part of this Section, the calculation to determine the average delay per pixel is explained, after which the results from the testbeam are presented. 

Each track can have an associated cluster on the DUT depending on whether or not a cluster is found on the DUT during the tracking sequence of Kepler. If a track has an associated cluster on the DUT, the delay of this cluster can be calculated. The delay is defined as the timestamp of the cluster minus the timestamp of the track. This delay consists of two parts, a constant part due to the signal propagation and formation of the coincidence setup, and a locally fluctuating part due to the difference in delay between the individual hits. In the latter, the information is present to calculate the average delay per pixel, while the constant offset can be determined and corrected in the analysis. 

For each track that has an associated cluster on the DUT, the cluster size on the DUT is checked. If the cluster size exceeds one pixel, the cluster is not used in the analysis to determine the delay. The reason to not include clusters that exceed a size of one pixel in the analysis, is explained later on. However, at this point it is important to make this selection to determine the average delay per pixel. The average cluster size on the DUT is on average one, which leaves enough statistics to do an analysis of the per-pixel delay. When a cluster on the DUT is found, the delay is calculated from the timestamp of the cluster and the trigger associated to the track. Since only one-hit clusters are considered, the position is determined from the column and row in which the hit on the DUT is measured. When all tracks are processed, a distribution of the delay for each pixel is constructed. 

\begin{figure}
\centering
\begin{minipage}{.47\textwidth}
\centering
\includegraphics[width=1\linewidth]{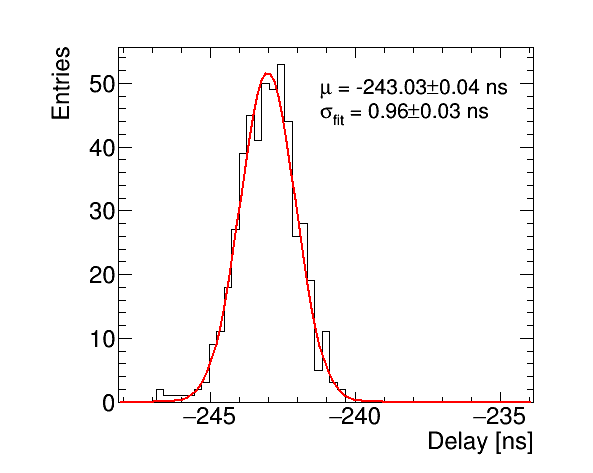}
\end{minipage}\qquad
\begin{minipage}{.47\textwidth}
\centering
\includegraphics[width=1\linewidth]{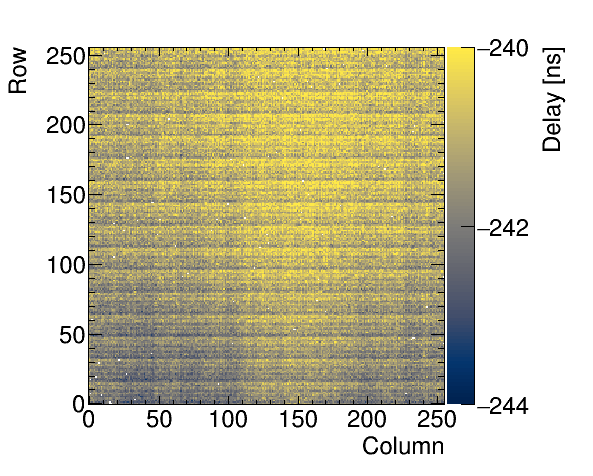}
\end{minipage}


\begin{minipage}[t]{.47\textwidth}
\centering
\captionsetup{width=.9\linewidth}
\captionof{figure}{The distribution of the delay with timewalk correction for pixel (200,200) of W0039I11. The average delay is indicated in the top right corner as well as the width of the distribution.}
\label{fig:Results/averageDelayOnePixel}
\end{minipage}\qquad
\begin{minipage}[t]{.47\textwidth}
\centering
\captionsetup{width=.9\linewidth}
\caption{The average delay per pixel of the pixel matrix of W0039I11 at a bias voltage of \SI{90}{\volt} measured with the Timepix3 Telescope. Runs 31440, 31441, 31442, and 31443 where used to determine this delay.}
\label{fig:Results/averageDelay50um}
\end{minipage}
\end{figure}

An example of such a distribution for a single pixel is shown in \autoref{fig:Results/averageDelayOnePixel}. Due to the limited interaction length of the \SI{50}{\micro\meter} sensor with the particles, the average deposited charge is insufficient to not suffer from timewalk. Therefore, only for the \SI{50}{\micro\meter} sensor, a timewalk correction as well as a cut on the charge values is made to determine the average delay. Instead of taking the mean of the delay as the average delay, a normal distribution is fitted to this distribution. Otherwise outliers will influence the value of the average delay. The different heights of the neighbouring bins in the delay distribution are due to the non-uniformity of the TDC bins that is discussed earlier.

Due to the difference in width between the time bins of the fToA and the trigger, a bin belonging to one fToA value consists of six different time bins from the trigger. Therefore, if a hit is assigned to a certain fToA value, there is still a uniform distribution of the six trigger bins that can be hit within the fToA time bin. This means that for the measurement in \autoref{fig:Results/averageDelayOnePixel}, just three different fToA bins are hit. 

The error that is assigned to the average delay per pixel is determined from a combination of errors. The standard deviation of the timestamp of the fToA timestamp is $1.56/\sqrt{12}=$ \SI{0.45}{\nano\second}, while the standard deviation from the timestamp of the trigger signal is $0.26/\sqrt{12}=$ \SI{0.075}{\nano\second}. Therefore the standard deviation associated to the calculated value of the delay is \SI{0.457}{\nano\second}. If there were no other sources of uncertainty, this standard deviation would be equal to the standard deviation $\sigma_{fit}$ of the normal distribution that is fitted to the delay distribution. However, this is not the case. Therefore the standard deviation (in \si{\nano\second}) associated to the value of the average delay for a pixel is given by
\begin{equation}
	\sigma_{\mu}=\sqrt{\sigma_{fit}^2-\left(\frac{1.56}{\sqrt{12}}\right)^2-\left(\frac{0.26}{\sqrt{12}}\right)^2}\approx \sqrt{\sigma_{fit}^2-0.457^2}.
\end{equation}

Using the data taken at the testbeam, the delay distribution for each pixel of the pixel matrix can be determined. After a normal distribution is fitted to these distributions, the average delay of the pixel matrix is determined. The distribution of this average delay for W0039I11 is shown in \autoref{fig:Results/averageDelay50um}. A structure with a period of sixteen rows can be observed in this Figure, along with a global structure over the chip. 

\subsection*{Global timing structure}
The global structure is due to a combination of the power distribution over the chip and a increasing delay along the column. The delay along the column is a result of the time the electrical signal takes to reach the pixels that are higher up in the pixel matrix. To further investigate the period that is visible along the rows, the global structure needs to be excluded. To quantify the global structure on a per-pixel basis, the average delay of a sixteen by sixteen pixel area is taken surrounding each pixel. This area is centred on the pixel for which the global delay is calculated, and thus also includes just half or a quarter of a pixel at the edges and corners. Therefore, the it is calculated according to
\begin{equation}
	\Delta_{global;x,y}=\frac{\sum_{i,j}\Delta_{x+i,y+j}w_{i,j}}{\sum_{i,j}w_{i,j}}.
\end{equation}
Here, $\Delta_{global;x,y}$ is the global delay for pixel $(x,y)$, $w_{i,j}$ is the weight assigned to pixel ${i,j}$ within the area that was defined, and $i,j\in \{\text{-}8,\text{-}7,...,7,8\}$. The weight $w_{i,j}$ assigned to the pixels is 1, except for pixels on the edge of the defined area. For those pixels the weight is $1/2$, or $1/4$ if it is on the corner of the area. If $0\leq x+i\leq 255, \text{or}\ 0\leq y+j \leq 255$ the pixel is outside of the pixel matrix, and thus the weight assigned to $w_{i,j}$ is 0. 

\begin{figure}
\centering
\begin{minipage}{.47\textwidth}
\centering
\includegraphics[width=1\linewidth]{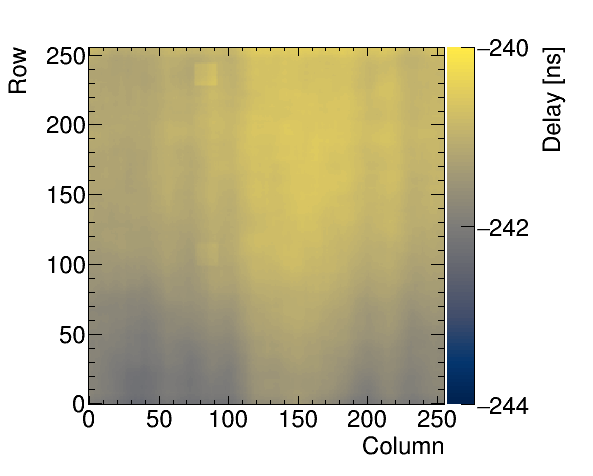}
\end{minipage}\qquad
\begin{minipage}{.47\textwidth}
\centering
\includegraphics[width=1\linewidth]{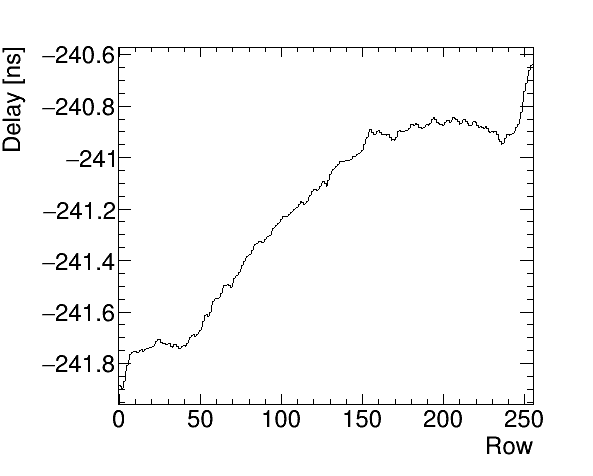}
\end{minipage}


\begin{minipage}[t]{.47\textwidth}
\centering
\captionsetup{width=.9\linewidth}
\captionof{figure}{The global delay of W0039I11 measured with the Timepix3 Telescope.}
\label{fig:Results/globalDelay50um}
\end{minipage}\qquad
\begin{minipage}[t]{.47\textwidth}
\centering
\captionsetup{width=.9\linewidth}
\caption{The average global delay of each row of W0039I11 measured with the Timepix3 Telescope.}
\label{fig:Results/globalDelayRow50um}
\end{minipage}
\end{figure}

The global delay calculated from the measurement shown in \autoref{fig:Results/averageDelay50um} is shown in \autoref{fig:Results/globalDelay50um}. The delay from the coincidence setup is the largest contribution to the offset of \SI{-241.9}{\nano\second} (lowest value of the delay within the pixel matrix). Another effect in this global structure is the decrease in delay along the column. The input and output for each double column is positioned at the side of the periphery (below row 0). Therefore, it physically takes a longer time for the signal to propagate from or to pixels that are located at the top row. \autoref{fig:Results/globalDelayRow50um} shows the average global delay for each row. The quick divergence between row 0 and 8, and row 247 and 255 is due to the averaging method. In this range, a part of the area over which the algorithm averages the delay, is outside of the pixel matrix. Therefore, the total number of pixels over which the average value is determined drops from 256 pixels at row 246 to 136 pixels at row 255. This decrease, forces the average value closer to the average delay at that specific row and column, thus explaining the sudden divergence of the global delay near the edge. From this Figure the average delay difference between row 0 and row 255 is determined to be \SI{1.23}{\nano\second}. This value coincides with the maximum difference that was allowed during the design phase of Timepix3 which was \SI{1.56}{\nano\second}.

\subsection*{Local timing structure}
\begin{figure}
\centering
\begin{minipage}{.47\textwidth}
\centering
\includegraphics[width=1\linewidth]{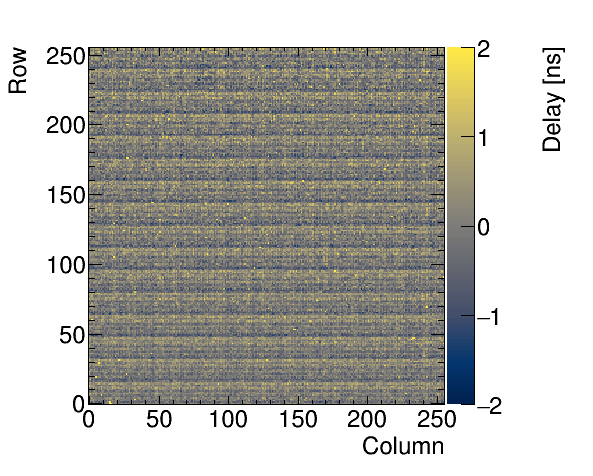}
\end{minipage}\qquad
\begin{minipage}{.47\textwidth}
\centering
\includegraphics[width=1\linewidth]{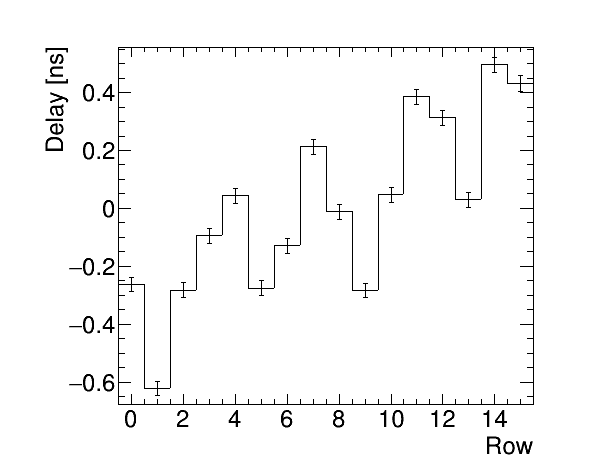}
\end{minipage}


\begin{minipage}[t]{.47\textwidth}
\centering
\captionsetup{width=.9\linewidth}
\captionof{figure}{The local delay of W0039I11 measured with the Timepix3 Telescope. }
\label{fig:Results/localDelay50um}
\end{minipage}\qquad
\begin{minipage}[t]{.47\textwidth}
\centering
\captionsetup{width=.9\linewidth}
\caption{The average delay for each row within the building blocks of the Timepix3 pixel matrix (W0020J07). }
\label{fig:Results/localDelaySuperSuperPixel}
\end{minipage}
\end{figure}

\begin{figure}
\centering
\includegraphics[width=1\linewidth]{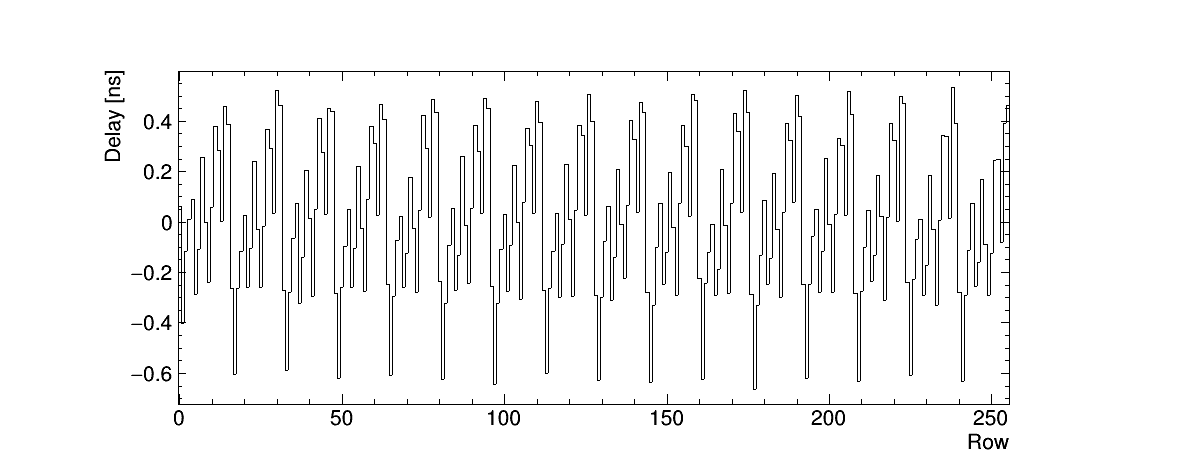}
\captionof{figure}{The average local delay of each row of W0039I11 determined from the measurements at the testbeam.}
\label{fig:Appendix/W0039I11RowProj}
\end{figure}

Now that the global delay of the chip is known, this delay can be subtracted from the average delay to find the local delay of the chip. This local delay for the same measurement is shown in \autoref{fig:Results/localDelay50um}. The differences in the delay between pixels, is only due to per-pixel differences, and not due to global effects since these have been excluded. The structure with a period of sixteen pixels is present in the local delay and not in the global delay. To highlight this structure, the average of row 16 up to and including row 237 is shown in \autoref{fig:Results/localDelaySuperSuperPixel}. Because the first and last period of sixteen rows include the eight rows on both sides of the chip for which the calculation of the global structure is not as precise as for the other rows these rows are excluded. The row projection of the complete chip is shown in \autoref{fig:Appendix/W0039I11RowProj}. This period of sixteen pixels coincides with the four super-pixels that were designed as the building block of the pixel matrix. The routing of the wires was designed for these four super-pixels after which it was copied for the complete pixel matrix. Therefore it is logical that after sixteen pixels (four rows per super-pixel times four super-pixels) the same structure is visible as the sixteen rows before. Even the design of the individual pixels can be seen in this period of sixteen rows. Every four rows, corresponding to a single super-pixel, a structure repeats itself. However, this structure is still convoluted with the delay caused by the propagation of the signals within the building block. Though, it can still clearly be seen that every second row of each super-pixel has on average a smaller delay, and is thus quicker compared to the other rows within the super-pixel. 

\begin{figure}[h]
\centering
\begin{minipage}{.47\textwidth}
\centering
\includegraphics[width=1\linewidth]{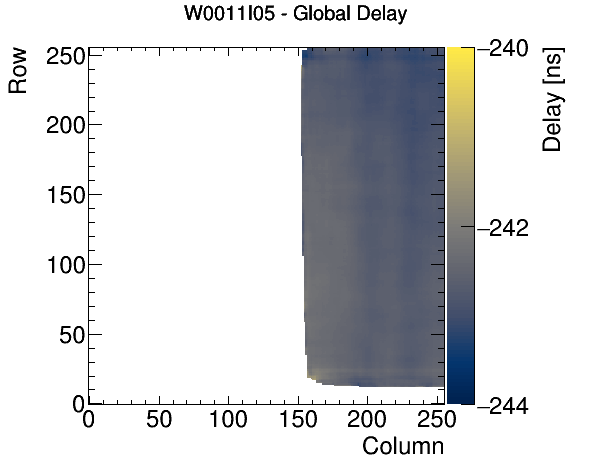}
\end{minipage}\qquad
\begin{minipage}{.47\textwidth}
\centering
\includegraphics[width=1\linewidth]{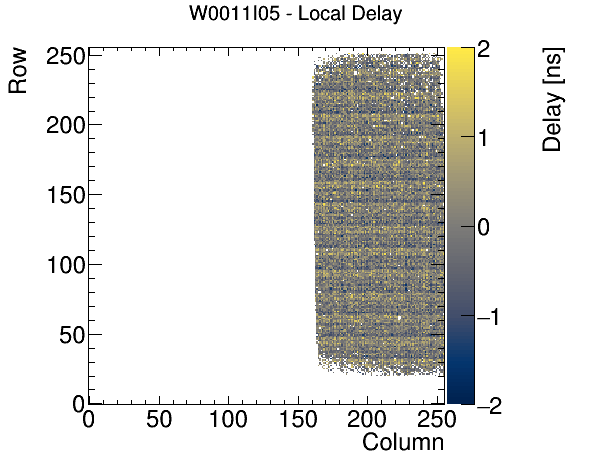}
\end{minipage}


\begin{minipage}[t]{.47\textwidth}
\centering
\captionsetup{width=.9\linewidth}
\captionof{figure}{The global delay of W0011I05 determined from the measurements at the testbeam.}
\label{fig:Appendix/W0011I05GlobalDelay}
\end{minipage}\qquad
\begin{minipage}[t]{.47\textwidth}
\centering
\captionsetup{width=.9\linewidth}
\caption{The local delay of W0011I05 determined from the measurements at the testbeam.}
\label{fig:Appendix/W0011I05LocalDelay}
\end{minipage}
\end{figure}

\begin{figure}[h]
\centering
\begin{minipage}{.47\textwidth}
\centering
\includegraphics[width=1\linewidth]{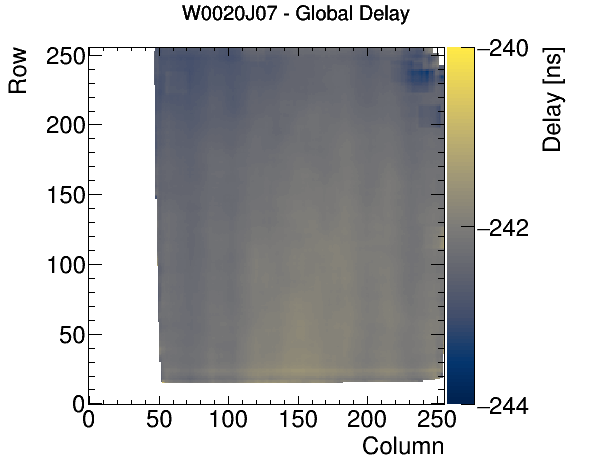}
\end{minipage}\qquad
\begin{minipage}{.47\textwidth}
\centering
\includegraphics[width=1\linewidth]{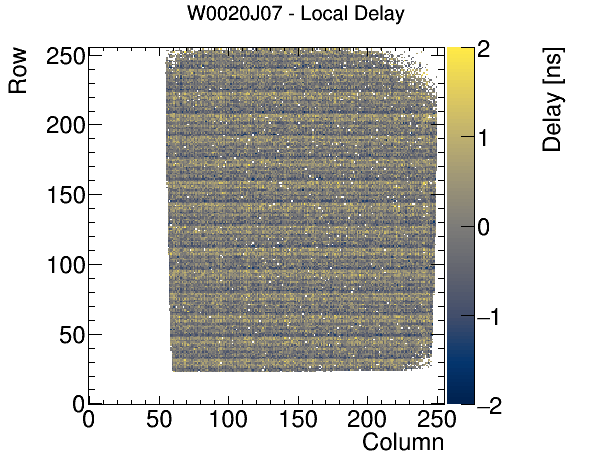}
\end{minipage}


\begin{minipage}[t]{.47\textwidth}
\centering
\captionsetup{width=.9\linewidth}
\captionof{figure}{The global delay of W0020J07 determined from the measurements at the testbeam.}
\label{fig:Appendix/W0020J07GlobalDelay}
\end{minipage}\qquad
\begin{minipage}[t]{.47\textwidth}
\centering
\captionsetup{width=.9\linewidth}
\caption{The local delay of W0020J07 determined from the measurements at the testbeam.}
\label{fig:Appendix/W0020J07LocalDelay}
\end{minipage}
\end{figure}

The results from the testbeam for W0011I05 is shown in \autoref{fig:Appendix/W0011I05GlobalDelay} and \autoref{fig:Appendix/W0011I05LocalDelay}, while the results from the testbeam for W0020J07 are shown in \autoref{fig:Appendix/W0020J07GlobalDelay} and \autoref{fig:Appendix/W0020J07LocalDelay}. In these Figures both the global and the local delay for these detectors are shown. The main difference between these two chips and W0039I11, which is presented in \autoref{fig:Results/globalDelay50um} and \autoref{fig:Results/localDelay50um}, is the periodic structure visible in the local delay. For W0039I11 the delay is decreasing along a column, while for the other two detectors the delay is increasing. This difference was unexpected and it is not yet known what determines this difference between the detectors. However, the structure within the building block, as well as the global delay, exhibits the same features on all detectors, indicating the similarity between the three detectors.

\begin{figure}
\centering
\includegraphics[width=.85\linewidth]{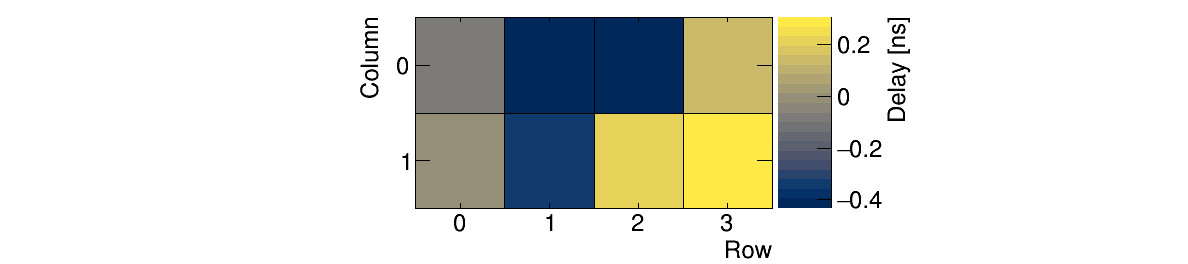}
\captionof{figure}{The average delay for the eight pixels that make up a single super-pixel (W0039I11).}
\label{fig:Results/SuperPixelAverage}
\end{figure}

As became evident from the local delay structure of the Timepix3, the delay of each pixel is mostly determined from the delay present in the building block of the pixel matrix. Though, per pixel there is still a variation on this average value. This variation is due to production variation throughout the pixels. Because of the production process of the components of the individual pixels, slight variations in different parameters such as width or capacitance of the wires within the ASIC change the delay of the pixels. However, by averaging all the pixels of a specific row of the building block, these variations are averaged out and the structure due to the design of the building block becomes evident. To further indicate the delay due to the design of the super-pixel, the average delay for the super-pixel is shown in \autoref{fig:Results/SuperPixelAverage}. Note that this Figure is rotated clock-wise compared to the previous Figures.

As can be observed from this Figure, the pixels closer to the centre of the super-pixel have on average a smaller delay compared to the outer pixels. These pixels are physically closer to the start-input of the VCO explaining the smaller start-up time. However, the frequency of the VCO is fixed at \SI{640}{\mega\hertz} and thus does not depend on the start-up time of the VCO. Therefore, if a pixel has a start-up time of \SI{0.7}{\nano\second}, the very last fToA bin will have a reduced width of $1.56$ \si{\nano\second}$-0.7$ \si{\nano\second} $=0.86$ \si{\nano\second}. This also implies that the first fToA bin is actually \SI{0.7}{\nano\second} wider. This effect can be observed in a normal measurement by looking at the distribution of the hits throughout the fToA bins. Because the first fToA bin is longer, and the last fToA bin is shorter, the number of hits that will arrive in the first fToA bin are more than the average, and the number of hits that will arrive within the last fToA bin are below average. The distribution of the hits between the first and the last bin for a pixel therefore give an indication of the magnitude of the delay of that pixel.

\subsection{Confirming the timing structures using the laser setup}
Besides the possibility to determine the average delay per pixel using the data taken at the testbeam, it is also possible to determine this delay with the laser setup that is described in \autoref{sec:LaserSetup}. An advantage of the laser setup is the higher charge that can be injected in one pixel. The only limitation on the injected charge is the repulsion of the created charge cloud on itself. If too many electrons are present in this cloud, the repulsion will increase the size of the charge cloud beyond the pitch of the Timepix3 before the charge carriers are collected by the electronics. However, compared to the average charge of a particle traversing the detector and creating around 15,000 electrons, the charge generated by the laser can increase to above 30,000 electrons. At this number of electrons the timewalk is negligible and thus makes the measurements more precise.

To determine the average delay per-pixel, the measured timestamp of the charge cloud created by the laser is compared to an external trigger that is generated by the pulse generator that drives the laser. To determine the average delay of the pixels, the laser is not synchronised to the system clock of the Timepix3, such that the time of arrival of the laser pulses is uniformly distributed throughout the \SI{25}{\nano\second} period. The same approach as with the testbeam is used to determine the average delay for each pixel. However, because there are no tracks in this setup, each cluster on the Timepix3 can directly be processed, without the need to check the properties that were associated to the track. Again, it is first checked if the cluster consists of one or more than one pixels. If the cluster has a size that is greater than one, it is not used to determine the average delay of the pixels. 

\begin{figure}
\centering
\begin{minipage}{.47\textwidth}
\centering
\includegraphics[width=1\linewidth]{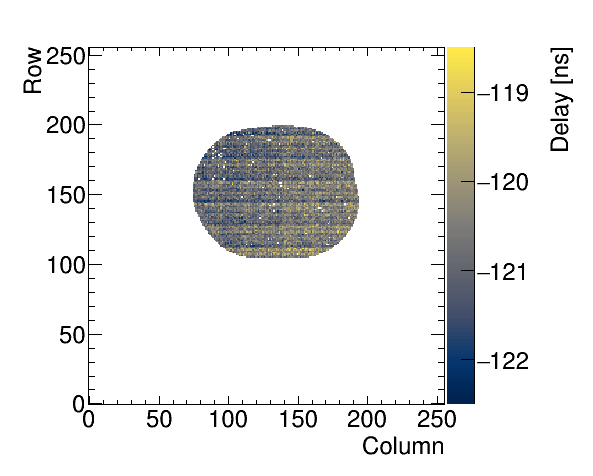}
\end{minipage}\qquad
\begin{minipage}{.47\textwidth}
\centering
\includegraphics[width=1\linewidth]{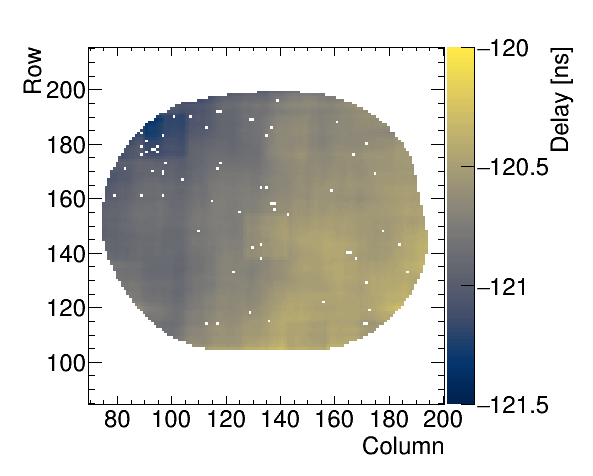}
\end{minipage}


\begin{minipage}[t]{.47\textwidth}
\centering
\captionsetup{width=.9\linewidth}
\captionof{figure}{The average delay of the pixel matrix of W0020J07 measured with the laser setup.}
\label{fig:Results/averageDelayLaser}
\end{minipage}\qquad
\begin{minipage}[t]{.47\textwidth}
\centering
\captionsetup{width=.9\linewidth}
\caption{The global delay of W0020J07 measured with the laser setup.}
\label{fig:Results/globalDelayLaser}
\end{minipage}
\end{figure}

After processing all the clusters of a measurement, and determining their delay, a distribution of the delay is found for each pixel. A normal distribution is fitted to determine the average delay for each pixel. The delay distribution for the pixels determined with the laser setup has similar characteristics to that of the delay distribution of the testbeam. The average delay that is determined with the laser setup for the W0020J07 is shown in \autoref{fig:Results/averageDelayLaser}. Note that the average delay is now around $\text{-}120$ \si{\nano\second}, instead of $\text{-}242$ \si{\nano\second} that it was for the measurement from the testbeam. This is due to the overall shorter length of the cables used to supply the trigger signal and the absence of any additional electronics between the creation of the trigger signal and the TDC.

The delay can only be calculated for a fraction of the pixel matrix with the laser setup. This is due to the metallization that is usually on top of the silicon. This metallization is there to shield the silicon from any stray light that would cause continuous signals in the sensor, and thus ensure that the detector measures charged particles in a room that has normal lighting. However, this also implies that the light from the laser at \SI{680}{\nano\meter} can also not reach the silicon. Therefore, we etched part of the metallization using an acid mixture. The part of the metallization that was removed directly corresponds to the pixels for which the average delay is calculated. In total the metallization is removed for 9373 pixels of the pixel matrix, corresponding to a fraction of 0.14 of the total pixels.

A problem arises due to the small area for which the delay can be determined. To accurately determine the global delay of a pixel, an area of sixteen by sixteen pixels surrounding the pixel should have a value for the delay, otherwise the global delay is determined by the average of less pixels, and thus converges more to the delay of the pixel itself. Therefore the local delay can only be determined for a smaller fraction of the pixels. The global delay for all the pixels shown in \autoref{fig:Results/averageDelayLaser}, is shown in \autoref{fig:Results/globalDelayLaser}. The variations due to power distribution are less apparent due to the limited area for which the delay is calculated. However, the decrease of the delay along the column, is still visible. The overall magnitude of the delay variations over the chip is also smaller due to the limited area. The white spots that are visible are pixels that were either masked or unresponsive during the measurement.

\begin{figure}
\centering
\begin{minipage}{.47\textwidth}
\centering
\includegraphics[width=1\linewidth]{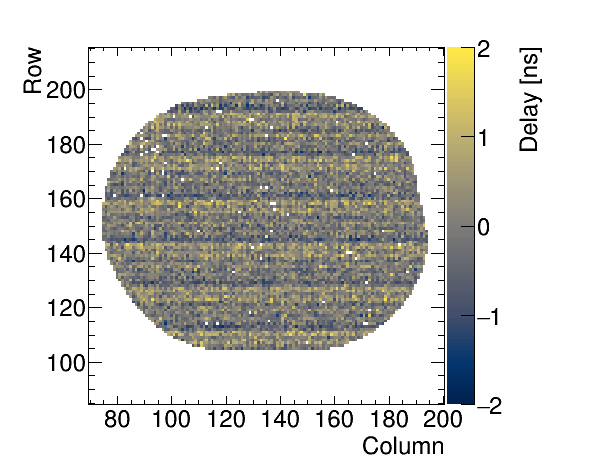}
\end{minipage}\qquad
\begin{minipage}{.47\textwidth}
\centering
\includegraphics[width=1\linewidth]{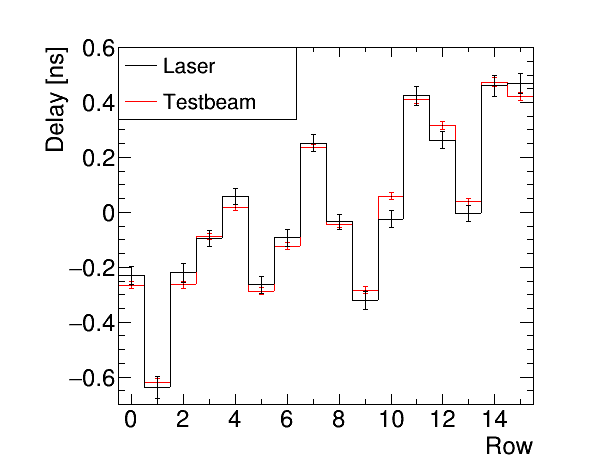}
\end{minipage}


\begin{minipage}[t]{.47\textwidth}
\centering
\captionsetup{width=.9\linewidth}
\captionof{figure}{The local delay of W0020J07 measured with the laser setup.}
\label{fig:Results/localDelayLaser}
\end{minipage}\qquad
\begin{minipage}[t]{.47\textwidth}
\centering
\captionsetup{width=.9\linewidth}
\caption{The average delay for the rows of the building block of the pixel matrix (W0020J07). The black line indicates the measurement with the laser setup and the red line indicates the measurement with the Timepix3 Telescope. }
\label{fig:Results/localDelaySuperSuperPixelLaser}
\end{minipage}
\end{figure}

Now that the global delay is known, the local delay of the pixel matrix can be calculated and is shown in \autoref{fig:Results/localDelayLaser}. Once again the sixteen period structure is present, which is as expected. On first glance the local delay measured with the laser setup looks similar to the local delay that is measured with the testbeam. To further indicate this similarity, the average local delay of each row within the building block is shown in \autoref{fig:Results/localDelaySuperSuperPixelLaser}, with the testbeam and laser data overlaid (note that this is not the same chip as in \autoref{fig:Results/localDelaySuperSuperPixel}). The bigger error bars on the data from the laser setup are due to the smaller number of pixels that were measured. The overall structure measured with the laser setup is similar to that of the structure measured with the testbeam. However, especially at bin ten, there is a small difference between the two methods. Small differences could arise due to the different operating temperatures, and it is thus also suspected that the small difference is due to these different operating conditions such as temperature. From this, it is concluded that both the testbeam and the laser setup produce the same results. Therefore, the laser setup is considered as a viable method to determine time related aspects of the Timepix3 ASIC.

\subsection{Cause of the local delay structure}
Now that it is concluded that the laser setup produces the same results as the testbeam, and is thus a viable method to investigate time related aspects of the Timepix3 ASIC, the laser setup can be used to determine the cause of the difference in delay for the different pixels. It is already concluded that the difference in delay between the different pixels is related to the position within the super-pixel and even within the building block of the pixel matrix. However, now that a working laser setup is developed, different measurements can be conducted to investigate the nature of the difference in the local delay. 

From the average delay within a super-pixel, the assumption can be made that pixels closer to the VCO have a smaller delay (see \autoref{fig:Results/SuperPixelAverage}). The VCO is physically located in the middle of both directions of the super-pixel \citep{poikela2015readout}. Therefore, the wire connecting the digital front-end of the pixels located at row 0 and 3, is longer than the pixels at row 1 and 2, and thus the time it takes the electrical signal to traverse the length of the wire is also longer for the pixels located at row 0 and 3 assuming that the capacitance of the wire is proportional with the length of the wire. However, a hit from just one of the pixels within the super-pixel is enough to start the VCO, and thus supply the \SI{640}{\mega\hertz} clock to all eight pixels. This implies that when two pixels within the same super-pixel are hit at the exact same time, the quickest one of the two pixels will start the VCO. When this happens, the delay of the slowest pixels does not determine the delay of both hits, but the delay of the quickest pixel does. This is the reason that clusters that consisted of two or more pixels are excluded during the analysis of the delay of the pixel matrix.

However, this effect is just limited to the pixels within the same super-pixel. When two pixels are hit that are located in neighbouring super-pixels, both pixels have to start their own VCO, and thus both hits have a different delay. 

This effect can not be investigated at the testbeam due to the random time of arrival within the system clock as well as the random position of the particles . Therefore, there are almost no clusters that have the exact same charge for each pixel within a cluster that consists of at least two pixels. An equal charge is required for these hits, such that timewalk does not influence the timestamp of these hits. However, the laser spot can be positioned such that the charge collected in two pixels is equal. This positioning can even be performed at the intersection between four different pixels within a super-pixel. The intensity of the laser can also be increased such that the desired charge is generated in each of the four pixel, and thus there is no influence of timewalk on the timestamp of these hits. 

\begin{figure}

\begin{minipage}{.5\linewidth}
\centering
\subfloat[]{\label{fig:Results/pixel2}\includegraphics[width=1\linewidth]{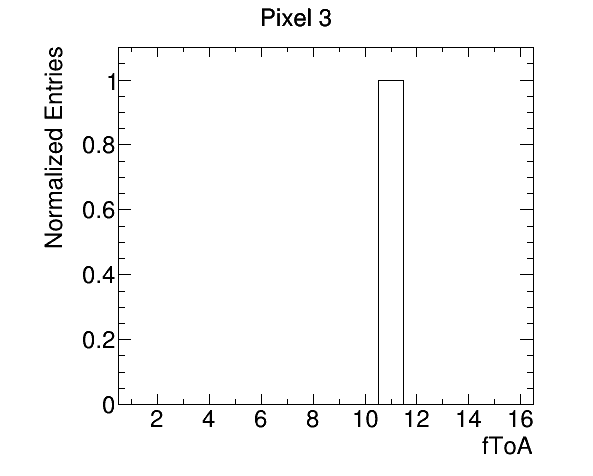}}
\end{minipage}%
\begin{minipage}{.5\linewidth}
\centering
\subfloat[]{\label{fig:Results/pixel4}\includegraphics[width=1\linewidth]{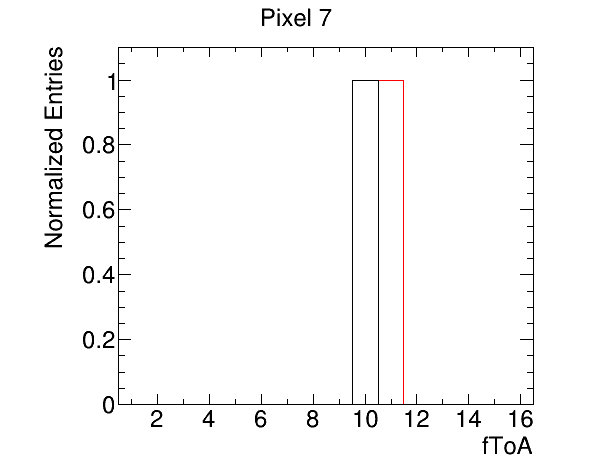}}
\end{minipage}\par\medskip
\begin{minipage}{.5\linewidth}
\centering
\subfloat[]{\label{fig:Results/pixel1}\includegraphics[width=1\linewidth]{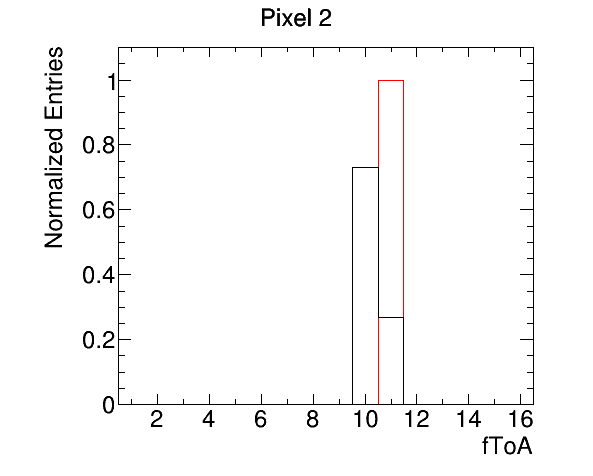}}
\end{minipage}%
\begin{minipage}{.5\linewidth}
\centering
\subfloat[]{\label{fig:Results/pixel3}\includegraphics[width=1\linewidth]{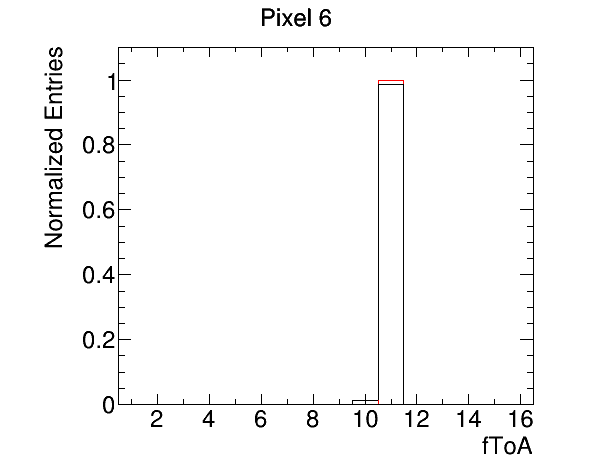}}
\end{minipage}
\caption{The fToA distribution for a laser induced hits with the same absolute delay with respect to the \SI{40}{\mega\hertz} clock. The black line indicates hits that are solely on a single pixel, and the red line indicates hits that are measured on all four pixel simultaneously.}
\label{fig:Results/mainResults}
\end{figure}

Using \autoref{fig:Results/SuperPixelAverage}, the optimal position is determined to measure the difference in the timing information within a super-pixel. The difference between the pixels at row 2 and 3 is the biggest difference between the pixels within a super-pixel. To be more precise, pixels (122,110), (122,111), (123,110), and (123,111) on W0020J07 are chosen. These pixels will be referred to as pixel 2, 3, 6, and 7 respectively from now on. From the data of the average delay of the pixels of this chip, it is determined that pixel 3 has the smallest delay, followed by pixel 6, pixel 2, and finally pixel 7. First the laser is positioned at each pixel independently, and the intensity of the laser is tuned such that on average a charge of 24,550 electrons is liberated in a single pixel. At this point the laser is synchronised to the \SI{40}{\mega\hertz} clock of the Timepix3 such that each laser pulse arrives at the same time within the \SI{25}{\nano\second} period. The delay of the offset is also adjusted such that the hits on pixel 3 all arrived in fToA bin 11. With these settings the fToA distribution of the four pixels is determined. This fToA distribution is shown in \autoref{fig:Results/mainResults} as the black line. The fraction of the number of hits that arrives in fToA bin 11 is proportional to the delay of that pixel. So for example, pixel 7 has highest delay, and thus none of the hits arrive in bin 11, but they all arrive in bin 10. This is because the difference in delay with respect to pixel 3 is larger than one fToA bin, and thus they never arrive in bin 11 (note that fToA is inversely proportional to time).

After scanning the four pixels individually, the laser intensity is quadrupled such that all four pixels would be hit at the same time with an equal charge of around 24,550 electrons. The fToA distributions of the hits for this configuration are shown in red in \autoref{fig:Results/mainResults}. Notice that all the hits for all pixels are now in fToA bin 11. This is because the VCO is started by the fastest pixel (pixel 3), and thus the delay of the other pixels does not influence the timestamp of the hits that are measured by those pixels. This confirms that the delay is mainly caused by the difference in the time it takes the pixels to start the VCO, and not due to a frequency shift of the VCO.

\section{Missing fToA bin}
\begin{figure}
\centering
\includegraphics[width=.47\linewidth]{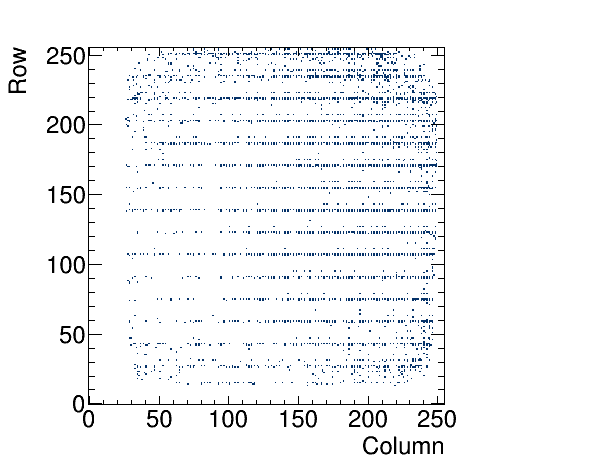}
\captionof{figure}{The pixels of W0039I11 that never measured a hit in fToA bin fifteen are indicated in dark blue. A structure is visible which reassembles the structure of that of the average delay of the pixels.}
\label{fig:Results/MissingBin50um}
\end{figure}

The Timepix3 is designed in such a way that a single PLL provides a control voltage that is distributed to each super-pixel within the pixel matrix to ensure a frequency of each VCO of \SI{640}{\mega\hertz} (see \autoref{Theory/Timepix3/Clocks}). However, a frequency shift of the VCO has been observed during both the testbeam and the laser measurements. During these measurements it was observed that the fifteenth fToA bin was never filled for specific pixels. \autoref{fig:Results/MissingBin50um} shows a distribution of the pixels from the \SI{50}{\micro\meter} sensor that never measured a hit in the fifteenth fToA bin during the testbeam measurements. The pixels from column 0 till 30 have too little hits to determine whether or not the missing fifteenth bin is due to statistics or due to it never reaching the bin. The missing bins are located mainly on specific rows spaced sixteen pixels apart. However, these bins are positioned in the middle of the super-pixel and not on the intersection between super-pixels like the abrupt jump in delay is. 

There are three possibilities for a missing fifteenth bin. First of all the frequency of the VCO could still be \SI{640}{\mega \hertz}, and thus the fifteenth bin is never registered meaning that the hits from the fifteenth bin are either missing or are assigned to a different fToA value. A second possibility is that the start-up of the VCO takes more than \SI{1.56}{\nano\second} while the frequency of the VCO is correct, and thus the effective width of the first fToA bin is at least \SI{3.12}{\nano\second}. The third possibility is that the frequency of the VCO is lower than \SI{640}{\mega\hertz}, meaning that it takes less than sixteen fToA bins to fill one \SI{25}{\nano\second} period, and thus corresponding to a maximum frequency of the VCO of \SI{600}{\mega\hertz}, indicating a shift in the VCO frequency, and thus possibly a drop of the control voltage of the VCO, over the pixel matrix.

\begin{figure}
\centering
\begin{minipage}{.47\textwidth}
\centering
\includegraphics[width=1\linewidth]{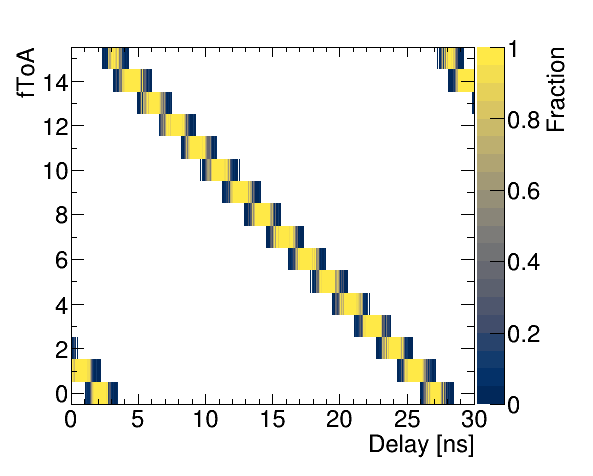}
\end{minipage}\qquad
\begin{minipage}{.47\textwidth}
\centering
\includegraphics[width=1\linewidth]{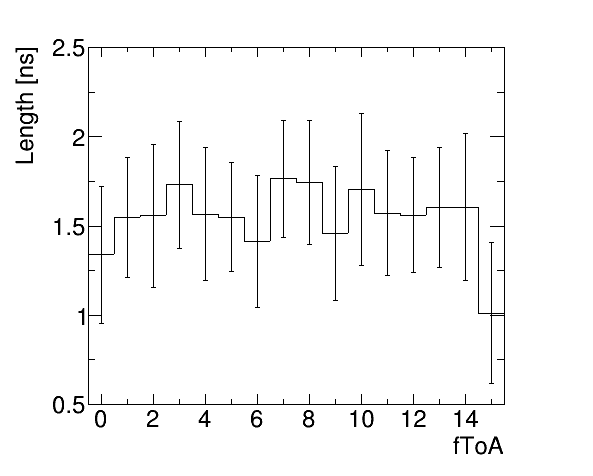}
\end{minipage}


\begin{minipage}[t]{.47\textwidth}
\centering
\captionsetup{width=.9\linewidth}
\captionof{figure}{The relative delay between the different fToA bins for a relatively slow pixel (120,119) on W0020J07.}
\label{fig:Results/fToABinsLength}
\end{minipage}\qquad
\begin{minipage}[t]{.47\textwidth}
\centering
\captionsetup{width=.9\linewidth}
\caption{The width of each fToA bin for pixel (120,119) on W0020J07, calculated from \autoref{fig:Results/fToABinsLength}.}
\label{fig:Results/fToALength120_119}
\end{minipage}
\end{figure}

To determine which of these options is causing the absence of the fifteenth fToA bin, the laser setup is utilized. The laser is first synchronised to the \SI{40}{\mega\hertz} clock of the Timepix3, such that the same fToA bin can be hit by each laser pulse consistently. The delay between the \SI{40}{\mega\hertz} and the laser pulse is incremented in steps of \SI{200}{\pico\second} such that the laser pulse arrives later within the \SI{25}{\nano\second} period. In this way the width of each fToA bin can be determined. Such a scan for a pixel that does exhibit the fifteenth fToA bin is shown in \autoref{fig:Results/fToABinsLength}. On the horizontal axis the delay between the \SI{40}{\mega\hertz} clock of the Timepix3 and the laser pulse is plotted, and on the vertical axis the fToA bin of the hits is shown. At a delay of zero, the hits are measured in fToA bin 1 due to an offset in the delay. Around a delay of \SI{3}{\nano\second}, the first fToA bin (bin 15) of the period of \SI{25}{\nano\second } is hit. Because the fToA is constructed by counting the number of periods until the rising edge of the \SI{40}{\mega\hertz}, the fToA value is inversely proportional to the delay. This relation can also be observed in the pattern in this Figure. From this measurement, the width of the individual fToA bins can be derived by fitting each bin to a combination of two error functions, given by
\begin{equation}
    f(x)=\frac{1}{2}\left[\erf\left(\frac{x-\mu_l}{\sigma_l}\right)-\erf\left(\frac{x-\mu_r}{\sigma_r}\right)\right].
\end{equation}
Here $\mu_{l/r}$ is the position of the left/right error function, and $\sigma_{l,r}$ is the width of the left/right error function. \autoref{fig:Results/fToALength120_119} shows the width of each fToA bin derived from the measurement shown in \autoref{fig:Results/fToABinsLength}. The error displayed is the error from the fit on the value of $\mu_l$ and $\mu_r$. From this Figure, it becomes apparent that the first fifteen bins are indeed of equal width, however the fifteenth bin (last bin) is smaller compared to the others. This is due to the start-up time that causes the delay that is discussed earlier. Because of this start-up time, the width of the first bin is the start-up time plus the normal \SI{1.56}{\nano\second}, while the last bin is the normal width of \SI{1.56}{\nano\second} minus the start-up time.

A fToA bin size scan also indicates why a pixel is either quicker or slower. If the fifteenth bin is on average smaller, the timestamp generated by that pixel is slower. While if the first fToA bin (bin 0) is on average smaller, the timestamp is quicker. Therefore, the pixel shown in \autoref{fig:Results/fToABinsLength}, is a relatively slow pixel, which indeed is indicated by the matrix delay scan (see \autoref{fig:Results/localDelayLaser}).

\begin{figure}
\centering
\begin{minipage}{.47\textwidth}
\centering
\includegraphics[width=1\linewidth]{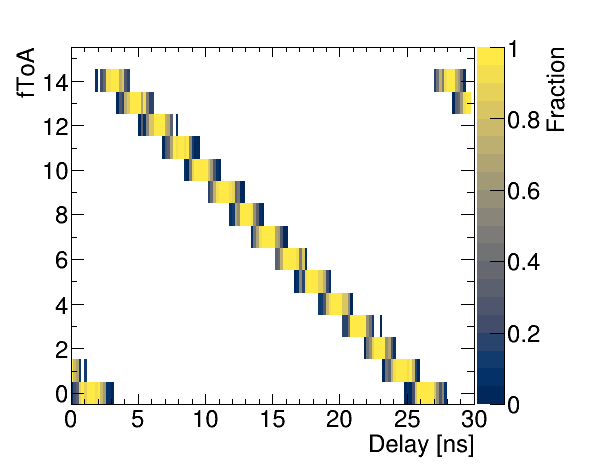}
\end{minipage}\qquad
\begin{minipage}{.47\textwidth}
\centering
\includegraphics[width=1\linewidth]{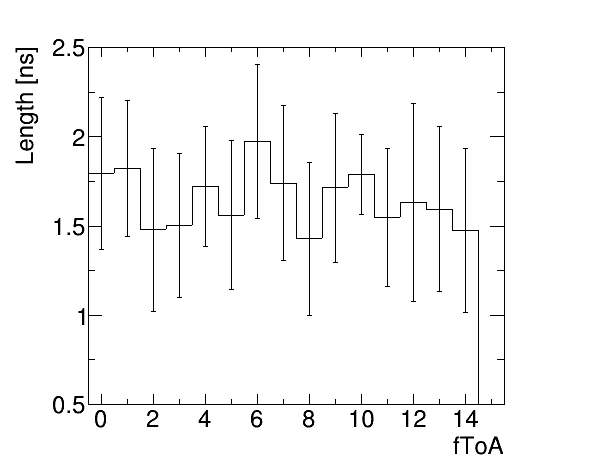}
\end{minipage}


\begin{minipage}[t]{.47\textwidth}
\centering
\captionsetup{width=.9\linewidth}
\captionof{figure}{The relative delay between the different fToA bins for a pixel (120,119) on W0020J07 that does not measure hits in fToA bin 15.}
\label{fig:Results/fToABinsLengthMissing}
\end{minipage}\qquad
\begin{minipage}[t]{.47\textwidth}
\centering
\captionsetup{width=.9\linewidth}
\caption{The width of each fToA bin for pixel (133,155) on W0020J07, calculated from \autoref{fig:Results/fToABinsLength}.}
\label{fig:Results/fToALength133_155}
\end{minipage}
\end{figure}

Now that it is known how the fToA bins are distributed within a period of the system clock for a normal pixel, an fToA scan is made for one of the pixels that did not exhibit the fifteenth fToA bin. The results of this scan is shown in \autoref{fig:Results/fToABinsLengthMissing}, as well as the corresponding width of the fToA bins in \autoref{fig:Results/fToALength133_155}. As can be observed in this Figure, the fifteenth bin is indeed never filled during this fToA bin size scan, thus indicating that the missing bin during the testbeam and the laser measurements is not due to the lack of statistics. One can also see that the first fToA bin is not twice the width that is should be (\SI{3.12}{\nano\second}), but that the average width of the different bins is larger than the \SI{1.56}{\nano\second}, indicating a shift in the frequency of the VCO instead of a start-up delay of the VCO that is longer than one fToA bin. However, if this missing bin is only due to a frequency shift of the VCO, the missing bin would be present in each pixel of the super-pixel that exhibits a missing bin. This on the other hand is not the case as one can see in \autoref{fig:Results/MissingBin50um}. Therefore, the missing fifteenth fToA bin is due to a frequency shift in the VCO combined with a different start-up time of the VCO for different pixels. This is further indicated by the position of the pixels that are missing the fifteenth fToA bin. These pixels are mainly located in the centre of a super-pixel indicating that the quicker start-up time of the VCO is causing these specific pixels to never reach the last fToA bin.

\section{Time resolution after timing corrections}
\label{sec:Results/TimeRes}

Now that it is known what causes the per-pixel difference in delay, and a method is developed to measure these per-pixel differences, the measured time difference in the average delay can be used to improve the time resolution of the Timepix3. The time resolution is defined as the width of the normal distribution that describes the distribution of the track time minus the hit time for all pixels combined. The best time resolution that can be reached with the Timepix3 is defined by the width of the fToA bins: $1.56/\sqrt{12}=0.45$ \si{\nano\second}. However, due to effects such as timewalk and per-pixel differences in the delay this time resolution will be worse than the naively expected \SI{0.45}{\nano\second}. Nonetheless, now that a timewalk correction can be applied and the per-pixel difference of the delay is measured for each pixel, an offline correction can be applied to increase the timing performance. 

\begin{figure}
\centering
\includegraphics[width=.48\linewidth]{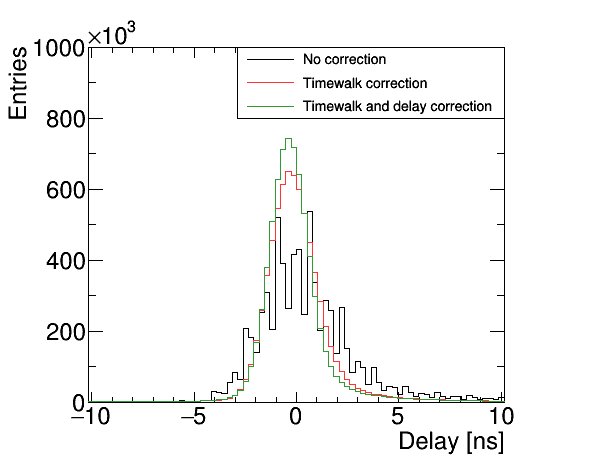}
\captionof{figure}{The time distribution of all the hit of run 31257 with W0039I11 with a bias of \SI{90}{\volt}. The time resolution of the uncorrected data is $1.6793\pm0.0004$ \si{\nano\second}, the time resolution of the data with timewalk correction is $1.1338\pm0.0004$ \si{\nano\second}, and the time resolution of the data with both a timewalk correction as well as a correction for the difference in the average delay per pixel is $0.9988\pm0.0003$ \si{\nano\second}.}
\label{fig:Results/TimeDist1}
\end{figure}

The time distribution without any offline correction for a single run of the testbeam for W0020J07 at a bias voltage of \SI{150}{\volt} is shown in \autoref{fig:Results/TimeDist1}. The normal distribution that is shown in this Figure indicates the fit that is used to determine the time resolution of this chip. The asymmetry in the time distribution is due to the timewalk effect of low charge hits. To mitigate this timewalk effect, the delay due to timewalk of each hit is calculated according to the measured charge of the hit. This delay can be subtracted from the measured delay to correct for timewalk. By doing so, the asymmetry in the time distribution disappears, and the time distribution becomes narrower, indicating an increase in timing performance. The time distribution after the timewalk correction has been applied is shown in \autoref{fig:Results/TimeDist1} as well. A further increase in the timing performance can be achieved by correcting the per-pixel differences in the average delay. The time distribution after this per-pixel correction of the differences in delay is shown in \autoref{fig:Results/TimeDist1} as well. By applying this per-pixel correction of the differences in delay, the timing performance is increased as well. 

By correcting for the per-pixel difference of the difference in delay, one indirectly corrects for the different width of specific fToA bins of specific pixels because the difference in the average delay is due to the difference in start-up time of the VCO as discussed before. A better correction can be applied by changing the time of a hit depending on the fToA bin it was measured in. However, to do so, one needs to know the width of each fToA bin for each pixel. This can be measured using the laser setup as discussed before with the delay scans, however one such a scan takes up to an hour to perform for a single pixel, which makes this method unsuitable. However, now that it is known what causes the difference in the delay, the data of the testbeam can be used to correct for the difference in width of the fToA bins. For each pixel, the relative width of the different fToA bins can be determined from the fToA distribution of the hits (assuming that the hits arrive randomly within a period of the system clock). Such a correction is not yet applied in this work, but would increase the timing performance more than the per-pixel correction of the difference in delay.

\begin{figure}
\centering
\begin{minipage}{.47\textwidth}
\centering
\includegraphics[width=1\linewidth]{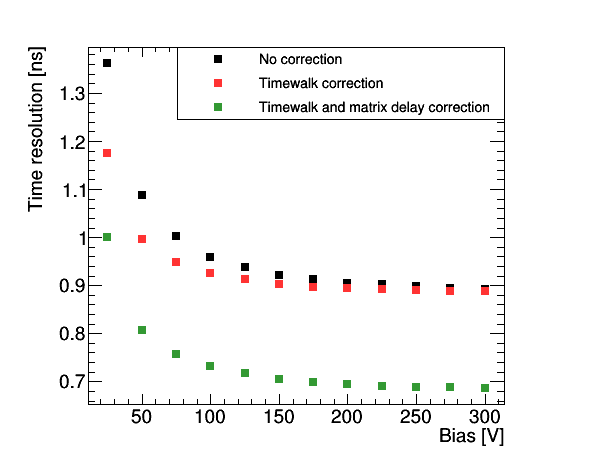}
\end{minipage}\qquad
\begin{minipage}{.47\textwidth}
\centering
\includegraphics[width=1\linewidth]{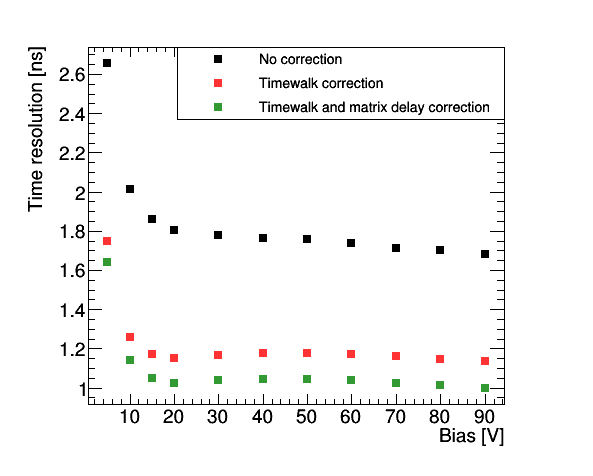}
\end{minipage}


\begin{minipage}[t]{.47\textwidth}
\centering
\captionsetup{width=.9\linewidth}
\captionof{figure}{The timer resolution of W0020J07 for a range of bias voltages. The top curve indicates the time resolution without any correction, the middle curve indicates the time resolution after a timewalk correction, and the bottom curve indicates the time resolution after a timewalk correction and a correction for the per-pixel difference in the average delay. }
\label{fig:Results/TimeRes200}
\end{minipage}\qquad
\begin{minipage}[t]{.47\textwidth}
\centering
\captionsetup{width=.9\linewidth}
\caption{The timer resolution of W0039I11 for a range of bias voltages. The top curve indicates the time resolution without any correction, the middle curve indicates the time resolution after a timewalk correction, and the bottom curve indicates the time resolution after a timewalk correction and a correction for the per-pixel difference in the average delay. }
\label{fig:Results/TimeRes50}
\end{minipage}
\end{figure}

During the testbeam, two bias scans, one for W0020J07 and one for W0039I11, have been performed. These bias scans consists of increasing the bias voltage in steps in between different runs such that the time resolution can be determined for each bias voltage. These two scans are shown in \autoref{fig:Results/TimeRes200} for W0020J07, and in \autoref{fig:Results/TimeRes50} for W0039I11. In these two Figures the time resolution of a run without any correction, with an offline timewalk correction, and with an offline timewalk correction and a per-pixel correction of the difference in delay are shown. For both detectors increasing the bias voltage consistently resulted in an increase of the timing precision. This is due to the increasing magnitude of the electric field within the silicon, resulting in a faster collection of the charge carriers. However, when the bias voltage is increased further the charge carriers are reaching their saturation velocity and thus the increase of the bias voltage does not significantly increase the timing performance further. One can also see that applying solely applying a timewalk correction to W0020J07 at a high voltage does not increase the timing performance significantly, while for W0039I11 applying a timewalk correction increases the time resolution significantly. This is due to the on average lower deposited charge in the thinner sensor (W0039I11). Only at low charge, a timewalk correction can increase the time resolution due to the higher delay induced by timewalk. One can also see that a per-pixel correction of the difference in delay consistently increases the time resolution by around 150-200 \si{\pico\second}, indicating that, besides timewalk, the per-pixel differences result in a large decrease of the timing performance of Timepix3. 

The best time resolution that is achieved is $686.4\pm0.2$ \si{\pico\second} with W0020J07 at a bias voltage of \SI{300}{\volt}. This time resolution is not yet naively expected \SI{451}{\pico\second} based on the size of the time bins, indicating that the time resolution should still be improved further.

\section{Cross-talk between pixels}

\begin{figure}
\centering
\begin{minipage}{.47\textwidth}
\centering
\includegraphics[width=1\linewidth]{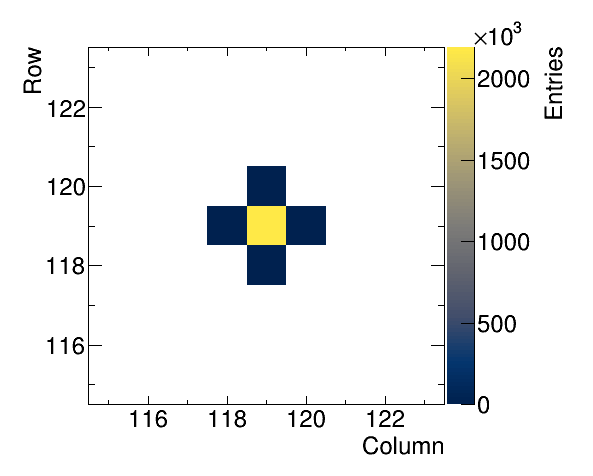}
\end{minipage}\qquad
\begin{minipage}{.47\textwidth}
\centering
\includegraphics[width=1\linewidth]{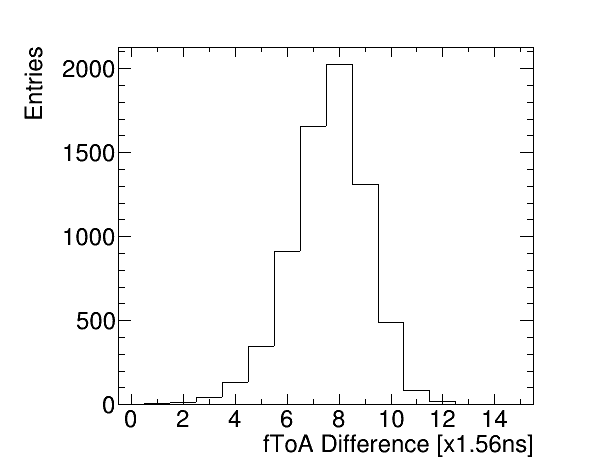}
\end{minipage}


\begin{minipage}[t]{.47\textwidth}
\centering
\captionsetup{width=.9\linewidth}
\captionof{figure}{Number of hits versus the position of the hits for a focused laser spot ($\sigma=$ $6.79\pm0.09$ \si{\micro\meter}).}
\label{fig:Results/CrossTalkHitmap}
\end{minipage}\qquad
\begin{minipage}[t]{.47\textwidth}
\centering
\captionsetup{width=.9\linewidth}
\caption{The delay distribution between the main hit and the low ToT hits surrounding the main hit.}
\label{fig:Results/CrossTalkTimeDifference}
\end{minipage}
\end{figure}

During the laser measurements it became evident that it was more difficult than expected to focus the laser beam to a single pixel. Only by ensuring that the charge generated in the sensor was below 30,000 electrons, a single pixel could be hit. When the main hit had a charge higher than that, the neighbouring pixels started to measure hits as well for just a fraction of the main hits. An example of this behaviour is shown in \autoref{fig:Results/CrossTalkHitmap}. In this Figure, the laser spot is aligned in the centre of pixel (119,119). Therefore, all the laser pulses are measured in this pixel, while just a fraction of these laser pulses are measured in the surrounding pixels. 

After further investigation into the collected charge of these hits, it was noticed that a single pixel collected all the charge, while the pixels surrounding this pixel only had hits with a ToT of 1. By increasing the charge deposited in this central pixel, the charge collected in the surrounding pixels does not increase, while if these hits are induced by the charge generated by the laser, the number of electrons that are collected should be increasing as well. A ToT of 1 also indicates that the time-over-threshold can be anywhere from \SI{0}{\nano\second} to \SI{50}{\nano\second}, because even if the time-over-threshold is just \SI{1}{\nano\second}, the time at which it was over threshold can still coincide with the rising edge of the system clock and thus be counted as ToT 1 (as described in \autoref{sec:Timepix3/ToAToT}). This specific value of ToT combined with the lower fraction of hits in these pixels therefore could indicate a time-over-threshold that is below \SI{25}{\nano\second}.

Another observation is the time between the main hit and the hits with ToT of 1. This time difference is on average eight fToA bins (see \autoref{fig:Results/CrossTalkTimeDifference}), corresponding to \SI{12.5}{\nano\second}, while the low ToT hits should suffer more from timewalk than \SI{12.5}{\nano\second} (see \autoref{fig:Results/TimewalkCurve50um} for an indication of the time delay from timewalk). Therefore, these hits seem to not be induced by the charge carriers generated in the detector. This led to the belief that these hits are induced via some sort of mechanism in the detector. One of the explanations is the induction of charge in the neighbouring pixels due to the movement of the main charge cloud within the weighting field inside the detector. However, this theory was quickly rejected after it became apparent that this effect is still visible when the charge carriers are injected directly inside the analogue front-end using test pulses. Therefore, the hits in the neighbouring pixels should be generated by a process that is independent of the charge moving through the silicon.

A second possible explanation could be capacitive coupling between the various electronic components in the Timepix3 ASIC. Therefore, test pulses are used to inject a charge in the analogue electronics of a single pixel on a Timepix3 ASIC that does not have the silicon detecting layer attached to it. Using test pulses a maximum number of 22,500 electrons is injected. However, below and at this number of electrons there is no sign of capacitive induction in the surrounding pixels. To further exclude capacitive coupling within the ASIC as a possibility to induce the low ToT hits in a normal measurement, all eight pixels surrounding one central pixel are pulsed with test pulses. The idea behind this is that capacitive coupling is additive for different components. Therefore, the eight different pixels should produce at least four times the capacitive coupling (four pixels at a distance of one pixel and four pixels at a distance of $\sqrt{2}$ pixels). Even with this configuration it was not possible to reproduce the low ToT hits that were observed during the laser measurement. Therefore, the capacitive coupling of the ASIC is excluded as a possible origin of the low ToT hits.

\begin{figure}
\centering
\begin{minipage}{.47\textwidth}
\centering
\includegraphics[width=1\linewidth]{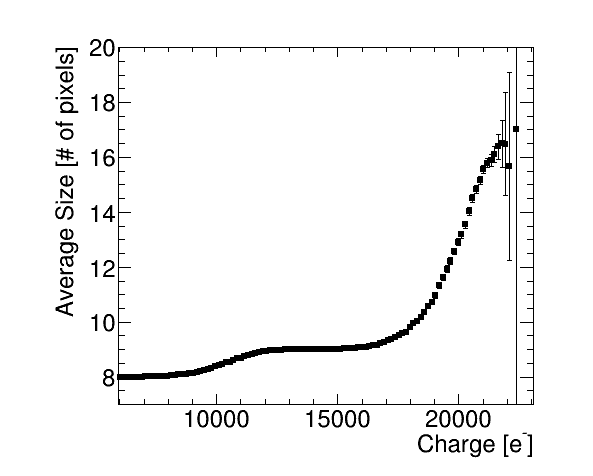}
\end{minipage}\qquad
\begin{minipage}{.47\textwidth}
\centering
\includegraphics[width=1\linewidth]{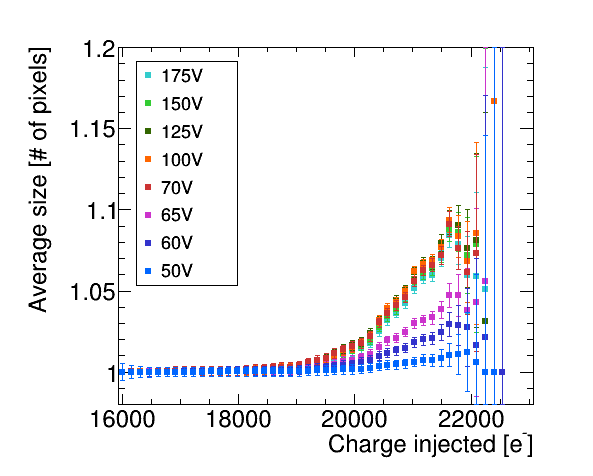}
\end{minipage}


\begin{minipage}[t]{.47\textwidth}
\centering
\captionsetup{width=.9\linewidth}
\captionof{figure}{The average cluster size for eight pixels that are pulsed with test pulses at different charges. The horizontal axis only shows the charge for a single pixel of the eight.}
\label{fig:Results/CrossTalk8Pixels}
\end{minipage}\qquad
\begin{minipage}[t]{.47\textwidth}
\centering
\captionsetup{width=.9\linewidth}
\caption{The average cluster size for a single pixel that is pulsed with test pulses for a range of bias voltages on W0020J07 (depletion voltage of \SI{115}{\volt}).}
\label{fig:Results/CrossTalkEfield}
\end{minipage}
\end{figure}

A third possibility still remains, the hits can be induced by the capacitive induction induced by the build-up of charge before the amplifier in the analogue front-end of the pixel. This build-up of charge, and thus voltage, is therefore also present in the implant in the sensor. This explains why this effect is not visible when a bare sensor is pulsed with test pulses. To further confirm if this is the cause of the low ToT hits, a test pulse scan with different number of charges is made. For this scan eight pixels surrounding one central pixel are pulsed again. For each different charge the average size of the cluster is determined. The result is shown in \autoref{fig:Results/CrossTalk8Pixels}. In this Figure, the average charge of one of the eight pixels is plotted against the average size of the cluster of pixels that is hit. One can see in this Figure that at first the only pixels that register a hit are the eight pixels that are pulsed. From around 10,000 electrons till 12.500 electrons the average length of the coupled hits grows from \SI{0}{\nano\second} to \SI{25}{\nano\second}. From that point on the coupled hits in the centre pixel is always measured and grows further in ToT magnitude. From 15,000 electrons, the ring of pixels surrounding the eight pixels starts to experience coupled hits as well. However, these are only the twelve pixels that have a distance of 1 to the closest pixels that is pulsed. The other four pixels in the corners, are further away from the pulsed pixels and thus require more charge in the pulsed pixels before they as well experience capacitively coupled hits. However, the charge injected with test pulses lacks the magnitude to induce hits in these corner pixels. Therefore, the curve shown in \autoref{fig:Results/CrossTalk8Pixels} eventually saturates at $8+1+12=21$ pixels.

\begin{figure}
\centering
\includegraphics[width=.6\linewidth]{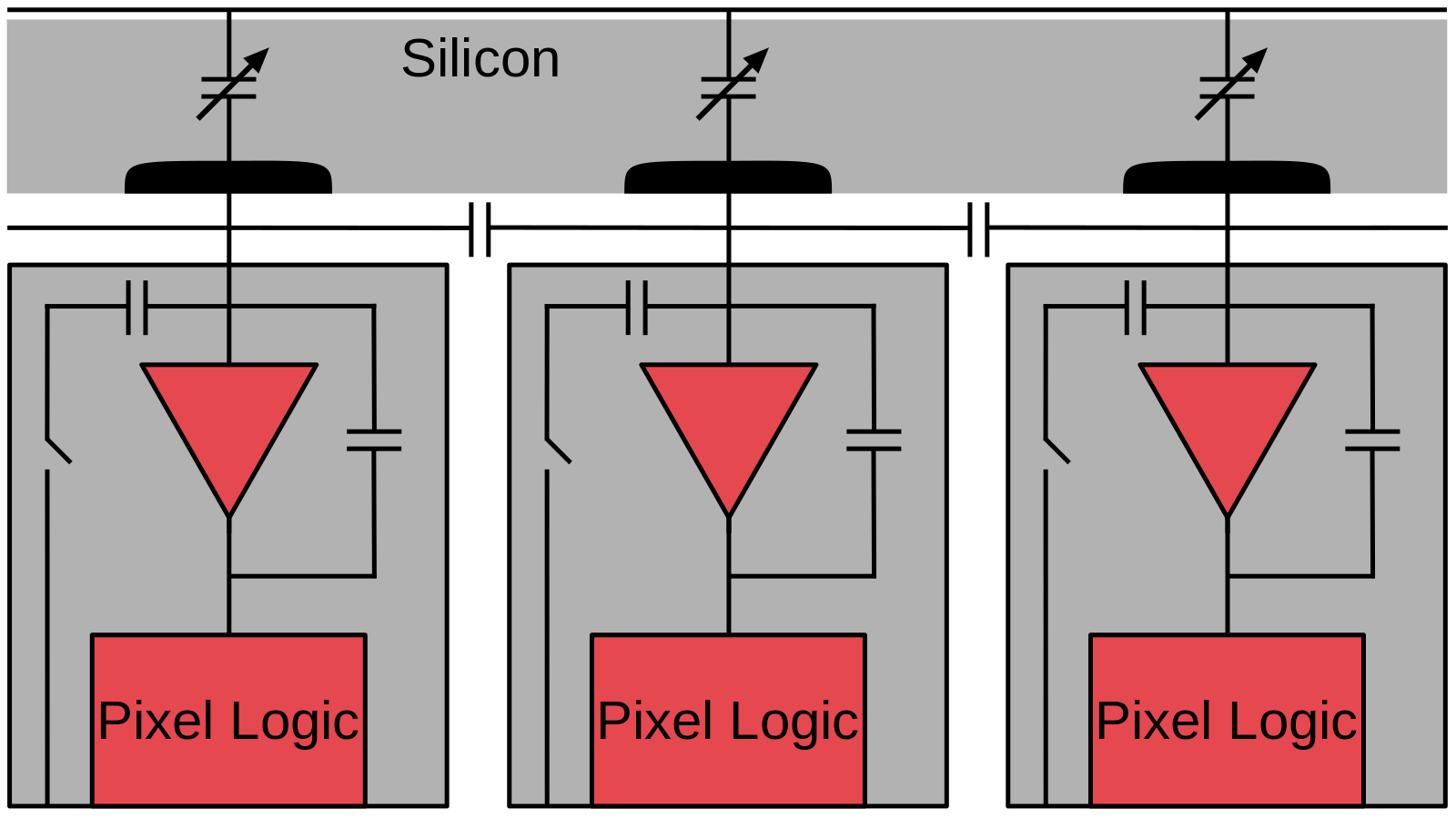}
\captionof{figure}{A schematic diagram of the various capacitors that are in the electronic circuit directly attached to the input pad of the Timepix3. }
\label{fig:Results/CrossTalkCapacitance}
\end{figure}

By looking at how the fraction of coupled hits behaves when the electric field within the silicon is changed, a further indication of the origin of these coupled hits can be obtained. By changing the bias voltage of the sensor, the magnitude of the electric field within the sensor changes as well. This in turn changes the capacitance between the back of the sensor, which is connected to the bias supply, and the implants. Therefore, the only capacitance within the circuitry that changes when the bias voltage is changed is the coupling to the back of the sensor. The location of the various capacitors in the analogue front-end as well as the capacitance between the implant and the back of the sensor is illustrated in \autoref{fig:Results/CrossTalkCapacitance}. The wire connecting to just before the amplifier, and going next to the pixel logic, is the test pulse injection wire. The capacitive coupling between the individual pixels is also indicated in this Figure. This capacitance is parasitic capacitance between the various electrical components, between the bump bonds, and between the implants of the pixels. 

For a range of bias voltages, a single pixel is fired four million times at different intensities. For these test pulses the injected charge varied between 15.000 and 22.500 electrons. For each different injected charge, the average size of the cluster that measured a hit is determined. This size should be at least one, because in every case at least the pixel at which the test pulse is injected should measure a hit. The result is shown in \autoref{fig:Results/CrossTalkEfield}. One can see from these results that when the bias voltage decreases, the capacitive coupling to the neighbouring pixels also decreases. Because the coupling to the neighbouring pixels decreases as the bias voltage is decreased, but the capacitance between the neighbouring pixels is constant. Therefore, the capacitance between the implant and the back of the sensor must be increasing. When the capacitance between the implant and the back of the sensor is relatively high compared to that of the coupling between the electronics, the current generated by the difference in voltage between the pixel and the back of the sensor (or the neighbouring pixels) is generated more easily when the capacitance is higher. Therefore, the current will be generated at the back of the sensor, instead of at the neighbouring pixels. However, when the capacitance between the implant and the back of the sensor is lowered, it becomes easier to generate a current at the neighbouring pixels instead of at the back of the sensor. This current can then be registered as a hit, when the charge crosses the threshold. The capacitance of the silicon sensor scales as $V^{-1/2}$, and thus explains the increasing coupling to the neighbouring pixels when the bias voltage $V$ is increased. 

The generated current in the neighbouring pixels, is generated by capacitive coupling. This implies that there is no net charge transferred between the pixels. When the charge generated in the sensor arrives at the implant, the voltage for that pixel will rise, inducing a positive current in the neighbouring pixels. After the charge from the hit starts to leave the analogue front-end of the pixel, the voltage starts to decrease, the current in the neighbouring pixels will be negative, thus in total not depositing any charge within the neighbouring pixels. 

This bipolar pulse also explains the relatively quick time between the main hit and the low ToT hits. Due to the bipolar pulse, the ToT is not an indication for the charge in the pixel. The build-up of the voltage for these low ToT hits, is most likely high compared to what is expected from the ToT. Therefore, the threshold is crossed quicker than what is expected from the measured ToT, thus explaining the relatively quick time between the initial hit and the coupled hits that were observed (see \autoref{fig:Results/CrossTalkTimeDifference}).

Such a hit should not be included in the total charge of a cluster in an analysis such as the one applied by Kepler (see \autoref{sec:Testbeam/Kepler}). By including the ToT of these capacitively coupled hits, the total charge of a cluster will be larger than the actual charge of the cluster liberated by a particle. Therefore, one should first determine the charge of the individual pixel, and when this charge exceeds 20,000 electrons, the 1 ToT hits should be disregarded during the clustering sequence.

\chapter{Conclusions and outlook}
\label{sec:Conclusion}

With the increasing performance of silicon hybrid pixel detectors, there is a growing interest in quantifying the systematics of the electronics of the pixel matrix of such detectors. In this thesis the systematics of the Timepix3 have been studied. For this analysis test pulses, testbeam, and laser beam data was used. For the testbeam experiment the Timepix3 telescope that was placed at the SPS beam at CERN was used. The laser system that was used has been assembled and designed for these studies. It is envisioned that it will also be used for later analysis of other pixel detectors.

The laser setup is a more easily accessible method to measure the systematics of a silicon pixel detector compared to the testbeam method. This laser setup consists of only a few components which makes it a \textit{table-top} setup, and can easily be employed in any lab, while the testbeam method relies on an accelerator complex which is only available at specific locations. The laser method is also more versatile compared to the testbeam method, by having the freedom to change the intensity as well as the time of arrival of the photons.

Both methods proved to provide similar results for the average delay of the pixel matrix of Timepix3 for a single chip. A slight difference is observed between the two methods, however this is most likely due to the different operating conditions and the limited timing precision of the laser diode that is currently used. Therefore, a further improvement on the current laser setup at Nikhef can be achieved by switching to a more precise laser diode, or replacing the current optical system with a dedicated two-photon absorption setup. Such a setup will provide a smaller spot size and utilizes a laser that has a pulse length of less than a picosecond, decreasing the pulse length significantly compared to the current laser.

The delay of the pixel matrix is determined to consist of two parts, a global and a local delay. The global delay, that shows variations in the order of \SI{2}{\nano\second}, is dominated by power distribution and signal propagation time differences between different parts of the pixel matrix. The local delay, that shows variations in the order of \SI{3}{\nano\second}, is dominated by local effects of the electronics at the pixel and super-pixel level. The variations are large compared to the naively expected time resolution of \SI{451}{\pico\second} and should therefore be corrected for.

The origin of the local delay of the pixels of Timepix3 is determined to be due to the difference in start-up time of the fast oscillator, as well as a frequency shift of the fast oscillator over the pixel matrix. These two effects combined also result in the fifteenth fToA bin never being hit for some pixels. This is confirmed by delay scans of the fToA bins for specific pixels with the laser setup. It is also observed that the timing jitter on the laser is clearly visible in these delay scans of the fToA bins. Therefore, if the time bins in future silicon pixel detectors, such as Timepix4, will be smaller than the current \SI{1.56}{\nano\second}, this jitter will obstruct the possibility to consistently inject a charge within one time bin. This will not influence the capability of determining the average pixel delay, due to the possibility to determine this delay using the average delay.

After applying a correction for the global and local delay structures we have achieved a timing resolution of $686.4\pm0.2$ \si{\pico\second}. This is slightly worse than the naively expected time resolution of \SI{451}{\pico\second}. Further improvements are suggested in this work but are unfeasible due to the estimated time it would take to derive a proper calibration.

An investigation on the start-up time and conditions of the fast oscillator employed by the Timepix3 is performed as well. This investigation focuses on the situation where an equal charge is injected in multiple pixels at the same moment in time. This confirmed that the difference in the local delay of the pixels is determined by the difference in start-up time of the fast oscillator. This also confirmed that when two different pixels are hit within the same super-pixel in the same period of the system clock, the slowest pixel encounters an already running fast oscillator and thus changes the timing behaviour of the second hit. In further applications of Timepix3, if precise timing is necessary, it should be noted that the timestamp has to be corrected for this effect. 

During the investigation on the origin of the delay of the pixel matrix, capacitively coupled hits have been observed in Timepix3. The origin of these coupled hits is due to the parasitic capacitance between neighbouring pixels. When the voltage of a neighbouring pixel is high enough, a current can be induced in the pixel itself, while it does not collect any charge from the silicon itself. These hits can be measured in pixels surrounding a pixel that collects more than 20,000 electrons, though the precise charge depends on the applied bias voltage. Therefore, when working with a high number of charge carriers with Timepix3, one should be aware that capacitively coupled hits could occur. These hits, if not spotted, can influence for example the total charge of a cluster in the offline analysis of a measurement.

To summarise, three methods are used to investigate the systematics of a silicon pixel detector, the Timepix3. The first method relies on the use of test pulses, the second method uses a testbeam facilities to investigate the Timepix3 using charged particles, and the third method relies on a laser setup that we build. Both the testbeam and the laser proved to provide similar results and can be used in the future to quantify the systematics of future silicon pixel detectors such as Timepix4. Compared to the testbeam method, the laser method provides the possibility to inject charge carriers at a specific time, which gives the option to investigate additional systematics of the pixel matrix of Timepix3.

At the beginning of 2020 the Timepix4 readout chip will become available. This chip will have a reduced fToA bin of \SI{200}{\pico\second}, with respect to the current time bin size of \SI{1.56}{\nano\second} in the Timepix3. The current laser setup will not have a sufficient timing performance to be able to perform similar detailed studies on the Timepix4 chip. Hence, the laser setup will need to be improved. This can be achieved by a decrease of the pulse length of the laser in order to decrease the timing jitter. Besides a decrease in the pulse length, a decrease in the current timing jitter of the laser system due to signal propagation in cables and the generation of the control voltage of the laser can be achieved by switching to instrumentation with a quicker rise time.

\appendix

\sloppy
\printbibliography[heading=bibintoc]
\fussy

\end{document}